\newcommand{\MSbar}{\overline{\rm MS}}
\newcommand{\deltaMSbar}{\delta_{\overline{\rm MS}}}
\newcommand{\DRbar}{\overline{\rm DR}}
\newcommand{\DRbarprime}{\overline{\rm DR}'}
\newcommand{\lnbar}{{\overline{\rm ln}}}
\newcommand\beq{\begin{eqnarray}}
\newcommand\eeq{\end{eqnarray}}
   
\newcommand{\propB}{{\rm B}}

\newcommand{\propM}{{\rm M}}
\newcommand{\propV}{{\rm V}}

\newcommand{\propY}{{\rm Y}}

\newcommand{\Fbar}{{\overline F}}
\newcommand{\sigmabar}{\overline\sigma}

\def\newcdot{\kern.06em{\cdot}\kern.06em}

\documentclass[twocolumn,%showpacs,preprintnumbers,
amsmath,%amssymb,aps,
prd,nofootinbib,floatfix%,preprint
]{revtex4}

\allowdisplaybreaks
\usepackage{axodraw}
\usepackage{graphicx}% Include figure files
\usepackage{bm}% bold math

\begin{document}
\renewcommand{\theequation}{\arabic{section}.\arabic{equation}}

\title{Fermion self-energies and 
pole masses at two-loop order in a general
renormalizable theory with massless gauge bosons}

\author{Stephen P. Martin}
\affiliation{
Physics Department, Northern Illinois University, DeKalb IL 60115 USA\\
{\rm and}
Fermi National Accelerator Laboratory, PO Box 500, Batavia IL 60510}

%DRAFT: \today

\phantom{.}
% The previous line unfortunately must be present to counteract a
% revtex4 bug. Don't ask me why!

\begin{abstract} I present the two-loop self-energy functions and pole
masses for fermions in an arbitrary renormalizable field theory, in the
approximation that vector bosons are treated as massless. The calculations
are done simultaneously in the mass-independent $\MSbar$, $\DRbar$, and
$\DRbarprime$ renormalization schemes, with a general covariant gauge
fixing, and treating Majorana and Dirac fermions in a unified way.  As
examples, I discuss the two-loop strong interaction corrections to the
gluino, neutralino, chargino, and quark pole masses in minimal
supersymmetry.  All other two-loop contributions to the fermion pole
masses in softly-broken supersymmetry can also be obtained as special
cases of the results given here, neglecting only the electroweak symmetry
breaking scale compared to larger mass scales in two-loop diagrams that
involve $W$ or $Z$ bosons.

\end{abstract}

%\pacs{}

\maketitle

%% The arXiv's use of hypertex conflicts with revtex4's use of
%% \tableofcontents in single column format. To avoid this problem,
%% include a file OOREADME.XXX with the word nohypertex in it when
%% you submit to the arXiv. 

\tableofcontents

\section{Introduction}\label{sec:introduction}
\setcounter{equation}{0}
\setcounter{footnote}{1}

The CERN Large Hadron Collider and a future electron-positron linear
collider should discover and, together, thoroughly explore
\cite{LHCILC} the mechanism behind electroweak symmetry breaking. The
small ratio of the scale of electroweak symmetry breaking to the Planck
mass scale suggests that supersymmetric particles will also be found at
these next-generation experiments. 
If so, then a primary goal of both
experimental and theoretical research will be to unravel the mechanism
behind supersymmetry breaking. The most important clues will be the masses
of the superpartners and the Higgs scalar bosons.  Therefore it is
important to be able to compute the physical masses accurately in terms of
the underlying Lagrangian parameters, including at least the leading
two-loop effects.

In this paper, I will present results for the two-loop contributions to
fermion self-energy functions and physical pole masses in a general
renormalizable field theory, in terms of the running renormalized
couplings and masses. The approach used is intended to be as flexible as
possible, so that a common framework of calculation can be used to treat
both Majorana and Dirac fermions, including chiral interactions, in both
supersymmetric and non-supersymmetric theories. As a simplifying
approximation, vector bosons will be treated as massless in the two-loop
parts in this paper.  In the Standard Model and extensions of it that do
not enlarge the gauge group, this amounts to neglecting the effects of 
electroweak symmetry breaking compared to the masses of heavier particles
in two-loop diagrams that have $W$ and/or $Z$ boson propagators. (The 
effects of non-zero $W$ and $Z$ boson
masses can be included as usual in the one-loop part.)  This will likely
be a good approximation for the pole masses of the top quark and most of
the supersymmetric particles, because of the exclusions of light squarks,
sleptons, and gluinos already achieved by the CERN LEP $e^+e^-$ collider
\cite{LEPSUSY} and the Fermilab Tevatron $p\overline p$ collider
\cite{squarksDzero,squarksCDF}.

The mass defined by the position of the complex pole in the propagator is
a gauge-invariant and renormalization scale-invariant quantity
\cite{Tarrach:1980up}-\cite{Gambino:1999ai}.  The pole mass in principle
does suffer from ambiguities \cite{poleambiguities} due to infrared
physics associated with the QCD confinement scale, but these are probably
relatively too small to cause a practical problem for strongly-interacting
superpartners. The pole mass should be closely related in a calculable way
to the kinematic observable mass reported by experiments \cite{massdefs}.  
In recent years, many important higher-order calculations of self-energy
functions and pole masses in the Standard Model have been performed,
including two-loop \cite{Gray:1990yh}-\cite{Jegerlehner:2003py} and
three-loop \cite{Chetyrkin:1999qi}-\cite{Melnikov:2000qh} contributions
for quarks, and two-loop results for electroweak vector bosons
\cite{Chang:1981qq}-\cite{Jegerlehner:2001fb}. In addition, there are
important two-loop results for top and bottom quarks
\cite{quarkpoleSUSYa,quarkpoleSUSYb,Bednyakov:2005kt} and the gluino
\cite{Yamada:2005ua} in low-energy supersymmetry.  The general treatment
of the present paper will confirm and extend the results of those papers.

The notation and strategy used here are very similar to those found in my
previous papers on scalar self-energy functions and pole masses at
two-loop order in a general theory \cite{Martin:2003it,Martin:2005eg}. In
the next section, I review the conventions used, discuss the formalism for
self-energy functions and pole masses for fermions in a two-component
notation, and review the methods used for numerical evaluation of the
required two-loop integrals.

\section{Notations and setup}\label{sec:notations}
\setcounter{equation}{0}
\setcounter{footnote}{1}

\subsection{Notations for fields, interactions, and indices} 

In this paper, the spacetime metric tensor is
\beq
\eta^{\mu\nu} = \mbox{diag}(-1,+1,+1,+1) .
\eeq
I use a two-component notation for fermions, as in 
ref.~\cite{Martin:1997ns}
and similar to that found in ref.~\cite{WessBaggerbook}. Left-handed 
spinor
fields $\psi_{\alpha}$ always carry undotted spinor indices 
$\alpha,\beta,\ldots 
= 1,2$, and 
right-handed
spinor fields $\psi^{\dagger}_{\dot\alpha}$ always carry daggers 
and dotted spinor 
indices $\dot\alpha,\dot\beta,\ldots
= 1,2$, with
\beq
\psi^{\dagger}_{\dot\alpha} \equiv
(\psi_{\alpha})^\dagger .
\eeq
However, the spinor indices are most often suppressed, as described 
below.
The spinor indices are raised and lowered
with the two-index antisymmetric symbol with components
$\epsilon^{12} = - \epsilon^{21} = \epsilon_{21} = -\epsilon_{12} = 1$,
and the same set of sign conventions for the corresponding dotted
spinor indices. Thus
\beq
&&\psi_\alpha = \epsilon_{\alpha\beta} \psi^\beta\,, \qquad
\psi^\alpha = \epsilon^{\alpha\beta} \psi_\beta\,,  
\\
&&\psi^\dagger_{\dot{\alpha}} = \epsilon_{\dot{\alpha}\dot{\beta}}
\psi^{\dagger\dot{\beta}}\,, \qquad
\psi^{\dagger\dot{\alpha}} = \epsilon^{\dot{\alpha}\dot{\beta}}
\psi^{\dagger}_{\dot{\beta}}\,.
\eeq
Spinor bilinears can be combined to form vector quantities using the
matrices
$\sigma_{\mu\,\alpha{\dot{\beta}}}$
and $\sigmabar_\mu^{\dot{\alpha}\beta}$ defined by
\beq
&&
\!\!\!\!\!\!\!\!\!\!
\sigmabar_0 = \sigma_0 = \begin{pmatrix}1&0\\ 0&1\end{pmatrix},
\qquad
\sigmabar_1 = -\sigma_1 = \begin{pmatrix}0&1\\ 1&0\end{pmatrix},
\nonumber
\\
&&
\!\!\!\!\!\!\!\!\!\!
\sigmabar_2 = -\sigma_2 = \begin{pmatrix} 0&-i\\ i&0\end{pmatrix},
\quad
\sigmabar_3 = -\sigma_3 = \begin{pmatrix}1&0\\ 0&-1\end{pmatrix}
.
\phantom{x}
\label{eq:pauli}
\eeq
When constructing Lorentz tensors from fermion fields, the heights of
spinor indices must be consistent in the sense that lowered
indices must only be contracted with raised indices.
As a convention, indices contracted like
$
{}^\alpha{}_\alpha
$
and
$
{}_{\dot{\alpha}}{}^{\dot{\alpha}} ,
$
can be suppressed. For example,
\beq
\xi\chi &\equiv & \xi^\alpha\chi_\alpha ,
\\
\xi^\dagger \chi^\dagger &\equiv & 
\xi^\dagger_{\dot\alpha} \chi^{\dagger\dot \alpha}
,
\\
\xi^\dagger\sigmabar^\mu\chi &\equiv &  \xi^\dagger_{\dot{\alpha}}
\sigmabar^{\mu\dot{\alpha}\beta}\chi_\beta ,
\\
\xi\sigma^\mu \chi^\dagger &\equiv & \xi^{{\alpha}}
\sigma^{\mu}_{\alpha \dot \beta} \chi^{\dagger\dot \beta} .
\eeq
The behavior of the spinor products under hermitian
conjugation (for quantum field operators) or complex conjugation (for
classical fields) is as follows:
\beq
&&(\xi\chi)^\dagger= \chi^\dagger\xi^\dagger ,
\label{eq:conbil}
\\
&&(\xi\sigma^\mu\chi^\dagger)^\dagger=\chi\sigma^\mu\xi^\dagger ,
\label{eq:conbilsig}
\\
&&(\xi^\dagger \sigmabar^\mu \chi)^\dagger = \chi^\dagger 
\sigmabar^\mu \xi .
\label{eq:conbilsigbar}
\eeq
The following identities also hold:
\beq
&&{[\sigma^\mu\sigmabar^\nu + \sigma^\nu \sigmabar^\mu ]_\alpha}^\beta
= -2\eta^{\mu\nu} \delta_{\alpha}^{\beta} ,
\label{eq:ssbarsym}
\\
&&[\sigmabar^\mu\sigma^\nu + \sigmabar^\nu \sigma^\mu
]^{\dot{\alpha}}{}_{\dot{\beta}}
= -2\eta^{\mu\nu} \delta^{\dot{\alpha}}_{\dot{\beta}} ,
\label{eq:sbarssym}
\\
&&{\rm Tr}[\sigma^\mu \sigmabar^\nu ] =
{\rm Tr}[\sigmabar^\mu \sigma^\nu ] = -2 \eta^{\mu\nu} .
\label{trssbar}
\eeq
In terms of two-component fermion notation, a single Dirac fermion
is given in the chiral representation by 
\beq
\Psi = \begin{pmatrix} \xi_\alpha \\ \chi^{\dagger\dot\alpha}
       \end{pmatrix},
\eeq
where $\xi$ is the two-component
fermion that describes the left-handed part of $\Psi$ and $\chi$ is 
the two-component fermion that describes the
conjugate of the right-handed part of $\Psi$. The Dirac matrices
are
\beq
\gamma_\mu = \begin{pmatrix} 0 & \sigma_\mu \\ \sigmabar_\mu & 0
\end{pmatrix}
.
\eeq

In this paper, I consider a general renormalizable field theory, 
containing\footnote{A complex scalar can be written as two real scalars,
and a Dirac fermion as two Weyl fermions, so this entails no loss of 
generality.}
a set of real scalars $R_i$, two-component
Weyl fermions $\psi_I$, and vector bosons $V^\mu_a$. 
Scalar field indices
are $i,j,k,\ldots$, fermion flavor indices are $I,J,K,\ldots$, and
$a,b,c,\ldots$ run over the adjoint representation of the gauge group,
while $\mu,\nu,\ldots$ are space-time vector indices.
Repeated indices of all types are summed over unless otherwise noted.

The masses and couplings are evaluated by taking the fields in the
Lagrangian in a squared-mass eigenstate basis, after the Higgs fields are
assumed to have been expanded around their vacuum expectation values as
determined by the loop-corrected effective potential (so that tadpole
graphs do not contribute). The kinetic part of the renormalized tree-level
Lagrangian is then written as:
\beq
{\cal L}_{\rm kin} &=&
-\frac{1}{2} \partial_\mu R_i \partial^\mu R_i
- \frac{1}{2} m^2_i R_i^2
\nonumber \\
&&
-i\psi^{\dagger I} \overline \sigma^\mu \partial_\mu \psi_I -
\frac{1}{2} (m^{IJ} \psi_I \psi_J + {\rm c.c.})
\nonumber \\
&&
-\frac{1}{2} (\partial_\mu V_\nu^a - \partial_\nu V_\mu^a)
  \partial^\mu V^{\nu}_a 
-\frac{1}{2} m^2_a V_\mu^a V^\mu_a
.\phantom{xxx}
\eeq
The non-gauge interactions of the scalar and fermion fields are given by
the renormalized Lagrangian:
\beq
{\cal L}_{\rm int} &=&
-\frac{1}{6} \lambda^{ijk} R_i R_j R_k
-\frac{1}{24} \lambda^{ijkm} R_i R_j R_k R_m
\nonumber \\
&&
-\frac{1}{2} (y^{JKi} \psi_J \psi_K R_i + {\rm c.c.}).
\eeq
where $\lambda^{ijk}$ and $\lambda^{ijkm}$ are real couplings and the 
Yukawa
couplings $y^{JKi}$ are symmetric complex matrices on the indices $J,K$,
for each $i$. Raising or lowering of fermion indices implies
complex conjugation of the Lagrangian parameters, so
\beq
m_{IJ} \equiv (m^{IJ})^*,\qquad y_{JKi} \equiv (y^{JKi})^* .
\eeq
Actually, without loss of generality, $m^{IJ}$ can be taken to have only
real and non-negative entries, but the index height convention is
maintained for clarity. The heights of real scalar and vector indices
have no significance, and in any given equation are chosen for
convenience.

The scalar squared masses $m_i^2$ and the fermion squared masses $m_{IK}
m^{KJ} = m^2_I \delta_I^J$ are taken to have been diagonalized (by an
appropriate rotation of the fields if necessary). However, the fermion
mass matrix $m^{IJ}$ is not necessarily diagonal; instead it must have
non-zero entries only when $I$ and $J$ label 
two-component 
fermions with
the same squared mass and in conjugate representations of the 
unbroken gauge group.
In particular, when dealing with Dirac fermions, it is most useful to work
in a basis in which the corresponding matrix $m^{IJ}$ contains $2\times 2$
blocks of the form $\begin{pmatrix} 0 & m\cr m & 0 \end{pmatrix}$ on the
diagonal.

Next consider the gauge interactions of the theory.
Let $T^a$ be the Hermitian generator matrices
of the gauge group for a (possibly reducible) representation $R$.
They are labeled by an adjoint representation index $a$ corresponding
to the vector bosons of the theory, $V_a^\mu$.
They satisfy $[T^a, T^b] = i f^{abc} T^c$, where $f^{abc}$ are
the totally antisymmetric structure constants of the gauge group.
Then results below are written in terms of the invariants:
\beq
{(T^a T^a)_i}^j &=& C(i) \delta_i^j,\\
{\rm Tr}[T^a T^b] &=& I(R) \delta^{ab},\\
f^{acd} f^{bcd} &=& C(G) \delta^{ab},
\eeq
which define the quadratic Casimir invariant for the representation
carrying the index $i$, the total Dynkin index summed over the
representation $R$, and the Casimir invariant of the adjoint 
representation
of the group,
respectively. When the gauge group contains several simple or $U(1)$
factors with distinct gauge couplings $g_a$,
the corresponding invariants are written $C_a(i)$, $I_a(R)$, and
$C_a(G)$. The normalization is such that for $SU(N)$, $C(G) = N$ and each
fundamental
representation
has $C(i) = (N^2-1)/2N$ and contributes $1/2$ to $I(R)$ for each Weyl 
fermion or complex scalar.
For a $U(1)$ gauge group, $C(G) = 0$ and a representation with charge $q$
has $C(i) = q^2$
and contributes $q^2$ to $I(R)$.
The two-loop results given below will be presented
in terms of these group theory invariants for the representations carried
by the scalar and fermion degrees of freedom.

The preceding paragraph
applies to the two-loop parts, in which the gauge group
is treated as unbroken and $m_a^2 = 0$.
In the one-loop parts of the self-energy functions and the fermion pole
masses, the effects of non-zero vector boson masses will be retained.
This means that the gauge group cannot be treated as unbroken,
and the interactions of the vector bosons with
the fermions have a more general form. They
can be written as:
\beq
{\cal L}_{\rm gauge} = 
-g^{aJ}_I V_\mu^a \psi^{\dagger I} \overline\sigma^\mu \psi_J 
\eeq
where 
$g^{aJ}_I$ are couplings obtained by going to the tree-level mass 
eigenstate
basis for the fermions and vector bosons.
In the special case of an unbroken gauge symmetry, one has
$g^{aJ}_I = g_a [T^a]_I{}^J$.

The computations in this paper are performed with a
vector boson propagator obtained by covariant
gauge fixing in the usual way:
\beq
-i\delta_{ab} \bigl ( 
\eta^{\mu\nu} + k^\mu k^\nu {\cal L}_{m_a^2} \bigr )
\left [ 1/(k^2 + m_a^2) \right ] 
,\phantom{xx}
\label{eq:vectorprop}
\eeq
where for later convenience I use the notation
\beq
{\cal L}_x f(x) \equiv [f(x) - f (\xi x)]/x ,
\eeq
with the appropriate limit for massless vectors:
\beq
\lim_{x\rightarrow 0}\left [{\cal L}_x f(x)\right ] 
= (1-\xi) f'(0).
\eeq
Here $\xi = 0, 1,$ and $3$ correspond to the Landau, Feynman, and
Fried-Yennie gauge-fixing choices, respectively. The self-energy
functions depend on $\xi$, but the pole masses do not.
For the two-loop computations below, the vector bosons are
treated as massless, so the propagators are
\beq
-i\delta_{ab} [ \eta^{\mu\nu}/k^2  - (1-\xi)k^\mu k^\nu/(k^2)^2 
]
.
\label{eq:vectorpropmassless}
\eeq
Infrared divergences are dealt with by first 
computing with a finite
vector boson mass, and later taking the massless vector limit.
All contributions involving gauge boson loops
implicitly include the corresponding contributions of ghost loops.

\subsection{Regularization and renormalization}

For each Feynman diagram, the integrations over internal momenta are
regulated by continuing to $d = 4 - 2 \epsilon$ dimensions, according to
\begin{eqnarray}
\int d^4 k \rightarrow (2 \pi \mu)^{2 \epsilon} \int d^d k .
\end{eqnarray}
In the dimensional regularization scheme, the vector bosons also have $d$
components, while in the dimensional reduction scheme they have $d$
ordinary components and $2\epsilon$ additional components known as epsilon
scalars.
This means that
the 4-dimensional metric in the vector propagator of
eq.~(\ref{eq:vectorprop}) is replaced by
\beq
\eta^{\mu\nu}/(k^2 + m_a^2) &\rightarrow&
g^{\mu\nu}/(k^2 + m_a^2) 
\nonumber \\ &&
+ \hat g^{\mu\nu}/(k^2 + m_a^2 + m_\epsilon^2) ,
\label{eq:decomposemetric}
\eeq
where $g^{\mu\nu}$ is projected onto a formal $d$--dimensional
subspace, and $\hat g^{\mu\nu}$ onto the complementary
$2\epsilon$--dimensional subspace, and $m_\epsilon^2$ is the
epsilon scalar squared mass parameter. 
(In general, there should be a 
different
$m_\epsilon^2$ for each $a$, but it should cause no confusion to omit
the additional subscript
in this paper.)
Counterterms for
the one-loop sub-divergences and the remaining two-loop divergences are
added, according to the rules of minimal subtraction, to give finite
results, which then depend on the renormalization scale $Q$ given by
\begin{eqnarray}
Q^2 = 4 \pi e^{-\gamma} \mu^2 .
\end{eqnarray}
Logarithms of dimensionful quantities are always written in terms of
\beq
\lnbar X \equiv \ln (X/Q^2)
.
\eeq
The resulting renormalization schemes are known as $\MSbar$ \cite{MSbar}
and $\DRbar$ \cite{DRbar}, respectively, for the cases
in which $\hat g^{\mu\nu}$ is not and is included.

The epsilon-scalar squared mass parameter $m_\epsilon^2$ appearing in the
$\DRbar$ scheme is unphysical.
One could set $m_\epsilon^2$ equal to zero at any fixed renormalization
scale, but then it will be non-zero at other renormalization scales,
since it has a non-homogeneous beta function \cite{Jack:1994kd}.
Furthermore, under renormalization group evolution it will feed into the
ordinary scalar squared masses in the $\DRbar$ scheme.
Fortunately, a
redefinition (given in \cite{DRbarprime} at one-loop order, and at 
two-loop
order in \cite{effpot}) of the ordinary scalar squared masses
completely removes the dependence on the unphysical
epsilon scalar squared mass $m_\epsilon^2$ from the renormalization group
equations and the equations relating tree-level parameters to physical
observables in softly-broken supersymmetric theories. The resulting $\DRbarprime$ scheme \cite{DRbarprime}
is therefore an appropriate one for realistic models based on
supersymmetry,
such as the Minimal Supersymmetric Standard Model (MSSM).
In this paper, calculations will be presented simultaneously in all three
schemes, using the following two notational devices. First,
\begin{eqnarray}
\deltaMSbar \equiv \begin{cases}
1 & \text{for}\quad\MSbar \\
0 & \text{for}\quad\DRbar,\> \DRbarprime .
\end{cases}
\label{eq:deltaMSbar}
\end{eqnarray}
Second, terms that involve the unphysical parameter
$m_\epsilon^2$ should be construed below to apply
only to the $\DRbar$ scheme, not the $\DRbarprime$ or $\MSbar$ schemes.

\vspace{0.01in}

\subsection{Self-energy functions and pole masses for fermions using
two-component notation\label{subsec:twocompself}}

The full, loop-corrected
Feynman propagators with four-momentum $p^\mu$
are denoted as shown in Fig.~\ref{fig:fullprops},
which defines $C_I^J$, $D^{IJ}$, and $D_{IJ}$ as functions 
of 
the masses
and couplings of the theory and of the
external momentum invariant
\beq
s \equiv -p^2 .
\eeq
They are given, starting from tree level, as
\beq
D^{IJ} &=&  {m}^{IJ}/(p^2 + m_I^2) + \mbox{\ldots}
\\
D_{IJ} &=&  {m}_{IJ}/(p^2 + m_I^2) + \mbox{\ldots}
\\
C^J_I &=& \delta^J_I/(p^2 + m_I^2) + \mbox{\ldots} 
\eeq
with no sum on $I$ in each case.
In general, ${D}^{IJ}$ is a complex
symmetric matrix, and ${D}_{IJ}$ is obtained from it by
taking the complex conjugate of all Lagrangian parameters appearing in its
calculation,
but not taking the complex conjugates of
loop integral functions, whose
imaginary (absorptive) parts correspond to fermion decay widths to
multi-particle intermediate states.

The computation of the full propagators can be organized, as usual in
quantum field theory, in terms of one-particle irreducible self-energy
functions. These are defined in Fig.~\ref{fig:selfenergies}.
(The same remark applies for the relationship between the
functions $\Omega^{IJ}$,
$\Omega_{IJ}$
as for $D^{IJ}$, $D_{IJ}$.)
Then one has the matrix diagrammatic identity shown in 
fig.~\ref{fig:selfenergyident}.
%%%%%%%%%%%%%%%%%%%%%%%%%%%%%%%%%%%%%%%%%
\begin{figure*}[tb!]
\begin{center}
\begin{picture}(80,72)(0,8)
\ArrowLine(0,40)(28,40)
\ArrowLine(52,40)(80,40)
\GBox(28,28)(52,52){0.85}
\Text(74,48)[]{$\beta$}
\Text(6,48)[]{$\dot\alpha$}
\Text(74,32)[]{$I$}
\Text(6,32)[]{$J$}
\Text(40,72)[c]{$p$}
\LongArrow(24,64)(56,64)
%\Text(40,8)[c]{$ip\newcdot\sigmabar^{\dot\alpha\beta}\,({\bf C}^T)^I{}_J$}
\Text(40,8)[c]{$ip\newcdot\sigmabar^{\dot\alpha\beta}\, C^J_I$}
\end{picture}
\hspace{1.1cm}
\begin{picture}(80,72)(0,8)
\ArrowLine(28,40)(0,40)
\ArrowLine(80,40)(52,40)
\GBox(28,28)(52,52){0.85}
\Text(6,48)[]{$\alpha$}
\Text(74,50)[]{$\dot\beta$}
\Text(6,32)[]{$I$}
\Text(74,32)[]{$J$}
\Text(40,72)[c]{$p$}
\LongArrow(24,64)(56,64)
%\Text(48,8)[c]{$ip\newcdot\sigma_{\alpha\dot\beta}\,{\bf C}_I{}^J$}
\Text(48,8)[c]{$ip\newcdot\sigma_{\alpha\dot\beta}\,C_I^J$}
\end{picture}
\hspace{1.1cm}
\begin{picture}(80,72)(0,8)
\ArrowLine(0,40)(28,40)
\ArrowLine(80,40)(52,40)
\GBox(28,28)(52,52){0.85}
\Text(6,48)[]{$\dot\alpha$}
\Text(74,48)[]{$\dot\beta$}
\Text(6,32)[]{$I$}
\Text(74,32)[]{$J$}
%\Text(40,8)[c]{$-i\delta^{\dot\alpha}{}_{\dot\beta}\, {{\bf D}}^{IJ}$}
\Text(40,8)[c]{$-i\delta^{\dot\alpha}{}_{\dot\beta}\, D^{IJ}$}
\end{picture}
\hspace{1.1cm}
\begin{picture}(80,72)(0,8)
\ArrowLine(28,40)(0,40)
\ArrowLine(52,40)(80,40)
\GBox(28,28)(52,52){0.85}
\Text(6,48)[]{$\alpha$}
\Text(74,48)[]{$\beta$}
\Text(6,32)[]{$I$}
\Text(74,32)[]{$J$}
%\Text(40,8)[c]{$-i\delta_\alpha{}^\beta \,\overline{\bf D}_{IJ}$}
\Text(40,8)[c]{$-i\delta_\alpha{}^\beta \,D_{IJ}$}
\end{picture}
\end{center}
\caption{The full loop-corrected propagators for fermions in two-component 
notation are associated with functions $C_I^J(s)$,
$D^{IJ}(s)$, and
$D_{IJ}(s)$, as shown. Here $s = -p^2$ is the external momentum invariant.
The shaded boxes represent the sum of all connected
Feynman diagrams, and the external legs are included.}
\label{fig:fullprops}
\end{figure*}
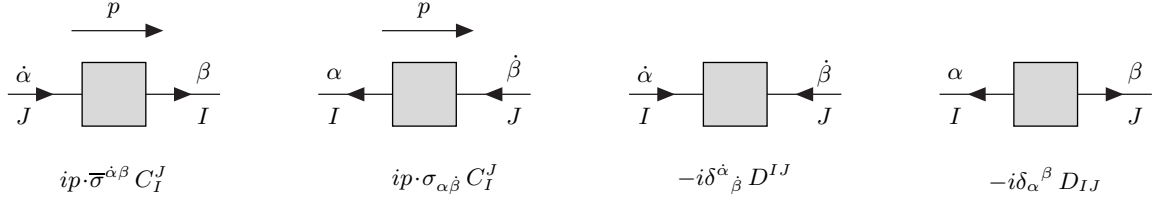
%%%%%%%%%%%%%%%%%%%%%%%%%%%%%%%%%%%%%%%%%
\begin{figure*}[tb!]
\begin{center}
\begin{picture}(80,72)(0,8)
\Text(40,72)[c]{$p$}
\LongArrow(24,64)(56,64)
\ArrowLine(0,40)(28,40)
\ArrowLine(52,40)(80,40)
\GCirc(40,40){12}{0.85}
\Text(74,48)[]{$\dot\beta$}
\Text(6,48)[]{$\alpha$}
\Text(6,32)[]{$J$}
\Text(74,32)[]{$I$}
%\Text(40,8)[c]{$-ip\newcdot\sigma_{\alpha\dot\beta}({\bf \Sigma}^T)^I{}_J$}
\Text(40,8)[c]{$ip\newcdot\sigma_{\alpha\dot\beta}\,\Sigma_I^J/s$}
\end{picture}
\hspace{1.1cm}
\begin{picture}(80,72)(0,8)
\Text(40,72)[c]{$p$}
\LongArrow(24,64)(56,64)
\ArrowLine(28,40)(0,40)
\ArrowLine(80,40)(52,40)
\GCirc(40,40){12}{0.85}
\Text(74,48)[]{$\beta$}
\Text(6,48)[]{$\dot\alpha$}
\Text(6,32)[]{$I$}
\Text(74,32)[]{$J$}
%\Text(40,8)[c]{$-ip\newcdot\sigmabar^{\dot\alpha \beta}{\bf \Sigma}_I{}^J$}
\Text(40,8)[c]{$ip\newcdot\sigmabar^{\dot\alpha \beta}\Sigma_I^J/s$}
\end{picture}
\hspace{1.1cm}
\begin{picture}(80,72)(0,8)
\ArrowLine(0,40)(28,40)
\ArrowLine(80,40)(52,40)
\GCirc(40,40){12}{0.85}
\Text(6,48)[]{$\alpha$}
\Text(74,48)[]{$\beta$}
\Text(6,32)[]{$I$}
\Text(74,32)[]{$J$}
\Text(40,8)[c]{$-i\delta_\alpha{}^\beta {\Omega}^{IJ}$}
\end{picture}
\hspace{1.1cm}
\begin{picture}(80,72)(0,8)
\ArrowLine(28,40)(0,40)
\ArrowLine(52,40)(80,40)
\GCirc(40,40){12}{0.85}
\Text(6,48)[]{$\dot\alpha$}
\Text(74,48)[]{$\dot\beta$}
\Text(6,32)[]{$I$}
\Text(74,32)[]{$J$}
\Text(40,8)[c]{$-i\delta^{\dot\alpha}{}_{\dot\beta}{\Omega}_{IJ}$}
\end{picture}
\end{center}
\caption{The self-energies for
fermions in two-component notation are associated with functions
$\Sigma^I_J(s)$ (for chirality-preserving propagation), and
$\Omega^{IJ}(s)$ and
$\Omega_{IJ}(s)$ (for chirality-violating propagation), defined as shown.
Here $s = -p^2$ is the external momentum invariant.
The shaded circles
represent the sum of all one-particle irreducible, connected
Feynman diagrams, and the external legs are not included.}
\label{fig:selfenergies}
\end{figure*}
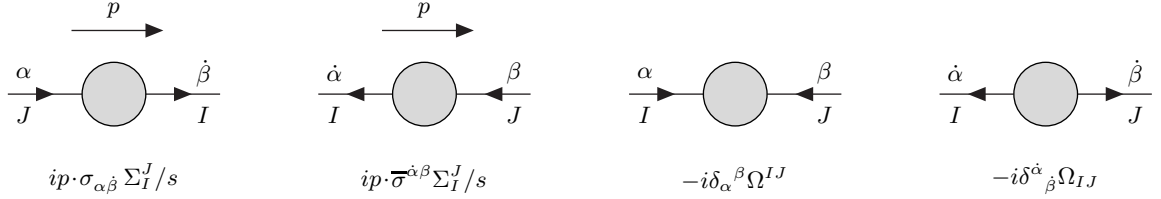
%%%%%%%%%%%%%%%%%%%%%
\begin{figure*}[tb!]
\beq
\begin{pmatrix}
\mbox{\begin{picture}(35,21)(0,10.5)
\SetScale{0.7}
\ArrowLine(0,25)(17.5,25)
\ArrowLine(32.5,25)(50,25)
\GBox(17.5,17.5)(32.5,32.5){0.85}
\end{picture}}
&\,\,\,\,
\mbox{\begin{picture}(35,21)(0,10.5)
%\mbox{\begin{picture}(50,30)(0,15)
\SetScale{0.7}
\ArrowLine(0,25)(17.5,25)
\ArrowLine(50,25)(32.5,25)
\GBox(17.5,17.5)(32.5,32.5){0.85}
\end{picture}}
\\
\mbox{\begin{picture}(35,21)(0,10.5)
%\mbox{\begin{picture}(50,30)(0,15)
\SetScale{0.7}
\ArrowLine(17.5,25)(0,25)
\ArrowLine(32.5,25)(50,25)
\GBox(17.5,17.5)(32.5,32.5){0.85}
\end{picture}}
&\,\,\,\,
\mbox{\begin{picture}(35,21)(0,10.5)
%\mbox{\begin{picture}(50,30)(0,15)
\SetScale{0.7}
\ArrowLine(17.5,25)(0,25)
\ArrowLine(50,25)(32.5,25)
\GBox(17.5,17.5)(32.5,32.5){0.85}
\end{picture}}
\end{pmatrix}
\>&=&\>
%\nonumber
%\\
\begin{pmatrix}
%\mbox{\begin{picture}(25,30)(0,15)
\mbox{\begin{picture}(17.5,21)(0,10.5)
\SetScale{0.7}
\ArrowLine(0,25)(30,25)
\end{picture}}
&\,\,\,\,
%\mbox{\begin{picture}(30,30)(0,15)
\mbox{\begin{picture}(20,21)(0,10.5)
\SetScale{0.7}
\ArrowLine(0,25)(15,25)
\ArrowLine(30,25)(15,25)
\end{picture}}
\\
%\mbox{\begin{picture}(30,30)(0,15)
\mbox{\begin{picture}(17.5,21)(0,10.5)
\SetScale{0.7}
\ArrowLine(15,25)(0,25)
\ArrowLine(15,25)(30,25)
\end{picture}}
&\,\,\,\,\,
%\mbox{\begin{picture}(25,30)(0,15)
\mbox{\begin{picture}(20,21)(0,10.5)
\SetScale{0.7}
\ArrowLine(27,25)(0,25)
\end{picture}}
\end{pmatrix}
\left [
\begin{pmatrix}
\mbox{\begin{picture}(9.8,21)(0,0)
\Text(4.9,10.5)[c]{$\bf 1$}
\end{picture}}
&\,\,\,\,
\mbox{\begin{picture}(9.8,21)(0,0)
\Text(4.9,10.5)[c]{$\bf 0$}
\end{picture}}
\\
\mbox{\begin{picture}(9.8,21)(0,0)
\Text(4.9,10.5)[c]{$\bf 0$}
\end{picture}}
&\,\,\,\,
\mbox{\begin{picture}(9.8,21)(0,0)
\Text(4.9,10.5)[c]{$\bf 1$}
\end{picture}}
\end{pmatrix}
-
\begin{pmatrix}
\mbox{\begin{picture}(35,21)(0,10.5)
%\mbox{\begin{picture}(50,30)(0,15)
\SetScale{0.7}
\ArrowLine(0,25)(17.5,25)
\ArrowLine(32.5,25)(50,25)
\GCirc(25,25){7.5}{0.85}
\end{picture}}
&\,\,\,\,
\mbox{\begin{picture}(35,21)(0,10.5)
%\mbox{\begin{picture}(50,30)(0,15)
\SetScale{0.7}
\ArrowLine(0,25)(17.5,25)
\ArrowLine(50,25)(32.5,25)
\GCirc(25,25){7.5}{0.85}
\end{picture}}
\\
\mbox{\begin{picture}(35,21)(0,10.5)
%\mbox{\begin{picture}(50,30)(0,15)
\SetScale{0.7}
\ArrowLine(17.5,25)(0,25)
\ArrowLine(32.5,25)(50,25)
\GCirc(25,25){7.5}{0.85}
\end{picture}}
&\,\,\,\,
\mbox{\begin{picture}(35,21)(0,10.5)
%\mbox{\begin{picture}(50,30)(0,15)
\SetScale{0.7}
\ArrowLine(17.5,25)(0,25)
\ArrowLine(50,25)(32.5,25)
\GCirc(25,25){7.5}{0.85}
\end{picture}}
\end{pmatrix}
\begin{pmatrix}
%\mbox{\begin{picture}(25,30)(0,15)
\mbox{\begin{picture}(17.5,21)(0,10.5)
\SetScale{0.7}
\ArrowLine(0,25)(30,25)
\end{picture}}
&\,\,\,\,
%\mbox{\begin{picture}(30,30)(0,15)
\mbox{\begin{picture}(20,21)(0,10.5)
\SetScale{0.7}
\ArrowLine(0,25)(15,25)
\ArrowLine(30,25)(15,25)
\end{picture}}
\\
%\mbox{\begin{picture}(30,30)(0,15)
\mbox{\begin{picture}(17.5,21)(0,10.5)
\SetScale{0.7}
\ArrowLine(15,25)(0,25)
\ArrowLine(15,25)(30,25)
\end{picture}}
&\,\,\,\,
%\mbox{\begin{picture}(25,30)(0,15)
\mbox{\begin{picture}(20,21)(0,10.5)
\SetScale{0.7}
\ArrowLine(27.5,25)(0,25)
\end{picture}}
\end{pmatrix}
\right ]^{-1}
\nonumber \\
\,&\>=\>&\,
%\nonumber
%\\
\left [
\begin{pmatrix}
%\mbox{\begin{picture}(25,30)(0,15)
\mbox{\begin{picture}(17.5,21)(0,10.5)
\SetScale{0.7}
\ArrowLine(0,25)(30,25)
\end{picture}}
&\,\,\,\,
%\mbox{\begin{picture}(30,30)(0,15)
\mbox{\begin{picture}(20,21)(0,10.5)
\SetScale{0.7}
\ArrowLine(0,25)(15,25)
\ArrowLine(30,25)(15,25)
\end{picture}}
\\
%\mbox{\begin{picture}(30,30)(0,15)
\mbox{\begin{picture}(17.5,21)(0,10.5)
\SetScale{0.7}
\ArrowLine(15,25)(0,25)
\ArrowLine(15,25)(30,25)
\end{picture}}
&\,\,\,\,
%\mbox{\begin{picture}(25,30)(0,15)
\mbox{\begin{picture}(20,21)(0,10.5)
\SetScale{0.7}
\ArrowLine(28,25)(0,25)
\end{picture}}
\end{pmatrix}^{-1}
-\,
\begin{pmatrix}
\mbox{\begin{picture}(35,21)(0,10.5)
%\mbox{\begin{picture}(50,30)(0,15)
\SetScale{0.7}
\ArrowLine(0,25)(17.5,25)
\ArrowLine(32.5,25)(50,25)
\GCirc(25,25){7.5}{0.85}
\end{picture}}
&\,\,\,\,
\mbox{\begin{picture}(35,21)(0,10.5)
%\mbox{\begin{picture}(50,30)(0,15)
\SetScale{0.7}
\ArrowLine(0,25)(17.5,25)
\ArrowLine(50,25)(32.5,25)
\GCirc(25,25){7.5}{0.85}
\end{picture}}
\\
\mbox{\begin{picture}(35,21)(0,10.5)
%\mbox{\begin{picture}(50,30)(0,15)
\SetScale{0.7}
\ArrowLine(17.5,25)(0,25)
\ArrowLine(32.5,25)(50,25)
\GCirc(25,25){7.5}{0.85}
\end{picture}}
&\,\,\,\,
\mbox{\begin{picture}(35,21)(0,10.5)
%\mbox{\begin{picture}(50,30)(0,15)
\SetScale{0.7}
\ArrowLine(17.5,25)(0,25)
\ArrowLine(50,25)(32.5,25)
\GCirc(25,25){7.5}{0.85}
\end{picture}}
\end{pmatrix}
\right ]^{-1}
\nonumber
\eeq
\caption{The diagrammatic version of the fermion self-energy identity
in eq.~(\ref{eq:selfenergyident}).}
\label{fig:selfenergyident}
\end{figure*}
To write this in terms of the self-energy functions, denote $N\times N$
matrices (where $N$ is the number of two-component left-handed fermion 
degrees of 
freedom, so that $I,J = 1,2,\ldots N$):
\beq
&&{\bf C}_I{}^J = ({\bf C}^T)^J{}_I \equiv C_I^J,
\\
&&{\bf D}^{IJ} \equiv D^{IJ},\qquad\qquad\quad
   \overline{\bf D}_{IJ} \equiv D_{IJ},
\\
&&{\bf \Sigma}_I{}^J = ({\bf \Sigma}^T)^J{}_I \equiv \Sigma_I^J,
\\
&&{\bf \Omega}^{IJ} \equiv \Omega^{IJ},\qquad\qquad\>
   \overline{\bf \Omega}_{IJ} \equiv \Omega_{IJ},
\\
&&{\bf m}^{IJ} \equiv m^{IJ},\qquad\qquad\quad
   \overline{\bf m}_{IJ} \equiv m_{IJ},
   \phantom{xxxx}
\eeq
Then fig.~\ref{fig:selfenergyident} implies that
the propagator functions obey
the $4N\times 4N$ matrix equation:
\begin{widetext}
\beq
\begin{pmatrix} 
i p \newcdot \sigmabar \, {\bf C}^T \phantom{x} & 
-i {\bf D}
\\[4pt]
-i\overline{{\bf D}} \phantom{x}& i p\newcdot \sigma \, {\bf C}
\end{pmatrix}
&=&
\begin{pmatrix}
 i p\newcdot \sigma\, [{\bf 1} - {\bf \Sigma}^T/s]\phantom{x}
&
i [{\bf m}+{\bf \Omega}]
\\[4pt]
i [\overline{\bf m} + \overline {\bf \Omega}]
&
i p\newcdot \sigmabar \,[{\bf 1} - {\bf \Sigma}/s]
\end{pmatrix}^{-1}.
\label{eq:selfenergyident}
\eeq
The pole mass can be found most easily by considering the rest 
frame of 
the fermion, in which the space components of the external momentum 
$p^\mu$ vanish. This reduces the spinor-index dependence to a triviality.
It follows from eq.~(\ref{eq:selfenergyident}) that the (complex, if
the fermion is unstable) 
poles of the full propagator
are the solutions for $s$ of the non-linear $N\times N$ matrix 
eigenvalue 
equation:
\beq
{\rm Det}\bigl [s {\bf 1} - 
({\bf 1} - {{\bf \Sigma}}/{s})^{-1}
(\overline{\bf m} + \overline{\bf \Omega})
({\bf 1} - {{\bf \Sigma}^T}/{s})^{-1}
({\bf m} + {\bf \Omega})
\bigr ] 
= 0 .
\label{eq:nonlineareigen}
\eeq
This can be solved iteratively by first expanding each of
the self-energy
functions in a Taylor series in $s$ about the 
tree-level squared masses $m_I^2$. 
Write the one- and two-loop
contributions to the self-energy functions as:
\beq
\Sigma_I^J &=& 
\frac{1}{16\pi^2} \Sigma_I^{(1)J} +
\frac{1}{(16\pi^2)^2} \Sigma_I^{(2)J} + \ldots
\\
\Omega^{IJ} &=& 
\frac{1}{16\pi^2} \Omega^{(1)IJ} +
\frac{1}{(16\pi^2)^2} \Omega^{(2)IJ} + \ldots
\\
\Omega_{IJ} &=& 
\frac{1}{16\pi^2} \Omega^{(1)}_{IJ} +
\frac{1}{(16\pi^2)^2} \Omega^{(2)}_{IJ} + \ldots
\eeq
where the superscripts 
$(1)$ and $(2)$ refer to the one- and two-loop contributions
respectively. Then define the quantities (with sums on $I'$, $J'$, $K$,
and $K'$,
but not on $I$ or $J$):
\beq
\Pi^{(1)J}_I &=&
\Sigma^{(1)J}_I m_J^2/s + m_{II'} \Sigma^{(1)I'}_{J'} m^{JJ'}/s
+ m_{II'} \Omega^{(1)JI'}
+ \Omega^{(1)}_{IJ'} m^{JJ'}
,
\label{eq:defPIoneloop}
\\
\Pi^{(2)J}_I &=&
\Sigma^{(2)J}_I m_J^2/s + m_{II'} \Sigma^{(2)I'}_{J'} m^{JJ'}/s
+ m_{II'} \Omega^{(2)JI'}
+ \Omega^{(2)}_{IJ'} m^{JJ'}
\nonumber \\ &&
+ \Bigl \lbrace 
\Sigma^{(1)K}_I \Sigma^{(1)J}_K m_J^2 
+ m_{II'} \Sigma^{(1)I'}_K \Sigma^{(1)K}_{J'} m^{JJ'} 
+ \Sigma^{(1)K}_I m_{KK'} \Sigma^{(1)K'}_{J'} m^{JJ'} 
\Bigr \rbrace/s^2
+ \Omega^{(1)}_{IK} \Omega^{(1)JK}
\nonumber \\ &&
+ \Bigl \lbrace
\Sigma^{(1)K}_I \Omega^{(1)}_{KJ'} m^{JJ'} 
+ \Sigma^{(1)K}_I m_{KK'} \Omega^{(1)JK'}
+ m_{II'} \Sigma^{(1)I'}_K \Omega^{(1)JK}
+ \Omega^{(1)}_{IK} \Sigma^{(1)K}_{J'} m^{JJ'}
\Bigr \rbrace/s ,
\eeq
which play a role analogous to the self-energy functions of scalar 
or vector bosons, with eq.~(\ref{eq:nonlineareigen}) taking the form
\beq
{\rm Det} \Bigl [
s \delta_I^J - \Bigl (
m_I^2 \delta_I^J 
+ \frac{1}{16\pi^2} \Pi^{(1)J}_I
+ \frac{1}{(16\pi^2)^2} \Pi^{(2)J}_I
\Bigr ) \Bigr ]  = 0 .
\label{eq:nonlineareigentwo}
\eeq
It follows
that the pole squared masses 
for the fermions are given by (with no sum on the index $I$):
\beq
s_{{\rm pole},I} \equiv M_I^2 - i\Gamma_I M_I 
= m_I^2 + \frac{1}{16 \pi^2} 
\Pi^{(1)I}_I 
+ \frac{1}{(16 \pi^2)^2} \Bigl [
\Pi^{(2)I}_I
+ \Pi^{(1)I}_I \Bigl (\frac{\partial \Pi^{(1)I}_I}{\partial s} \Bigr )
+ \sum_{J \not= I}  \Pi^{(1)J}_I \Pi^{(1)I}_J/(m_I^2 - m_J^2)
\Bigr ],
\phantom{xx}
\label{eq:polemasstwoloops}
\eeq
\end{widetext}
where one must put $s = m_I^2 + i \varepsilon$ (note with an
infinitesimal {\em positive} imaginary part; this is necessary
to give the correct 
{\em negative} imaginary part to the pole mass) 
everywhere on the right-hand side.
Terms that are of three-loop order have been dropped.

In writing eq.~(\ref{eq:polemasstwoloops}), it is assumed that
the fermions that mix with each other are not degenerate,
so that the last term is part of a well-defined perturbative expansion.
If (nearly) degenerate fermions do mix, then the appropriate version of
(nearly) degenerate perturbation theory should be used instead to 
solve eq.~(\ref{eq:nonlineareigentwo}). One
can also obtain a solution iteratively, by first taking $s = m_I^2$
as the argument of the self-energy functions in 
eq.~(\ref{eq:nonlineareigentwo}), solving for $s$ to obtain the next value
for the argument Re$(s)$ of the self-energy functions, and repeating until sufficient
numerical convergence is obtained. However, 
despite the formal gauge invariance of the pole mass, 
this iterated procedure does not give a gauge-invariant result
at two-loop order when massless gauge bosons are present,
because of the branch cut in the one-loop self-energy
that is present except in the Fried-Yennie gauge $\xi=3$. This is
because the pole mass result obtained by the iterative
procedure is not formally 
analytic in the gauge coupling for $\xi \not= 3$, as
explained in more detail in the analogous case for scalars in 
ref.~\cite{Martin:2005eg}.

For taking the limit $s\rightarrow m_I^2$ in 
eq.~(\ref{eq:polemasstwoloops}), 
it is convenient to define (again with no sum on $I$):
\beq
\widetilde \Pi^{(2)}_I \equiv \lim_{s \rightarrow m_I^2} \Bigl [
\Pi^{(2)I}_I
+ \Pi^{(1)I}_I \Bigl (\frac{\partial \Pi^{(1)I}_I}{\partial s} \Bigr )
\Bigr ],
\label{eq:defpitwotilde}
\eeq
since this combination is independent of the gauge-fixing parameter 
$\xi$, and free of logarithmic divergences of the form $\ln(1 - s/m_I^2)$
that do appear in the individual terms when there are massless
gauge bosons.
The results for one-loop self-energy functions and pole 
squared mass 
contributions 
$\Sigma_I^{(1)J}$,
$\Omega^{(1)IJ}$,
$\Omega^{(1)}_{IJ}$, and
$\Pi_I^{(1)J}$,
will be reviewed in section
\ref{sec:oneloop}. The two-loop contributions to
$\Sigma_I^{(2)J}$,
$\Omega^{(2)IJ}$,
$\Omega^{(2)}_{IJ}$, and
$\widetilde\Pi_I^{(2)}$
are presented in
section \ref{sec:twoloop}. 

\subsection{The Feynman diagrams}
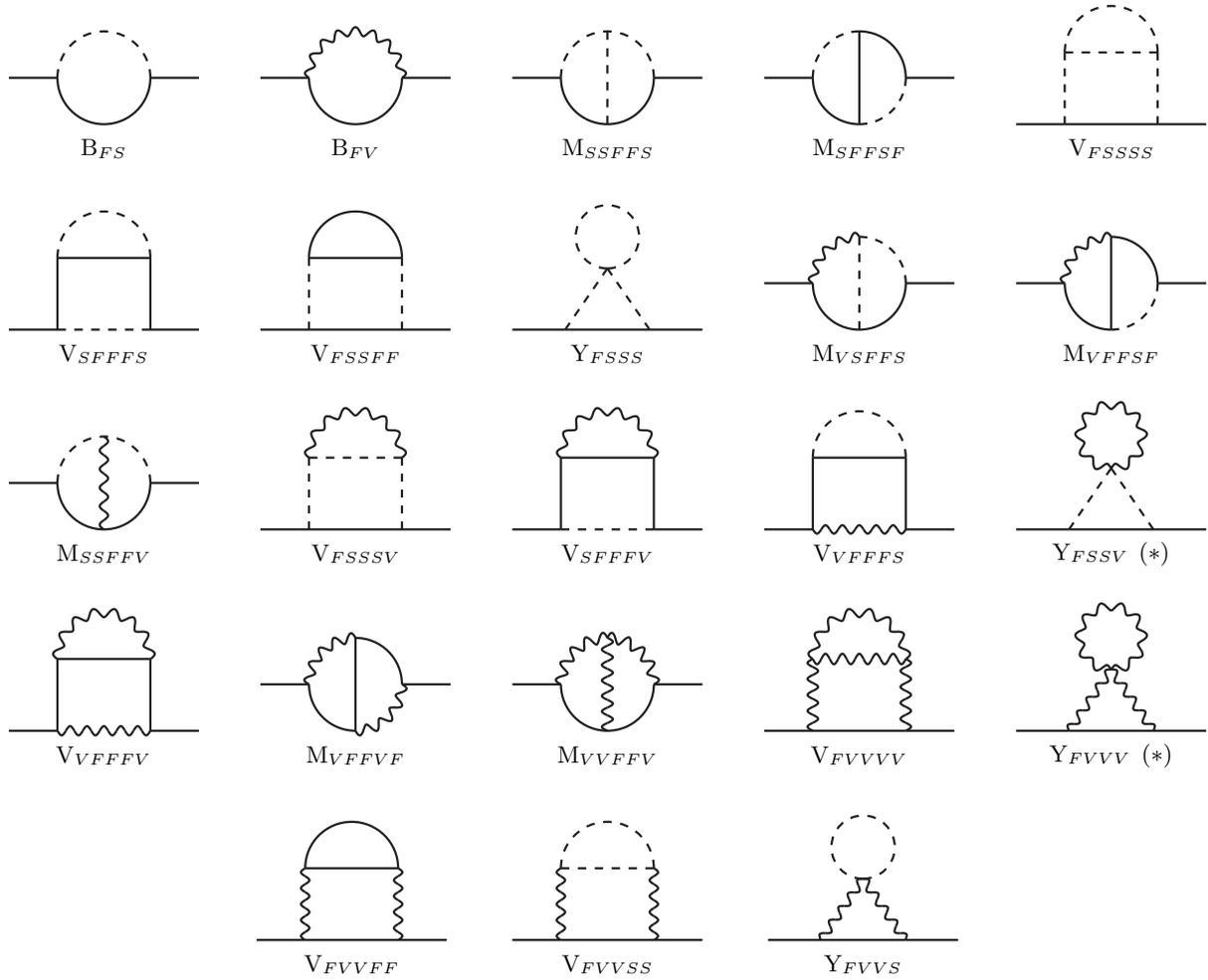
\begin{figure*}[p] 
%
%\begin{picture}(98,64)(-49,-35)
\begin{picture}(78,54)(-39,-27)
\SetScale{0.8}
\SetWidth{1.05}
\Line(-45,0)(-22,0)
\Line(45,0)(22,0)
\DashCArc(0,0)(22,0,180){4}
\CArc(0,0)(22,180,360)
\Text(0,-27)[]{$\propB_{FS}$}
\end{picture}
\hspace{0.16in}
%
%\begin{picture}(98,64)(-49,-35)
\begin{picture}(78,54)(-39,-27)
\SetScale{0.8}
\SetWidth{1.05}
\Line(-45,0)(-22,0)
\Line(45,0)(22,0)
\PhotonArc(0,0)(22,0,180){2}{8.5}
\CArc(0,0)(22,180,360)
\Text(0,-27)[]{$\propB_{FV}$}
\end{picture}
\hspace{0.16in}
%
%\begin{picture}(98,64)(-49,-35)
\begin{picture}(78,54)(-39,-27)
\SetScale{0.8}
\SetWidth{1.05}
\Line(-45,0)(-22,0)
\Line(45,0)(22,0)
\DashCArc(0,0)(22,0,180){4}
\CArc(0,0)(22,180,360)
\DashLine(0,-22)(0,22){4}
\Text(0,-27)[]{$\propM_{SSFFS}$}
\end{picture}
\hspace{0.16in}
%
%\begin{picture}(98,64)(-49,-35)
\begin{picture}(78,54)(-39,-27)
\SetScale{0.8}
\SetWidth{1.05}
\Line(-45,0)(-22,0)
\Line(45,0)(22,0)
\CArc(0,0)(22,0,90)
\DashCArc(0,0)(22,90,180){4}
\CArc(0,0)(22,180,270)
\DashCArc(0,0)(22,270,360){4}
\Line(0,-22)(0,22)
\Text(0,-27)[]{$\propM_{SFFSF}$}
\end{picture}
\hspace{0.16in}
%
%\begin{picture}(98,82)(-49,-39)
\begin{picture}(78,54)(-39,-27)
\SetScale{0.8}
\SetWidth{1.05}
\Line(-45,-22)(-22,-22)
\Line(45,-22)(22,-22)
\Line(-22,-22)(22,-22)
\DashCArc(0,12)(22,0,180){4}
\DashLine(-22,12)(22,12){4}
\DashLine(-22,-22)(-22,12){4}
\DashLine(22,-22)(22,12){4}
\Text(0,-27)[]{$\propV_{FSSSS}$}
\end{picture}

\vspace{0.32in}

%\begin{picture}(98,82)(-49,-39)
\begin{picture}(78,54)(-39,-27)
\SetScale{0.8}
\SetWidth{1.05}
\Line(-45,-22)(-22,-22)
\Line(45,-22)(22,-22)
\DashLine(-22,-22)(22,-22){4}
\DashCArc(0,12)(22,0,180){4}
\Line(-22,12)(22,12)
\Line(-22,-22)(-22,12)
\Line(22,-22)(22,12)
\Text(0,-27)[]{$\propV_{SFFFS}$}
\end{picture}
\hspace{0.16in}
%
%\begin{picture}(98,82)(-49,-39)
\begin{picture}(78,54)(-39,-27)
\SetScale{0.8}
\SetWidth{1.05}
\Line(-45,-22)(-22,-22)
\Line(45,-22)(22,-22)
\Line(-22,12)(22,12)
\CArc(0,12)(22,0,180)
\Line(-22,-22)(22,-22)
\DashLine(-22,-22)(-22,12){4}
\DashLine(22,-22)(22,12){4}
\Text(0,-27)[]{$\propV_{FSSFF}$}
\end{picture}
\hspace{0.16in}
%
%\begin{picture}(98,82)(-49,-39)
\begin{picture}(78,54)(-39,-27)
\SetScale{0.8}
\SetWidth{1.05}
\Line(-45,-22)(-20,-22)
\Line(45,-22)(20,-22)
\Line(-20,-22)(20,-22)
\DashLine(-20,-22)(0,7){4}
\DashLine(20,-22)(0,7){4}
\DashCArc(0,22)(15,-90,270){4}
\Text(0,-27)[]{$\propY_{FSSS}$}
\end{picture}
\hspace{0.16in}
%
%\begin{picture}(98,82)(-49,-39)
\begin{picture}(78,54)(-39,-27)
\SetScale{0.8}
\SetWidth{1.05}
\Line(-45,0)(-22,0)
\Line(45,0)(22,0)
\DashCArc(0,0)(22,0,90){4}
\PhotonArc(0,0)(22,90,180){2}{4.5}
\CArc(0,0)(22,180,360)
\DashLine(0,-22)(0,22){4}
\Text(0,-27)[]{$\propM_{VSFFS}$}
\end{picture}
\hspace{0.16in}
%
%\begin{picture}(98,82)(-49,-39)
\begin{picture}(78,54)(-39,-27)
\SetScale{0.8}
\SetWidth{1.05}
\Line(-45,0)(-22,0)
\Line(45,0)(22,0)
\DashCArc(0,0)(22,-90,0){4}
\CArc(0,0)(22,0,90)
\PhotonArc(0,0)(22,90,180){2}{4.5}
\CArc(0,0)(22,180,270)
\Line(0,-22)(0,22)
\Text(0,-27)[]{$\propM_{VFFSF}$}
\end{picture}

\vspace{0.29in}

%\begin{picture}(98,82)(-49,-39)
\begin{picture}(78,54)(-39,-27)
\SetScale{0.8}
\SetWidth{1.05}
\Line(-45,0)(-22,0)
\Line(45,0)(22,0)
\DashCArc(0,0)(22,0,180){4}
\CArc(0,0)(22,180,360)
\Photon(0,-22)(0,22){2}{5}
\Text(0,-27)[]{$\propM_{SSFFV}$}
\end{picture}
\hspace{0.16in}
%
%\begin{picture}(98,82)(-49,-39)
\begin{picture}(78,54)(-39,-27)
\SetScale{0.8}
\SetWidth{1.05}
\Line(-45,-22)(-22,-22)
\Line(45,-22)(22,-22)
\Line(-22,-22)(22,-22)
\PhotonArc(0,12)(22,0,180){2}{7.5}
\DashLine(-22,12)(22,12){4}
\DashLine(-22,-22)(-22,12){4}
\DashLine(22,-22)(22,12){4}
\Text(0,-27)[]{$\propV_{FSSSV}$}
\end{picture}
\hspace{0.16in}
%
%\begin{picture}(98,82)(-49,-39)
\begin{picture}(78,54)(-39,-27)
\SetScale{0.8}
\SetWidth{1.05}
\Line(-45,-22)(-22,-22)
\Line(45,-22)(22,-22)
\DashLine(-22,-22)(22,-22){4}
\PhotonArc(0,12)(22,0,180){2}{7.5}
\Line(-22,12)(22,12)
\Line(-22,-22)(-22,12)
\Line(22,-22)(22,12)
\Text(0,-27)[]{$\propV_{SFFFV}$}
\end{picture}
\hspace{0.16in}
%
%\begin{picture}(98,82)(-49,-39)
\begin{picture}(78,54)(-39,-27)
\SetScale{0.8}
\SetWidth{1.05}
\Line(-45,-22)(-22,-22)
\Line(45,-22)(22,-22)
\Line(-22,12)(22,12)
\DashCArc(0,12)(22,0,180){4}
\Photon(-22,-22)(22,-22){-2}{5.5}
\Line(-22,-22)(-22,12)
\Line(22,-22)(22,12)
\Text(0,-27)[]{$\propV_{VFFFS}$}
\end{picture}
\hspace{0.16in}
%
%\begin{picture}(98,82)(-49,-39)
\begin{picture}(78,54)(-39,-27)
\SetScale{0.8}
\SetWidth{1.05}
\Line(-45,-22)(-20,-22)
\Line(45,-22)(20,-22)
\Line(-20,-22)(20,-22)
\DashLine(-20,-22)(0,7){4}
\DashLine(20,-22)(0,7){4}
\PhotonArc(0,22)(15,-90,270){-2}{10.5}
\Text(0,-27)[]{$\propY_{FSSV}\>\> (*)$}
\end{picture}

\vspace{0.3in}

%\begin{picture}(98,82)(-49,-39)
\begin{picture}(78,54)(-39,-27)
\SetScale{0.8}
\SetWidth{1.05}
\Line(-45,-22)(-22,-22)
\Line(45,-22)(22,-22)
\Photon(-22,-22)(22,-22){-2.2}{5.5}
\PhotonArc(0,12)(22,0,180){2.2}{7.5}
\Line(-22,12)(22,12)
\Line(-22,-22)(-22,12)
\Line(22,-22)(22,12)
\Text(0,-27)[]{$\propV_{VFFFV}$}
\end{picture}
\hspace{0.16in}
%
%\begin{picture}(98,82)(-49,-39)
\begin{picture}(78,54)(-39,-27)
\SetScale{0.8}
\SetWidth{1.05}
\Line(-45,0)(-22,0)
\Line(45,0)(22,0)
\CArc(0,0)(22,0,90)
\PhotonArc(0,0)(22,90,180){2.3}{4.5}
\CArc(0,0)(22,180,270)
\PhotonArc(0,0)(22,270,360){2.3}{4.5}
\Line(0,-22)(0,22)
\Text(0,-27)[]{$\propM_{VFFVF}$}
\end{picture}
\hspace{0.16in}
%
%\begin{picture}(98,82)(-49,-39)
\begin{picture}(78,54)(-39,-27)
\SetScale{0.8}
\SetWidth{1.05}
\Line(-45,0)(-22,0)
\Line(45,0)(22,0)
\PhotonArc(0,0)(22,0,90){2.5}{4.5}
\PhotonArc(0,0)(22,90,180){2.5}{4.5}
\CArc(0,0)(22,180,360)
\Photon(0,-22)(0,22){2.6}{6}
\Text(0,-27)[]{$\propM_{VVFFV}$}
\end{picture}
\hspace{0.16in}
%
%\begin{picture}(98,82)(-49,-39)
\begin{picture}(78,54)(-39,-27)
\SetScale{0.8}
\SetWidth{1.05}
\Line(-45,-22)(-22,-22)
\Line(45,-22)(22,-22)
\Line(-22,-22)(22,-22)
\Photon(-22,-22)(-22,12){2.5}{4.5}
\Photon(22,-22)(22,12){-2.5}{4.5}
\Photon(-22,12)(22,12){-2.5}{5.5}
\PhotonArc(0,12)(22,0,180){2.2}{7.5}
\Text(0,-27)[]{$\propV_{FVVVV}$}
\end{picture}
\hspace{0.16in}
%
%\begin{picture}(98,82)(-49,-39)
\begin{picture}(78,54)(-39,-27)
\SetScale{0.8}
\SetWidth{1.05}
\Line(-45,-22)(-20,-22)
\Line(45,-22)(20,-22)
\Line(-20,-22)(20,-22)
\Photon(-20,-22)(0,7){2}{4.5}
\Photon(20,-22)(0,7){-2}{4.5}
\PhotonArc(0,22)(15,-90,270){-2}{10.5}
\Text(0,-27)[]{$\propY_{FVVV} \>\> (*)$}
\end{picture}

\vspace{0.34in}

%\begin{picture}(98,82)(-49,-39)
\begin{picture}(78,54)(-39,-27)
\SetScale{0.8}
\SetWidth{1.05}
\Line(-45,-22)(-22,-22)
\Line(45,-22)(22,-22)
\Line(-22,-22)(22,-22)
\CArc(0,12)(22,0,180)
\Line(-22,12)(22,12)
\Photon(-22,-22)(-22,12){2}{4.5}
\Photon(22,-22)(22,12){-2}{4.5}
\Text(0,-27)[]{$\propV_{FVVFF}$}
\end{picture}
\hspace{0.18in}
%
%\begin{picture}(98,82)(-49,-39)
\begin{picture}(78,54)(-39,-27)
\SetScale{0.8}
\SetWidth{1.05}
\Line(-45,-22)(-22,-22)
\Line(45,-22)(22,-22)
\Line(-22,-22)(22,-22)
\DashCArc(0,12)(22,0,180){4}
\DashLine(-22,12)(22,12){4}
\Photon(-22,-22)(-22,12){2}{4.5}
\Photon(22,-22)(22,12){-2}{4.5}
\Text(0,-27)[]{$\propV_{FVVSS}$}
\end{picture}
\hspace{0.18in}
%
%\begin{picture}(98,82)(-49,-39)
\begin{picture}(78,54)(-39,-27)
\SetScale{0.8}
\SetWidth{1.05}
\Line(-45,-22)(-20,-22)
\Line(45,-22)(20,-22)
\Line(-20,-22)(20,-22)
\Photon(-20,-22)(0,7){2}{4.5}
\Photon(20,-22)(0,7){-2}{4.5}
\DashCArc(0,22)(15,-90,270){4}
\Text(0,-27)[]{$\propY_{FVVS}$}
\end{picture}
\caption{\label{fig:alldiagrams}
The one-loop and two-loop Feynman diagrams for fermion
self-energies in the approximation of this paper. Dashed lines
stand for scalars, solid lines for fermions, and wavy lines for massless
vector bosons. Diagrams involving vector boson loops also include the
corresponding ghost loop diagrams. The label for each diagram refers to a
corresponding function obtained as the result of the two-loop
integration. All counterterm diagrams for each diagram are included in
these functions, rendering them ultraviolet finite.  For each diagram, 
fermion mass insertions (indicated by adding a bar to the
corresponding subscript $F$ in the name) are to be made
in all possible ways.  Diagrams indicated by $(*)$ vanish 
identically in
the $\MSbar$ scheme with massless vector bosons, but not in the $\DRbar$ 
scheme with non-zero
epsilon-scalar masses.}
\end{figure*}

The one-loop and two-loop Feynman diagrams needed for the results
just mentioned
are shown in fig.~\ref{fig:alldiagrams}. They are labeled according to  
a system described in ref.~\cite{Martin:2003it}. 

\subsection{Two-loop basis integrals}

\begin{figure*}[p]
\begin{flushleft}
\begin{picture}(73,54)(-36.5,-25)
\SetScale{0.67}
\SetWidth{1.05}
\CArc(0,4)(20,0,360)
\Text(0,20)[]{$x$}
\Text(0,-26)[]{$A(x)$}
\Line(-45,-16)(45,-16)
\end{picture}
\begin{picture}(73,54)(-36.5,-25)
\SetScale{0.67}
\SetWidth{1.05}
\Line(-45,0)(-22,0)
\Line(45,0)(22,0)
\CArc(0,0)(22,0,360)
\Text(0,19)[]{$x$}
\Text(0.1,-9.6)[]{$y$}
\Text(0,-26.5)[]{$B(x,y)$}
\end{picture}
\begin{picture}(50,54)(-25,-25)
\SetScale{0.67}
\SetWidth{1.05}
\CArc(0,0)(22,0,360)  
\Line(-22,0)(22,0)
\Text(0,19)[]{$x$}
\Text(0,4)[]{$y$}
\Text(0,-10)[]{$z$}
\Text(0,-26.5)[]{$I(x,y,z)$}
\end{picture}
\begin{picture}(73,54)(-36.5,-25)
\SetScale{0.67}
\SetWidth{1.05}
\Line(-45,0)(-22,0)
\Line(45,0)(22,0)
\CArc(0,0)(22,0,360)
\Line(-22,0)(22,0)
\Text(0,19)[]{$x$}
\Text(0,4)[]{$y$}
\Text(0,-10)[]{$z$}   
\Text(0,-26.5)[]{$S(x,y,z)$}
\end{picture}
\begin{picture}(73,54)(-36.5,-25)
\SetScale{0.67}
\SetWidth{1.05}
\Line(-45,0)(-22,0)
\Line(45,0)(22,0)
\CArc(0,0)(22,0,360)
\Line(-22,0)(22,0)
\Vertex(0,22){3.5} 
\Text(0,21)[]{$x$}
\Text(0,4)[]{$y$} 
\Text(0,-10)[]{$z$}
\Text(0,-26.5)[]{$T(x,y,z)$}
\end{picture}
\begin{picture}(73,54)(-36.5,-25)
\SetScale{0.67}
\SetWidth{1.05}
\Line(-45,-14)(-22,-14)
\Line(45,-14)(22,-14)
\Line(-22,-14)(0,14)
\Line(22,-14)(0,14)
\CArc(11,0)(17.8045,-51.843,128.157)
\Line(-22,-14)(22,-14)
\Text(0,-5.3)[]{$x$}
\Text(-11.2,2.5)[]{$y$}
\Text(9,3.5)[]{$z$}
\Text(20.5,9.4)[]{$u$}
\Text(0,-26.5)[]{$U(x,y,z,u)$}
\end{picture}
\begin{picture}(73,54)(-36.5,-25)
\SetScale{0.67}
\SetWidth{1.05}
\Line(-45,0)(-22,0)
\Line(45,0)(22,0)
\CArc(0,0)(22,0,360) 
\Line(0,-22)(0,22)  
\Text(-13.5,13.5)[]{$x$}
\Text(13.5,13.5)[]{$y$}
\Text(-13.5,-13.5)[]{$z$} 
\Text(13.5,-13.5)[]{$u$} 
\Text(4.5,0)[]{$v$}
\Text(0,-26.5)[]{$M(x,y,z,u,v)$}
\end{picture}
\end{flushleft}
\caption{\label{fig:topologies}
Feynman diagram topologies for the one- and two-loop
self-energy basis integrals used in this paper. The letters $x,y,z,u,v$
refer to the squared masses of the corresponding propagators. The 
dot on the
$T$ diagram means that $T(x,y,z) = -\partial S(x,y,z)/\partial x$.
The precise definitions of these Euclideanized scalar integral functions,
and methods for their
evaluation, are described in \cite{evaluation,TSIL}.}
\end{figure*}
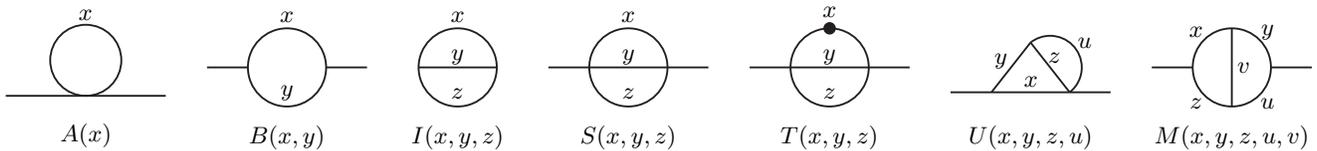

The results below will be written in terms of two-loop integral basis
functions, following the notation given in \cite{evaluation,TSIL}.
The one-loop and two-loop integral functions are reduced using
Tarasov's algorithm \cite{Tarasov:1997kx,Mertig:1998vk}
to a set of basis integrals $A(x)$, $B(x,y)$, $I(x,y,z)$, $S(x,y,z)$, 
$T(x,y,z)$, $U(x,y,z,u)$,
and $M(x,y,z,u,v)$, corresponding to the Feynman diagram
topologies shown in fig.~\ref{fig:topologies}.
Here $x,y,z,u,v$ are squared mass arguments.
The additional arguments $s$ and $Q^2$ are not shown explicitly, because
they are
the same for all functions in a given equation. The functions $A(x)$
and $I(x,y,z)$ do not depend on the external momentum at all,
with $A(x) = x(\lnbar x - 1)$ and $I(x,y,z) = S(x,y,z)|_{s=0}$.
Each of the basis integral functions contains counterterms that render them
ultraviolet finite. The precise definitions, and the
calculation of these functions and a publicly available
computer code (TSIL) for that purpose, are described in
\cite{evaluation,TSIL}.

Several shorthand notations will be used. 
As explained in refs.~\cite{evaluation,TSIL},
it is convenient to define:
\beq
{\overline T} (0,y,z) &\equiv& \lim_{x \rightarrow 0}
[T(x,y,z) + B(y,z) \lnbar x], 
\phantom{xxx}
\label{eq:defTbar}
\\
V(x,y,z,u) &\equiv& -\partial U(x,y,z,u)/\partial y
\label{eq:defV} .
\eeq
A prime on a squared-mass argument of
a function is used to denote a derivative with respect to that
argument, so:
\beq
B(x,y') &\equiv& \partial B(x,y)/ \partial y ,
\label{eq:defBprime}
\\
I(x',y,z) &\equiv& \partial I(x,y,z)/\partial x .
\label{eq:defIprime}
\eeq
A prime on a function itself indicates a derivative with 
respect to the external momentum invariant $s$, so:
\beq
B'(x,y) &\equiv& \partial B(x,y)/\partial s\\
\label{eq:defdBds}
U'(x,y,z,u) &\equiv& \partial U(x,y,z,u)/\partial s .
\label{eq:defdUds}
\eeq
But, note that below primes also appear on fermion indices, where
they are used for a completely different 
purpose; fermions are labeled by indices $I$ and $I'$ if they combine 
to have a common squared mass $m_I^2$.

Each of the functions in eqs.~(\ref{eq:defTbar})-(\ref{eq:defdUds}) 
can be reduced to combinations of other
basis functions; see 
eqs.~(3.1), %Bprime
(3.22), %V
(4.14), %dBds
(4.26), %dUds
(5.3), %Iprime
and (6.18)
of ref.~\cite{evaluation} for formulas in the notation of the present paper. 
However, this explicit reduction is not done below in cases where 
it would needlessly complicate the expressions. 

\section{Fermion self-energy functions and pole masses at one-loop 
order\label{sec:oneloop}}
\setcounter{equation}{0}
\setcounter{footnote}{1}

In this section, I review the results at one-loop order. The
chirality-preserving and chirality-violating parts of the fermion
self-energy function are respectively:
\begin{widetext}
\beq
\Sigma^{(1)J}_I &=& 
y_{IKi} y^{JKi} \propB_{FS}(m_K^2,m_i^2) 
%\nonumber \\ &&
            + g^{aK}_I g^{aJ}_K \propB_{FV} (m_K^2,m_a^2)
,
\\ 
\Omega^{(1)IJ} &=& 
y^{IKi} y^{JK'i}m_{KK'} \propB_{\Fbar S}(m_K^2, m_i^2)
%\nonumber \\ &&
            - g^{aI}_K g^{aJ}_{K'} m^{KK'} \propB_{\Fbar V} (m_K^2,m_a^2),
\label{eq:Omega1}
\eeq
where
\beq
\propB_{FS}(x,y) &=& [(y-x-s) B(x,y) - A(x) + A(y)]/2 ,
\phantom{xxxx}
\\
\propB_{\Fbar S}(x,y) &=& -B(x,y) ,
\\
\propB_{FV}(x,v) &=& (v-x-s) B(x,v) + A(v) - A(x) 
%\nonumber \\ &&
%\!\!\!\!\!\!\!\!\!\!\!\!\!
+ s \deltaMSbar 
+ {\cal L}_v \bigl [ \lbrace v (s+x) - (x-s)^2 \rbrace B(x,v)
%\nonumber \\ &&
%\!\!\!\!\!\!\!\!\!\!\!\!\!
+ (x-s) A(v) \bigr ]/2 ,
\phantom{xxx}
\\
\propB_{\Fbar V}(x,v) &=& 3 B(x,v) + \xi B(x, \xi v) - 2 \deltaMSbar .
\eeq
These follow from direct evaluation of the first two Feynman diagrams,
with and without mass insertions, in fig.~\ref{fig:alldiagrams}.
[The result for $\Omega^{(1)}_{IJ}$ follows from eq.~(\ref{eq:Omega1})
by replacing the coupling parameters by their complex conjugates.]
Here I have allowed for the possibility of general fermion-fermion-vector
interactions and vector masses arising from spontaneous breaking of
gauge symmetries. In the following, I will also make use of:
\beq
\propB_{FS}'(x,y) &= &
[(y-x-s) B'(x,y) - B(x,y)]/2 ,
\phantom{xxxx}
\\
\propB_{\Fbar S}'(x,y) &=& -B'(x,y) ,
\eeq
where the prime means a derivative with respect to $s$.
 
In the special case of massless vectors
corresponding to unbroken gauge symmetries, one makes the 
simplifications:\footnote{The minus sign in eq.~(\ref{eq:BFbarV}) 
occurs because
the left-handed fermions with labels $K$ and $K'$ 
necessarily occur in conjugate representations of the unbroken gauge group.} 
\beq
g^{aK}_I g^{aJ}_K \propB_{FV} (m_K^2,m_a^2)
&\rightarrow&
g_a^2 C_a(I) \delta_I^J \propB_{FV} (m_I^2,0) ,
\\
- g^{aI}_K g^{aJ}_{K'} m^{KK'} \propB_{\Fbar V} (m_K^2,m_a^2)
&\rightarrow&
g_a^2 C_a(I) m^{IJ} \propB_{\Fbar V} (m_I^2,0),
\label{eq:BFbarV}
\eeq
where
\beq
\propB_{FV}(x,0) &=& 
%-(x+s) B(0,x) -A(x) + s \deltaMSbar +
%(1-\xi) [(x+s) B(0,x) + A(x) -s]
\xi [s -(x+s) B(0,x) -A(x)]  - s + s \deltaMSbar 
,
\\
\propB_{\Fbar V}(x,0) &=& (3 + \xi) B(0,x) -2 \deltaMSbar .
\eeq

It follows that the quantity defined in eq.~(\ref{eq:defPIoneloop})
is
\beq
\Pi_I^{(1)J} =  \Pi_I^{(1,0)J} + \Pi_I^{(1,1)J},
\eeq
where the contribution from scalar exchange is:
\beq
\Pi^{(1,0)J}_I &=& 
   (y_{IKi} y^{JKi} m_J^2 + m_{II'} y^{I'Ki} y_{J'Ki} m^{JJ'}) 
                    \propB_{FS}(m_K^2, m_i^2)/s 
\nonumber \\ &&
   + (m_{II'} y^{I'Ki} y^{JK'i} m_{KK'} + y_{IKi} y_{J'K'i} m^{KK'} m^{JJ'})
                    \propB_{\Fbar S}(m_K^2, m_i^2)
,
\eeq
and the contribution from vector exchange is:
\beq
\Pi^{(1,1)J}_I &=& 
   (g_I^{aK} g_K^{aJ} m_J^2 + m_{II'} g_K^{aI'} g_{J'}^{aK}) 
                      \propB_{FV} (m_K^2, m_a^2)/s 
\nonumber \\ &&
   - (m_{II'} g_K^{aI'} g_{K'}^{aJ} m^{KK'} 
   + g_I^{aK} g_{J'}^{aK'} m_{KK'} m^{JJ'})
                      \propB_{\Fbar V} (m_K^2, m_a^2)
.
\eeq
In the special case of massless vector bosons, the latter expression
reduces to:
\beq
\Pi^{(1,1)J}_I &=& 2 g_a^2 C_a(I) \delta_I^J m_I^2
[\propB_{FV} (m_I^2, 0)/s + \propB_{\Fbar V} (m_I^2, 0)],
\eeq
with the well-known limit:
\beq
\lim_{s \rightarrow m_I^2} 
\Pi^{(1,1)J}_I = 
2 g_a^2 C_a(I) \delta_I^J m_I^2 [5 - \deltaMSbar - 3 \lnbar m_I^2] .
\eeq
The above expressions can be inserted in the
formula eq.~(\ref{eq:polemasstwoloops}) to obtain the one-loop 
contribution
(and part of the two-loop contribution) to the pole squared mass.

\section{Fermion self-energy functions and pole masses 
at two-loop order\label{sec:twoloop}}
\setcounter{equation}{0}
\setcounter{footnote}{1}

In this section I present the results for the two-loop contributions to 
the self-energy functions and pole squared 
masses of fermions
as defined in 
fig.~\ref{fig:selfenergies}. The results are
divided into parts due to diagrams with no vector propagators, one
vector propagator, and two vector propagators, with superscripts 
$(2,0)$, $(2,1)$ and $(2,2)$ respectively:
\beq
\Sigma^{(2)J}_I &=&
\Sigma^{(2,0)J}_I +
\Sigma^{(2,1)J}_I +
\Sigma^{(2,2)J}_I
,
\\
\Omega^{(2)IJ} &=&
\Omega^{(2,0)IJ} +
\Omega^{(2,1)IJ} +
\Omega^{(2,2)IJ}
,
\\
\widetilde\Pi^{(2)}_I &=&
\widetilde\Pi^{(2,0)}_I +
\widetilde\Pi^{(2,1)}_I +
\widetilde\Pi^{(2,2)}_I
.
\label{eq:defPitwocomps}
\eeq
In the next three subsections, these results are expressed in terms of the
basis integrals. The two-loop fermion pole squared masses then 
follow by plugging eq.~(\ref{eq:defPitwocomps}) and the results of section
\ref{sec:oneloop} into 
eq.~(\ref{eq:polemasstwoloops}).

\subsection{Contributions from diagrams with no vector propagators}

The fermion self-energy functions 
following from the two-loop diagrams of fig.~\ref{fig:alldiagrams}
without vector or ghost propagators
are:
\beq
\Sigma^{(2,0)J}_I &=& 
y_{ILi} y^{JKj} \Bigl [
 y_{KNi} y^{LNj}  
  \propM_{SFFSF} (m_i^2,m_K^2,m_L^2,m_j^2,m_N^2) 
\nonumber \\ &&
+ y^{K'Ni} y_{L'Nj} m_{KK'} m^{LL'}
  \propM_{S\Fbar\Fbar SF} (m_i^2,m_K^2,m_L^2,m_j^2,m_N^2) 
\nonumber \\ &&
+  y_{KNi} y_{L'N'j} m^{LL'} m^{NN'} 
  \propM_{SF\Fbar S\Fbar} (m_i^2,m_K^2,m_L^2,m_j^2,m_N^2) 
\nonumber \\ &&
+  y^{K'Ni} y^{LN'j} m_{KK'} m_{NN'} 
  \propM_{SF\Fbar S\Fbar} (m_j^2,m_L^2,m_K^2,m_i^2,m_N^2) 
\nonumber \\ &&
+ \lambda^{ijk} y_{KL'k} m^{LL'}
  \propM_{SSF\Fbar S}(m_j^2,m_i^2,m_K^2,m_L^2,m_k^2)
+ \lambda^{ijk} y^{K'Lk} m_{KK'}
  \propM_{SSF\Fbar S}(m_i^2,m_j^2,m_L^2,m_K^2,m_k^2)
\Bigr ]
\nonumber \\ &&
+ y_{IKi} y^{JKj} \Bigl \lbrace
\frac{1}{2} \lambda^{ijkk} 
  \propY_{FSSS} (m_K^2, m_i^2, m_j^2, m_k^2)
+ \frac{1}{2} \lambda^{ikn}\lambda^{jkn} 
  \propV_{FSSSS} (m_K^2, m_i^2, m_j^2, m_k^2, m_n^2)
\nonumber \\ &&
+ {\rm Re}[y^{LNi} y_{LNj}] 
  \propV_{FSSFF} (m_K^2, m_i^2,m_j^2,m_L^2,m_N^2)
\nonumber \\ &&
+ {\rm Re}[y^{LNi} y^{L'N'j} m_{LL'} m_{NN'}] 
  \propV_{FSS\Fbar\Fbar} (m_K^2, m_i^2,m_j^2,m_L^2,m_N^2)
\Bigr \rbrace
\nonumber \\ &&
+ y_{IKi} y^{JLi} \Bigl [
y^{KNj} y_{LNj} 
  \propV_{SFFFS} (m_i^2, m_K^2, m_L^2, m_N^2, m_j^2)
\nonumber \\ &&
+ y_{K'Nj} y^{L'Nj} m^{KK'} m_{LL'} 
  \propV_{S\Fbar\Fbar FS} (m_i^2, m_K^2, m_L^2, m_N^2, m_j^2)
\nonumber \\ &&
+ y^{KNj} y^{L'N'j} m_{LL'} m_{NN'} 
  \propV_{SF\Fbar\Fbar S} (m_i^2, m_K^2, m_L^2, m_N^2, m_j^2)
\nonumber \\ &&
+ y_{K'Nj} y_{LN'j} m^{KK'} m^{NN'} 
  \propV_{SF\Fbar\Fbar S} (m_i^2, m_L^2, m_K^2, m_N^2, m_j^2)
\Bigr ]
,
%\eeq
%
%\beq
%
\\
\Omega^{(2,0)IJ} &=&
y^{ILi} y^{JKj} \Bigl [
y_{KNi} y_{LN'j}  m^{NN'}
  \propM_{SFFS\Fbar} (m_i^2,m_K^2,m_L^2,m_j^2,m_N^2)
\nonumber \\ &&
+y_{KNi} y^{L'Nj} m_{LL'}
  \propM_{SF\Fbar SF} (m_i^2,m_K^2,m_L^2,m_j^2,m_N^2)
%\nonumber \\ &&
+y^{K'Ni} y_{LNj} m_{KK'}
  \propM_{SF\Fbar SF} (m_j^2,m_L^2,m_K^2,m_i^2,m_N^2)
\nonumber \\ &&
+y^{K'Ni} y^{L'N'j} m_{KK'} m_{LL'} m_{NN'}
  \propM_{S\Fbar\Fbar S\Fbar} (m_i^2,m_K^2,m_L^2,m_j^2,m_N^2)
\nonumber \\ &&
+ \lambda^{ijk} y_{KLk}
  \propM_{SSFFS}(m_i^2,m_j^2,m_L^2,m_K^2,m_k^2)
+ \lambda^{ijk} y^{K'L'k} m_{KK'} m_{LL'}
  \propM_{SS\Fbar\Fbar S}(m_i^2,m_j^2,m_L^2,m_K^2,m_k^2)
\Bigr ]
\nonumber \\ &&
+ y^{IKi} y^{JK'j} m_{KK'} \Bigl \lbrace
\frac{1}{2} \lambda^{ijkk}
  \propY_{\Fbar SSS} (m_K^2, m_i^2, m_j^2, m_k^2)
+ \frac{1}{2} \lambda^{ikn}\lambda^{jkn}
  \propV_{\Fbar SSSS} (m_K^2, m_i^2, m_j^2, m_k^2, m_n^2)
\nonumber \\ &&
+ {\rm Re}[y^{LNi} y_{LNj}]
  \propV_{\Fbar SSFF} (m_K^2, m_i^2,m_j^2,m_L^2,m_N^2)
\nonumber \\ &&
+ {\rm Re}[y^{LNi} y^{L'N'j} m_{LL'} m_{NN'}]
  \propV_{\Fbar SS\Fbar\Fbar} (m_K^2, m_i^2,m_j^2,m_L^2,m_N^2)
\Bigr \rbrace
{}
\nonumber \\ &&
+ y^{IKi} y^{JLi} \Bigl [
y_{KNj} y_{LN'j} m^{NN'}
  \propV_{SFF\Fbar S} (m_i^2, m_K^2, m_L^2, m_N^2, m_j^2)
\nonumber \\ &&
+ y_{KNj} y^{L'Nj} m_{LL'}
  \propV_{SF\Fbar FS} (m_i^2, m_K^2, m_L^2, m_N^2, m_j^2)
\nonumber \\ &&
+ y^{K'Nj} y_{LNj} m_{KK'}
  \propV_{SF\Fbar FS} (m_i^2, m_L^2, m_K^2, m_N^2, m_j^2)
\nonumber \\ &&
+ y^{K'Nj} y^{L'N'j} m_{KK'} m_{LL'} m_{NN'}
  \propV_{S\Fbar\Fbar\Fbar S} (m_i^2, m_K^2, m_L^2, m_N^2, m_j^2)
\Bigr ]
,
\eeq
where the functions corresponding to each diagram are:
\beq
\propM_{SFFSF}(x,y,z,u,v) &=& 
\bigl [
(u x - y z - s v) M (x,y,z,u,v)
+ y U (u, y, x, v) + z U (x, z, u, v) 
- u U (y, u, z, v) 
\nonumber \\ &&
- x U (z, x, y, v)
- S(x, u, v) + S(y, z, v) + s B (x,z) B (y,u)
\bigr ]/2
,
\\
\propM_{SFFS\Fbar}(x,y,z,u,v) &=& 
\bigl [
(x + u - v - s) M(x,y,z,u,v)
- U(y, u, z, v) - U(z, x, y, v)
+ B(x, z) B(y, u)
\bigr ]/2
,\phantom{xxx}
\\
\propM_{SF\Fbar SF}(x,y,z,u,v) &=& 
\bigl [
(x - y - v) M(x,y,z,u,v)
+ U(x, z, u, v) - U(y, u, z, v)
+ B(x, z) B(y, u)
\bigr ]/2
,
\\
\propM_{SF\Fbar S\Fbar}(x,y,z,u,v) &=& 
\bigl [
(u - y - s) M(x,y,z,u,v)
+ U(x, z, u, v) - U(z, x, y, v)
\bigr ]/2
,
\\
\propM_{S\Fbar\Fbar SF}(x,y,z,u,v) &=& 
\bigl [
(x - y - z + u) M(x,y,z,u,v)
+ U(x, z, u, v) + U(u, y, x, v)
- U(y, u, z, v) 
\nonumber \\ &&
- U(z, x, y, v)
\bigr ]/2
,
\\
\propM_{S\Fbar\Fbar S\Fbar}(x,y,z,u,v) &=& -M(x,y,z,u,v)
, 
\eeq
and
\beq
\propM_{SSFFS}(x,y,z,u,v) &=& 
\bigl [
(v-z-u) M(x,y,z,u,v) + U(z,x,y,v) + U(u,y,x,v) - B(x,z) B(y,u)
\bigr ]/2
,
\phantom{xxx}
\\
\propM_{SSF\Fbar S}(x,y,z,u,v) &=& 
\bigl [
(x-z-s) M(x,y,z,u,v) - U(y,u,z,v) + U(u,y,x,v) \bigr ]/2
,
\\
\propM_{SS\Fbar\Fbar S}(x,y,z,u,v) &=& 
-M(x,y,z,u,v)
,
\eeq
and
\beq
\propY_{FSSS}(x,y,z,u) &=& A(u) \left [
A(y) - A(z) + (y-x-s) B(x,y) - (z-x-s) B(x,z)
\right ]/2(y-z)
,
\\
\propY_{FSSS}(x,y,y,u) &=& A(u) \left [
1 + A(y)/y + B(x,y) + (y-x-s) B(x,y')
\right ]/2
,
\\
\propY_{\Fbar SSS}(x,y,z,u) &=& A(u) \left [
B(x,z) - B(x,y)
\right ]/(y-z)
,
\\
\propY_{\Fbar SSS}(x,y,y,u) &=& -A(u) B(x,y')
,
\eeq
and
\beq
\propV_{FSSSS}(x,y,z,u,v) &=& \bigl [ 
(s+x-y) U(x,y,u,v) - (s+x-z) U(x,z,u,v) - I(y,u,v) + I(z,u,v) \bigr ]/2 (y-z)
,
\phantom{xxx}
\\
\propV_{FSSSS}(x,y,y,u,v) &=& \bigl [ 
(y-x-s) V(x,y,u,v) - U(x,y,u,v) - I(y',u,v) \bigr ]/2 
,
\\
\propV_{\Fbar SSSS}(x,y,z,u,v) &=& \bigl [ 
U(x,y,u,v) - U(x,z,u,v) \bigr ]/(y-z)
,
\\
\propV_{\Fbar SSSS}(x,y,y,u,v) &=& -V(x,y,u,v)
,
\eeq
and
\beq
\propV_{SFFFS}(x,y,z,u,v) &=& \bigl [ 
  (s-x+y)(y+u-v) U(x,y,u,v) 
  -y S(x,u,v) + (y+u-v) I(y,u,v) 
\nonumber \\ &&
  + [A(u) - A(v)][(s-x +y) B(x,y)+ A(y)] \bigr ]/4(y-z) 
+ (y\leftrightarrow z)
,
\phantom{xxx}
\\
\propV_{SFFFS}(x,y,y,u,v) &=& \bigl [ 
  (s-x+y)(v-y-u) V(x,y,u,v) 
  +(s - x + 2 y + u - v) U(x,y,u,v)
\nonumber \\ &&
  -S(x,u,v) 
  + (y+u-v) I(y',u,v)  + I(y,u,v)
\nonumber \\ &&
  + [A(u) - A(v)][(s-x+y) B(x,y') + B(x,y) 
  + 1 + A(y)/y] 
\bigr ]/4
,
\\
\propV_{SFF\Fbar S}(x,y,z,u,v) &=& 
  \bigl [ y U(x,y,u,v) - z U(x,z,u,v) \bigr ]/(y-z)
,
\\
\propV_{SFF\Fbar S}(x,y,y,u,v) &=& U(x,y,u,v) - y V(x,y,u,v)
,
\\
\propV_{SF\Fbar FS}(x,y,z,u,v) &=& 
  \bigl [ (y+u-v) U(x,y,u,v) 
  + [A(u) - A(v)]B(x,y)  \bigr ]/2 (y-z) + (y \leftrightarrow z)
%    \bigl [ (y+u-v) U(x,y,u,v) - (z+u-v) U(x,z,u,v) 
%  + [A(u) - A(v)][B(x,y) - B(x,z)] \bigr ]/2 (y-z)
,
\\
\propV_{SF\Fbar FS}(x,y,y,u,v) &=& 
[U(x,y,u,v) + (v-y-u) V(x,y,u,v) + [A(u) - A(v)] B(x,y')]/2
,
\\
\propV_{SF\Fbar\Fbar S}(x,y,z,u,v) &=& 
  \bigl [ (s-x+y) U(x,y,u,v) - (s-x+z) U(x,z,u,v) 
  + I(y,u,v) - I(z,u,v) \bigr ]/2 (y-z)
,
\\
\propV_{SF\Fbar\Fbar S}(x,y,y,u,v) &=& 
[U(x,y,u,v) + (x-y-s) V(x,y,u,v) + I(y',u,v)]/2
,
\\
\propV_{S\Fbar\Fbar FS}(x,y,z,u,v) &=& 
  \bigl \lbrace (s-x+y)(y+u-v) U(x,y,u,v)  
  + (u - v + y) I(y,u,v) 
\nonumber \\ &&
  + [A(u) - A(v)] [(s-x+y) B(x,y) + A(y)]  
  \bigr \rbrace/4 y (y-z) + (y \leftrightarrow z)
\nonumber \\ &&
  + \lbrace
     2(u-v) [S(x,u,v) + x T(x,u,v)] + u (x+u-v-s) T(u,x,v) 
\nonumber \\ &&
     + v (s + u - v - x)T(v,x,u)     
     + (v-u)[A(x) + A(u) + A(v) -x - u - v + s/4] \rbrace/4 y z
,
\phantom{xxxx}
\\
\propV_{S\Fbar\Fbar FS}(x,y,y,u,v) &=& 
\Bigl [
y (y+u-v)(x-y-s) V(x,y,u,v) + (s v - s u + x u - x v + y^2) U(x,y,u,v)
\nonumber \\ &&
+ 
2 (u - v) [S(x,u,v) + x T(x, u, v)] + u (x + u - v -s) T(u, x, v) 
\nonumber \\ &&
+ v (s + u - v - x) T(v, x, u) 
+ y (y+u-v) I(y',u,v) 
\nonumber \\ &&
+ (v-u) [I(y,u,v) + A(x) + 2 A(v) -x -u-v + s/4]
\nonumber \\ &&
+ [A(u) - A(v)][y (s-x+y) B(x,y') 
+ (x-s) B(x,y) + y - u + v]
\Bigr ]/4 y^2
,
\\
\propV_{S\Fbar\Fbar\Fbar S}(x,y,z,u,v) &=& 
\bigl [ U(x,y,u,v) - U(x,z,u,v) \bigr ]/(y-z)
,
\\
\propV_{S\Fbar\Fbar\Fbar S}(x,y,y,u,v) &=& -V(x,y,u,v)
,
\eeq
and
\beq
\propV_{FSSFF}(x,y,z,u,v) &=&
\bigl [
(x-y+s)(y-u-v) U(x,y,u,v) + y S(x,u,v) + (u+v-y) I(y,u,v) 
\nonumber \\ &&
+ [A(u) + A(v)][ (s+x-y) B(x,y) - A(y)] \bigr ]/2 (y-z)
+ (y \leftrightarrow z)
,
\\
\propV_{FSSFF}(x,y,y,u,v) &=&
\bigl [
(x-y+s)(u+v-y) V(x,y,u,v) + 
(x-2y + u+v+s) U(x,y,u,v) 
\nonumber \\ &&
+ S(x,u,v)
+ (u+v-y) I(y',u,v) - I(y,u,v) 
\nonumber \\ &&
+ [A(u) + A(v)][ (s+x-y) B(x,y') - B(x,y) 
- A(y)/y -1] \bigr ]/2 
,
\\
\propV_{FSS\Fbar\Fbar}(x,y,z,u,v) &=&
\bigl [
(y-x-s) U(x,y,u,v) + I(y,u,v)
\bigr ]/(y-z) 
+ (y \leftrightarrow z)
,
\\
\propV_{FSS\Fbar\Fbar}(x,y,y,u,v) &=&
(x-y+s) V(x,y,u,v) + U(x,y,u,v) + I(y',u,v)
,
\\
\propV_{\Fbar SSFF}(x,y,z,u,v) &=&
\bigl [
(y-u-v) U(x,y,u,v) + [A(u) + A(v)] B(x,y)
\bigr ]/(y-z) 
+ (y \leftrightarrow z)
,
\\
\propV_{\Fbar SSFF}(x,y,y,u,v) &=&
(u+v-y) V(x,y,u,v) + U(x,y,u,v) + [A(u) + A(v)] B(x,y')
,
\\
\propV_{\Fbar SS\Fbar\Fbar}(x,y,z,u,v) &=&
2 \bigl [
U(x,z,u,v) - U(x,y,u,v)
\bigr ]/(y-z) 
,
\\
\propV_{\Fbar SS\Fbar\Fbar}(x,y,y,u,v) &=&
2 V(x,y,u,v)  
.
\eeq
Note that for diagrams with the $\propV$ or $\propY$ topology,
the limits of identical (or degenerate) squared masses in the
second and third arguments required separate expressions to avoid
the threats of vanishing denominators.
Also, the limit $y\rightarrow 0$ appropriate for massless
fermions is only needed when the corresponding propagator
has no mass insertion. For the case $\propV_{SFF\Fbar S}$,
this limit is trivial, since $y V(x,y,u,v)$ vanishes as 
$y \rightarrow 0$. The remaining non-trivial case involving
a massless internal fermion is
\beq
\propV_{SFFFS}(x,0,0,u,v) &=& \bigl [ 
  (s-x)(v-u) \overline{V}(x,0,u,v) 
  +(s - x + u - v) U(x,0,u,v)
  -S(x,u,v) 
\nonumber \\ &&
  + I(0,u,v)
  + 2 [A(u) - A(v)][s - A(x) - x B(0,x)]/(s-x) 
\bigr ]/4
,
\eeq
where the function $\overline{V}(x,0,u,v)$ is defined in eq.~(2.20)
of ref.~\cite{Martin:2003it}
and given in terms of the basis integral functions in 
eqs.~(A.11)-(A.13) of that paper.
Two useful special cases are:
\beq
\propV_{SFFFS}(x,0,0,u,u) &=& \bigl [ 
  (s-x) U(x,0,u,u) 
  -S(x,u,u) 
  + I(0,u,u)
\bigr ]/4
,
\eeq
and for $s=x$ with $u\not= v$,
\beq
\propV_{SFFFS}(x,0,0,u,v) |_{s=x} &=&  
\lbrace [A(u) - u] [A(v) - v] + u v + (u+v) I(0,u,v)/2 \rbrace/2 (u-v)
\nonumber \\ &&
+ \lbrace 2 (u-v+x) T(x,u,v) -3 v T(v, u, x) + 5 u T (u,x,v)
+ I(0,u,v) 
\nonumber \\ &&
-3 A(u) + A(v) 
- A(x) + 3 u - v
+ 2 [A(v) - A(u)] A(x)/x + 3x/4 \rbrace/8
.
\eeq

The corresponding contribution to the two-loop pole masses [see
eqs.~(\ref{eq:polemasstwoloops}), (\ref{eq:defpitwotilde}), and 
(\ref{eq:defPitwocomps})] is
given in terms of the above results by:
\beq 
\widetilde\Pi^{(2,0)}_{I} &=& 
\Sigma^{(2,0)I}_I + \Sigma^{(2,0)I'}_{I''} m_{II'} m^{II''}/m_I^2
+ m_{II'} \Omega^{(2,0)II'} + \Omega^{(2,0)}_{II'} m^{II'}
+ 
\Bigl (
y_{IKi} y^{JKi} y_{JLj} y^{ILj} m_I^2 
\nonumber \\  &&
+y^{I'Ki} y_{JKi} y^{JLj} y_{I''Lj} m_{II'} m^{II''} 
+y_{IKi} y^{JKi} y^{J'Lj} y_{I'Lj} m_{JJ'} m^{II'}
\Bigr )
\propB_{FS}(m_K^2, m_i^2) \propB_{FS}(m_L^2, m_j^2)/m_I^4
\nonumber \\ &&
+ y_{IKi} y_{JK'i} y^{JLj} y^{IL'j} m^{KK'} m_{LL'}
\propB_{\Fbar S}(m_K^2, m_i^2) \propB_{\Fbar S}(m_L^2, m_j^2)
+ \Bigl (
2 {\rm Re}[y^{I'Ki} y_{JKi} y^{JLj} y^{IL'j} m_{II'} m_{LL'}]
\nonumber \\ &&
+ y_{IKi} y^{JKi} \bigl [ y_{JLj} y_{I'L'j} m^{II'} m^{LL'} 
+y^{J'Lj} y^{IL'j} m_{JJ'} m_{LL'}\bigr ] 
\Bigr )
\propB_{FS}(m_K^2, m_i^2) \propB_{\Fbar S}(m_L^2, m_j^2)/m_I^2 
\nonumber \\ &&
+ \Bigl (
 \bigl [ |y^{IKi}|^2 + |m^{II'} y_{I'Ki}|^2/m_I^2 \bigr ]
 \propB_{FS} (m_K^2, m_i^2) 
 + 2 {\rm Re} \bigl [m_{II'} y^{I'Ki} y^{IK'i} m_{KK'} \bigr ] 
 \propB_{\Fbar S} (m_K^2, m_i^2)
\Bigr ) 
\nonumber \\ &&
\Bigl (
 \bigl [ |y^{ILj}|^2 + |m^{II''} y_{I''Lj}|^2/m_I^2 \bigr ]
 \bigl [\propB_{FS}' (m_L^2, m_j^2) -\propB_{FS}(m_L^2, m_j^2)/m_I^2 
 \bigr ]
\nonumber \\ &&
 + 2 {\rm Re}\bigl [m_{II''} y^{I''Lj} y^{IL'j} m_{LL'} \bigr ]
 \propB_{\Fbar S}'(m_L^2, m_j^2)
\Bigr )
,
\eeq
where $s =m_I^2 + i \varepsilon$ is taken everywhere on the right side.
There is no sum on the index $I$, but all other indices
(including $I'$ and $I''$) are summed over as usual.

\subsection{Contributions from diagrams with one vector propagator}

Next we consider the contributions coming from the two-loop diagrams
in fig.~\ref{fig:alldiagrams} that
involve exactly one vector line. It is
convenient to organize these in terms of certain linear combinations of
the quadratic Casimir group theory invariants for the fermions and scalars 
appearing in
the diagrams, as follows:
\beq
\Sigma^{(2,1)J}_I &=&
y_{IKi} y^{JKi} g_a^2 
\Bigl \lbrace
  \frac{1}{2} [C_a(K) + C_a(i) - C_a(I)] G_{FS}(m_K^2, m_i^2)
\nonumber \\ &&
  + [C_a(K) - C_a(i)] G_{FFS}(m_I^2, m_K^2, m_i^2)
  + C_a(I) H_{FFS}(m_I^2, m_K^2, m_i^2) 
\Bigr \rbrace
\nonumber \\ &&
+ \left (y_{IKi} y_{J'K'i} m^{JJ'} m^{KK'} +
   y^{I'Ki} y^{JK'i} m_{II'} m_{KK'} \right ) g_a^2
\Bigl \lbrace 
  [C_a(K) - C_a(i)] G_{\Fbar\Fbar S}(m_I^2, m_K^2, m_i^2) 
\nonumber \\ &&
  + C_a(I) H_{\Fbar\Fbar S}(m_I^2, m_K^2, m_i^2) 
\Bigr \rbrace
,
\\ {}
\Omega^{(2,1)IJ} &=& 
y^{IKi} y^{JK'i} m_{KK'} g_a^2 
\Bigl \lbrace
  \frac{1}{2} [C_a(K) + C_a(i) - C_a(I)] G_{\Fbar S}(m_K^2, m_i^2)
\nonumber \\ &&
  + [C_a(K) - C_a(i)] G_{F\Fbar S}(m_I^2, m_K^2, m_i^2)
  + C_a(I) H_{F\Fbar S}(m_I^2, m_K^2, m_i^2) \Bigr \rbrace
\nonumber \\ &&
(y^{IKi} y_{J'Ki} m^{JJ'} + y_{I'Ki} y^{JKi} m^{II'}) g_a^2
\Bigl \lbrace 
  [C_a(K) - C_a(i)] G_{\Fbar FS}(m_I^2, m_K^2, m_i^2) 
\nonumber \\ &&
  + C_a(I) H_{\Fbar FS}(m_I^2, m_K^2, m_i^2) 
\Bigr \rbrace
.
\eeq
The loop integral functions appearing here are:
\beq
G_{FS}(y,z) &=&
[y^2 - (z-s)^2] M(y,y,z,z,0) 
+ S(0,y,z)/2 
+ (3y - z +s) T(y,0,z)/2
+ (y+s) T(z,0,y) 
\nonumber \\ &&
+ 2 (y-z+s) {\overline{T}} (0,y,z)
+ s B(y,z)^2 
+ [ 3 (s-y+z) A(y) + 3 (s+y-z) A(z)/2 + 2 (y-z)^2 
\nonumber \\ &&
- 2 s (y+z)] B(y,z')
+ [(s+3y-z) A(y)/2y + (y+s-z/2) A(z)/z +6 (z-y-s)] B(y,z)
\nonumber \\ &&
-33s/8 - 3y/2 + 5z/2 + A(y)/2 - A(y)^2/2y + 2A(z) 
  -2y A(z)/z + 3 A(y) A(z)/z - 3 A(z)^2/2z
\nonumber \\ &&
+ \deltaMSbar 
[ y (z-y+s) B(y,z') + (z+y-s) B(y,z)/2 + (1/2 + y/z) A(z) - A(y)/2 +y-s/8 ]
\nonumber \\ &&
+ m_{\epsilon}^2 [ (y -z + s) B(y,z') - B(y,z) - A(z)/z - 1]
,
\\
G_{\Fbar S}(y,z) &=&
2 (y+z-s) M(y,y,z,z,0) 
+ 2 T(y,0,z) + 2 T(z,0,y) + 4 {\overline{T}} (0,y,z)
+ B(y,z)^2
+ [ 2 y - 2 z - 2s
\nonumber \\ &&
+ 3(s-y-z) A(y)/y +  3 A(z)] B(y,z')
+ [5 A(y)/y + 2 A(z)/z - 14] B(y,z)
+ 3 A(y) A(z)/y z 
\nonumber \\ &&
-3 A(y)/y - 2 A(z)/z -6
+ \deltaMSbar [(s-y-z) B(y,z') + B(y,z) + A(z)/z - 1]
+ 2 m_{\epsilon}^2 B(y,z')
,
\eeq
and
\beq
G_{FFS}(x,y,z) &=& 
[(y-z)^2 - x z + s x - s z] M(0,y,x,z,y)
+ [(y-z)^2 - (s + x) (y + z)/2] M(0,z,x,y,z)
\nonumber \\ &&
- x U(0,x,y,z)/2 + (y-z +s/2 + x/2) U(x,0,y,y)
+ (z-y) U(x,0,z,z) -y  U(y, z, x, y) 
\nonumber \\ &&
- y U(z, y, x, z)
+ 5 S(0,y,z)/4 + S(x,z,z)/2 - S(x,y,y)/2 
+ (z-y/2) T(z,0,y)
\nonumber \\ &&
+ (s - z + 3 y) T(y, 0, z)/4
+ [(y+z-s)(y-z) - 3(s - z + y) A(z)/4 
\nonumber \\ &&
+ 3(s + z - y) A(y)/2] B(y,z')
+ [(z/4 -y/2) A(z)/z + (s+3y-z) A(y)/4y -s/2 B(0,x) 
\nonumber \\ &&
-s/2] B(y,z)
+ (x+s) B(0,x) + 3 A(z)^2/4z + 3 A(y) A(z)/2z - A(y)^2/4y + A(x) 
\nonumber \\ &&
+ A(y)/4
- (3/2 + y/z) A(z) + 5 (y+z)/4 - 9s/16
\nonumber \\ &&
+ \deltaMSbar [(z-y+s) y B(y,z') + (5y - 3z + 3s) B(y,z)/2 
+ 3 A(y)/2 + (y/z - 3/2) A(z) + y - s/8]/2
\nonumber \\ &&
+ m_{\epsilon}^2 [1 + B(y,z) + (z-y-s) B(y,z') + A(z)/z]/2
,
\\
G_{\Fbar FS}(x,y,z) &=&
\lbrace 4 y M(0,y,x,z,y) 
+ (2 y + 2 z - x - s) M(0,z,x,y,z)
+ U(0, x, y, z) - 2 U(x, 0, z, z) 
\nonumber \\ &&
- 2 U(y, z, x, y)
-T(z,0,y) 
+ [B(0,x) + 1 - A(z)/z] B(y,z)
-4 B(0,x) + 1\rbrace/4
,
\\
G_{F\Fbar S}(x,y,z) &=&
(2x +y-z-s) M(0,y,x,z,y) 
+ (2 y - 2 z - x - s) M(0,z,x,y,z)/2
+ U(x, 0, y, y) 
\nonumber \\ &&
- U(x, 0, z,z) - U(y, z, x, y)
- U(z, y, x, z)/2
+ 2 T(y,0,z) - T(z,0,y)/2 
\nonumber \\ &&
+ [ 3(s-y-z) A(y)/2y - 3 A(z)/2 +y + 3z - s]B(y,z')
+ [7 A(y)/2y - A(z)/2z 
\nonumber \\ &&
- 5/2] B(y,z)
+ 3 A(y) A(z)/2 y z
- 3 A(y)/2 y
- A(z)/z -1/2
\nonumber \\ &&
+ \deltaMSbar [(s-y-z) B(y,z') + 5 B(y,z) + A(z)/z + 1]/2
-m^2_{\epsilon} B(y,z')
,
\\
G_{\Fbar\Fbar S}(x,y,z) &=&
\lbrace
2 (2 s - x + y - z) M(0, y, x, z, y) 
+(2 y - 2 z -x -s) M(0, z, x, y, z) 
+2 U(x, 0, y, y) 
\nonumber \\ &&
- 2 U(x, 0, z, z) 
- 2 U(y, z, x, y) - U(z, y, x, z)
- 2 T(y, 0, z) - T(z, 0, y) 
\nonumber \\ &&
+ [3 - 2 A(y)/y - A(z)/z] B(y,z) + 3 \rbrace/4
,
\eeq
and
\beq
&&
H_{FFS}(x,y,z) = 
[(y-z)^2 - x z + s x - s z] M(0, y, x, z, y) 
+ [(x+s)(y+z)/2 - (y-z)^2] M(0, z, x, y, z) 
\nonumber \\ && \qquad
+ (x+s) (z -x - y) V(0, x, y, z)
+ [(3+s/x)(z-y) + s] U(0,x,y,z)/2
+ (x/2 + y - z + s/2) U(x,0,y,y)
\nonumber \\ && \qquad
+ (y-z) U(x,0,z,z)
-y U(y,z,x,y) + y U(z,y,x,z)
+ [s x - 2 s y + 3 x y + 2 y^2  - x z - 2 y z] T(y,0,z)/4x
\nonumber \\ && \qquad
+ (y-z+s) \overline{T}(0,y,z)
+ (x y +y z-z^2 + z s) T(z,0,y)/2x
+ (3/4 + y/x - z/x) S(0,y,z)
- S(x,y,y)/2 
\nonumber \\ && \qquad
- S(x,z,z)/2 
+(x+y-z) I(x',y,z) 
+ (x-y+z) I(x,y,z)/2x
+ [(y-z)^2 - s (y+z) 
\nonumber \\ && \qquad
+ 3 (z-y+s) A(y)/2 
+ 3 (y-z+s) A(z)/4] B(y,z')
+ [(2 z - 2y -3s/2) B(0, x) 
+ (s + 3y - z) A(y)/4y 
\nonumber \\ && \qquad
+ (y/2z -1/4 ) A(z)  
  -5s/2 - 3y 
  + 3z] B(y,z)
+ [A(y) - A(z)][2 s B(0,x') - (1+s/x) B(0,x)/2]
\nonumber \\ && \qquad
- A(y)^2/4y - 3 A(z)^2/4z 
+3[A(y)-A(z)] A(x)/2x + 3 A(y) A(z)/2z
+ (z/2x - y/2x + 1/4) A(y)
\nonumber \\ && \qquad
+ (1 -y/z)(1 + z/2x) A(z)
-25s/16 - 3y/4 - s y/8x + y^2/2x + 5z/4 + s z/8x - z^2/2x
\nonumber \\ && \qquad
+ \deltaMSbar [2 y (z-y+s) B(y,z') + (5y-3z+3s) B(y,z)
+ (2y/z -3) A(z) + 3 A(y) + 2y - 3s/4]/4
\nonumber \\ && \qquad
+ m^2_{\epsilon} [(y-z+s) B(y,z') - B(y,z) - A(z)/z -1]/2
+(1-\xi) \lbrace
(s y - s x - 3 x^2 
- x y - s z + x z) U(0, x, y, z) 
\nonumber \\ && \qquad
+ 2 (s + x) (x + y - z) x V(0, x, y, z)
- 2 s x (x + y - z) U'(0, x, y, z) 
+ (s + 2 x - y + z) y T(y, 0, z) 
\nonumber \\ && \qquad
+ (z - y - s) z T(z, 0, y)
+ (s + y - z) x \overline{T}(0, y, z) 
+ 2 (x - y + z) S(0, y, z) 
+ (y - x - z) I(x, y, z)
\nonumber \\ && \qquad
+ 2 (z -x - y) x I(x', y, z) 
+ 2 x s [(y-z+s) B(y,z) + 2 A(y) - 2 A(z)] B'(0,x)
\nonumber \\ && \qquad
+ 2 x (s+y-z) B(0,x) B(y,z)
+ [A(x) + (s+3x) B(0,x)] [A(y) - A(z)]
+ (y-z) A(y) 
\nonumber \\ && \qquad
+ (y-2x-z) A(z)
+ (x-y+z) (y+z-s/4)
\rbrace/2x
,
\\
&& 
H_{F\Fbar S}(x,y,z) =
(2x + y -z -s) M(0, y, x, z, y) 
+ (x/2 + s/2 -y + z) M(0, z, x, y, z) 
+ U(x, 0, y, y) 
+ U(x, 0, z, z) 
\nonumber \\ && \qquad
- 4 U(0, x, y, z) 
- U(y, z, x, y) + U(z, y, x, z)/2 
+ 8 x V(0, x, y, z)
+ 2 T(y,0,z) 
+ T(z,0,y)/2
+ 2 \overline{T} (0,y,z)
\nonumber \\ && \qquad
+ [3 (s - y - z) A(y)/2y + 3 A(z)/2 -s + y - z ] B(y,z')
+ [7 A(y)/2 y + A(z)/2z - 2 B(0,x) - 15/2] B(y,z)
\nonumber \\ && \qquad
+ 3 [A(z)/z - 1] A(y)/2y  - A(z)/z - 7/2
+ \deltaMSbar [3 + A(z)/z + 5 B(y,z) + (s-y-z) B(y,z')]/2
+ m^2_{\epsilon} B(y,z')
\nonumber \\ && \qquad
+ (1-\xi) 
\lbrace
\overline{T} (0, y, z) 
- 2 s U'(0, x, y, z) 
- 2 x V(0, x, y, z)
- U(0, x, y, z) 
+ 2 [B(0, x) + s B'(0,x)] B(y, z)
\rbrace
,
\\
&& H_{\Fbar FS}(x,y,z) = \lbrace
4 y M(0, y, x, z, y) 
+ (x - 2 y - 2 z + s) M(0, z, x, y, z) 
- 5 U(0, x, y, z) +2 U(x, 0, z, z) 
- 2 U(y, z, x, y) 
\nonumber \\ && \qquad
+ 8 (x + y - z) V(0, x, y, z)
+ T(z, 0, y) 
+ [3 B(0,x) + A(z)/z - 1] B(y,z)
+8 [A(z) - A(y)] B(0,x')
- 1 
\rbrace/4
\nonumber \\ && \qquad
+ \deltaMSbar/2
+ (1 - \xi) 
\lbrace
(z-x-y) [x V(0,x,y,z) 
+ s U'(0,x,y,z)] 
+ (z-y) U(0,x,y,z)
-y T(y,0,z) + z T(z,0,y)
\nonumber \\ && \qquad
+ [(s+y-z) B(y,z) - A(y) + A(z)] x B'(0,x)
+ [A(z) - A(y)][(1+2x/s) B(0,x) + 2 A(x)/s]
\rbrace/2x
,
\\
&&H_{\Fbar\Fbar S}(x,y,z) = \lbrace
2 (2 s - x + y - z) M(0, y, x, z, y) +
(s + x - 2 y + 2 z) M(0, z, x, y, z) 
+ 2 U(x, 0, y, y) 
\nonumber \\ &&\qquad
+ 2 U(x, 0, z, z) 
+ U(z, y, x, z) 
- 2 U(y, z, x, y) -4 (s + x) V(0, x, y, z)
- 2 T(y, 0, z) + T(z, 0, y) 
+ 4 I(x',y,z) 
\nonumber \\ && \qquad
+ [1 - 2 A(y)/y + A(z)/z] B(y,z) + 1\rbrace/4
+ (1- \xi) [ (s + x) V(0, x, y, z) - U(0,x,y,z)
- s U'(0,x,y,z) 
\nonumber \\ && \qquad
-I(x',y,z)  + s B(y,z) B'(0,x)]
.
\eeq
Note that the terms involving 
functions $G_{FS}$ and $G_{\Fbar S}$ are the 
only parts
that contribute when the external fermion is neutral; they and the 
functions
$G_{FFS}$, $G_{F\Fbar S}$, $G_{\Fbar FS}$ and $G_{\Fbar\Fbar S}$
are 
each gauge-invariant and finite in the limit $s\rightarrow m_I^2$. 
In contrast, the functions
$H_{FFS}$, $H_{F\Fbar S}$, $H_{\Fbar FS}$ and $H_{\Fbar\Fbar S}$
are not by themselves gauge-invariant and have logarithmic
divergences as  
$s\rightarrow m_I^2$, but they combine with one-loop parts to
give a finite, gauge-invariant pole mass. This cancellation provides
a nice check on the calculations. The resulting contribution to the
pole squared mass  [see
eqs.~(\ref{eq:polemasstwoloops}), (\ref{eq:defpitwotilde}), and
(\ref{eq:defPitwocomps})]
is:
\beq 
\widetilde\Pi^{(2,1)}_{I} &=& 
\left [|y^{IKi}|^2 + |m^{II'} y_{I'Ki}|^2/m_I^2  
\right ] 
g_a^2 \Bigl \lbrace
\frac{1}{2} [C_a(K) + C_a(i) - C_a(I)] f_1 (m_I^2, m_K^2, m_i^2)
\nonumber \\ &&
+ [C_a(K) - C_a(i)] f_2 (m_I^2, m_K^2, m_i^2)
+ C_a(I)  f_3 (m_I^2, m_K^2, m_i^2)
\Bigr \rbrace
\nonumber \\ &&
+ 2 {\rm Re}\left [y^{IKi} y^{I'K'i} m_{II'} m_{KK'} \right ] 
g_a^2 \Bigl \lbrace
\frac{1}{2} [C_a(K) + C_a(i) - C_a(I)] f_4 (m_I^2, m_K^2, m_i^2)
\nonumber \\ &&
+ [C_a(K) - C_a(i)] f_5 (m_I^2, m_K^2, m_i^2)
+ C_a(I)  f_6 (m_I^2, m_K^2, m_i^2)
\Bigr \rbrace
,
\eeq
where:
\beq
f_1(x,y,z) 
&=& \lim_{s\rightarrow x} G_{FS}(y,z)
\\
&=& 
[y^2 - (x-z)^2] M(y,y,z,z,0) 
+ S(0,y,z)/2 
+ 2 (x+y-z) \overline{T} (0,y,z)
+ (x+3y-z) T(y,0,z)/2
\nonumber \\ &&
+ (x+y) T(z,0,y)
+ B(y,z') \lbrace 3 (x - y + z) A(y) + 3 (x + y - z) A(z)/2
          + 2 [(y-z)^2 - x (y+z)] \rbrace
\nonumber \\ &&
+ x B(y,z)^2
+ B(y,z) [ (x + 3 y - z) A(y)/2y + (2x + 2y -z) A(z)/2z 
                 + 6 (z-x-y) ]
\nonumber \\ &&
+ [1 - A(y)/y] A(y)/2
+ [2z - 2 y + 3 A(y) - 3 A(z)/2] A(z)/z
-33x/8 - 3y/2 + 5z/2
\nonumber \\ &&
+ \deltaMSbar [y (x-y + z) B(y,z') + (y+z-x) B(y,z)/2 + (1/2 + y/z) A(z)
- A(y)/2 + y -x/8]
\nonumber \\ &&
+ m^2_{\epsilon} [(x+y-z) B(y,z') - B(y,z) - A(z)/z - 1]
,
\\
f_2 (x,y,z) 
&=& \lim_{s\rightarrow x} [G_{FFS}(x,y,z) + 2 x G_{\Fbar FS}(x,y,z)]
\\
&=& 
(x + y - z)^2 M(0, y, x, z, y) 
+ [(y - z)^2 - x^2] M(0, z, x, y, z) 
+ (x + y - z) U(x, 0, y, y) 
\nonumber \\ &&
+ (z - x - y) U(x, 0, z, z) 
- (x + y) U(y, z, x, y) 
- y U(z, y, x, z)
+ 5 S(0, y, z)/4 
- S(x, y, y)/2 
\nonumber \\ &&
+ S(x, z, z)/2 
+ (x + 3 y - z) T(y, 0, z)/4 
+ (2 z -x - y) T(z, 0, y)/2 
\nonumber \\ &&
+ [3 (x - y + z) A(y)/2 
+ 3 (z - x - y) A(z)/4 + (z-y)(x - y - z)] B(y,z')
\nonumber \\ &&
+ [(x + 3 y - z) A(y)/4 y 
+ (z - 2 x - 2 y) A(z)/4 z] B(y,z)
- A(y)^2/4 y 
+ 3 A(y) A(z)/2 z 
\nonumber \\ &&
+ 3 A(z)^2/4 z
+ A(x) 
+ A(y)/4 
- (3/2 + y/z) A(z) 
+ 5 (y+z)/4 - x/16
\nonumber \\ &&
+ \deltaMSbar [
2 y (x - y + z) B(y, z')
+ (3 x + 5 y - 3 z) B(y,z)
+ (2y/z - 3) A(z) + 3 A(y) + 2 y -x/4]/4
\nonumber \\ &&
+ m^2_{\epsilon} [
(z-x-y) B(y,z') + B(y,z) + A(z)/z + 1]/2
,
\\
f_3 (x,y,z) 
&=& \lim_{s\rightarrow x} \Bigl \lbrace
H_{FFS}(x,y,z) + 2 x H_{\Fbar FS}(x,y,z)
+ 2 [\propB_{FV}(x,0) + x \propB_{\Fbar V}(x,0)] \propB'_{FS}(y,z)
\nonumber \\ &&
+ 2 [\propB'_{FV}(x,0) + x \propB'_{\Fbar V}(x,0)] \propB_{FS}(y,z)
- \propB_{FV}(x,0) \propB_{FS}(y,z)/x
\Bigr \rbrace
\\
&=& 
(x + y - z)^2 M(0, y, x, z, y) 
+ [x^2 - (y - z)^2] M(0, z, x, y, z) 
+ (x + y - z) U(x, 0, y, y) 
\nonumber \\ &&
+ (x + y - z) U(x, 0, z, z) 
- (x + y) U(y, z, x, y) 
+ y U(z, y, x, z)
+ [(z - y)/x - 5/4] S(0, y, z) 
\nonumber \\ &&
- S(x, y, y)/2 
- S(x, z, z)/2 
+ 2 (x + y - z) \overline{T} (0, y, z) 
+ (x^2 + x y - x z - y z + z^2) T(z, 0, y)/2 x 
\nonumber \\ &&
+ (x^2  - 2 y^2 + 2 y z - 3 x y - x z) T(y, 0, z)/4 x 
+ (3x + y - z) I(x, y, z)/2 x 
+ 3 (x + y - z) I(x', y, z) 
\nonumber \\ &&
+ [
(5z - 3x - 5y) A(x)/2 x 
+ (x + 3 y - z) A(y)/4 y 
+ (2 x + 2 y - z) A(z)/4 z
+ 6 (z - x - y) 
] B(y,z)
\nonumber \\ &&
+ [
x (y + z) - (y-z)^2 + 3 (y + z - x) A(x) 
+ 3 (x - y + z) A(y)/2 +3 (x + y - z) A(z)/4
] B(y,z')
\nonumber \\ &&
- A(x) A(y)/x 
- A(y)^2/4 y 
+ (1/x - 3/z) A(x) A(z) 
+ 3 A(y) A(z)/2z 
+ (x + 2 y - 2 z) A(y)/4x 
\nonumber \\ &&
- 3 A(z)^2/4z
+ (2 x y + 2 x z + y z - z^2) A(z)/2x z 
+ 3x/16 + 3y/8 - y^2/2x - 15z/8 + z^2/2x
\nonumber \\ &&
+ \deltaMSbar [2 (y^2 - 3 y z + 2 z^2 - x y - 2 x z) B(y,z')
+ (5 x + 3 y - z) B(y, z)
- (1 + 2y/z) A(z) + 5 A(y) 
\nonumber \\ &&
+ 4 z - 2 y - 3x/4]/4
\,
+ m^2_{\epsilon} [
(x+y -z) B(y,z') - B(y,z) - A(z)/z -1]/2
,
\\
f_4 (x,y,z) 
&=& \lim_{s\rightarrow x} G_{\Fbar S}(y,z)
\\
&=& 
2 (y + z - x) M(y, y, z, z, 0) 
+ 2 T(y, 0, z) 
+ 2 T(z, 0, y) 
+ 4 \overline{T}(0, y, z)
+ B(y,z) [B(y,z) 
+5 A(y)/y 
\nonumber \\ &&
+ 2 A(z)/z -14]
+ [ 2 (y-x-z)
+ 3(x - y - z) A(y)/y 
+ 3 A(z)
] B(y,z')
+ 3 A(y) A(z)/y z
\nonumber \\ &&
- 3 A(y)/y - 2 A(z)/z 
-6 
+ \deltaMSbar [(x-y-z) B(y,z') + B(y,z) + A(z)/z - 1] 
+ m^2_{\epsilon} 2 B(y,z')
,
\\
f_5 (x,y,z) 
&=& \lim_{s\rightarrow x} [G_{F\Fbar S}(x,y,z) + 2 G_{\Fbar \Fbar S}(x,y,z)]
\\
&=& 
2 (x + y - z) M(0, y, x, z, y) 
+ 2 (y - x - z) M(0, z, x, y, z) 
+ 2 U(x, 0, y, y) 
- 2 U(x, 0, z, z) 
\nonumber \\ &&
- 2 U(y, z, x, y) 
- U(z, y, x, z)
+ T(y, 0, z) 
- T(z, 0, y) 
+ [1 - A(z)/z] [1 - 3 A(y)/2 y]
\nonumber \\ &&
+ [3 (x - y - z) A(y)/2 y - 3 A(z)/2
-x + y + 3 z] B(y,z')
+ [5 A(y)/2y -A(z)/z -1] B(y,z)
\nonumber \\ &&
+ \deltaMSbar [(x-y-z) B(y,z') + 5 B(y,z) + A(z)/z +1]/2
- m^2_{\epsilon} B(y,z')
,
\\
f_6 (x,y,z) 
&=& \lim_{s\rightarrow x} \Bigl \lbrace
H_{F\Fbar S}(x,y,z) + 2 H_{\Fbar \Fbar S}(x,y,z)
+ 2 [\propB_{FV}(x,0) + x \propB_{\Fbar V}(x,0)] \propB'_{\Fbar S}(y,z)
\nonumber \\ &&
+ 2 [\propB'_{FV}(x,0) + x \propB'_{\Fbar V}(x,0)] \propB_{\Fbar S}(y,z)
+ \propB_{\Fbar V}(x,0) \propB_{\Fbar S}(y,z)
\Bigr \rbrace
\\
&=& 
2 (x + y - z) M(0, y, x, z, y) 
+ 2 (x - y + z) M(0, z, x, y, z) 
+ 2 U(x, 0, y, y) 
+ 2 U(x, 0, z, z) 
\nonumber \\ &&
- 2 U(y, z, x, y) 
+ U(z, y, x, z)
- 4 S(0, y, z)/x 
+(1 - 2y/x) T(y, 0, z) 
+ (1 - 2z/x) T(z, 0, y) 
\nonumber \\ &&
+ 4 \overline{T}(0, y, z) 
+ 2 I(x, y, z)/x 
+ 6 I(x', y, z) 
+ [3 (y - x - z) A(x)/x 
+ 3 (x - y - z) A(y)/2y 
\nonumber \\ &&
+ 3 A(z)/2
+ x - y + z 
] B(y,z')
+ [5 A(y)/2y + A(z)/z- A(x)/x -11] B(y,z)
\nonumber \\ &&
- [1 +  A(z)/z] 3 A(x)/x 
+ 3 A(y) A(z)/2y z
+ (2/x - 3/2y) A(y) 
+ (2/x + 1/z) A(z) 
\nonumber \\ &&
- 2(y+z) /x 
-1/2 
+ \deltaMSbar [(y-x-3z) B(y,z') + 3 B(y,z) - A(z)/z + 1]/2
+ m^2_{\epsilon} B(y,z')
.
\eeq
The limits $y \rightarrow 0$ (for massless internal 
fermions) can be
important when there is no corresponding mass insertion:
\beq
f_1 (x,0,z) 
&=& 
-(x - z)^2 M(0, 0, z, z, 0) 
- 5 (x - z) U(z, 0, 0, 0)/2
+ S(0, 0, z)/2 
+ x T(z, 0, 0) 
+ x B(0, z)^2 
\nonumber \\ &&
+ [(x/z - 2) A(z) + 7z/2 -3x/2] B(0, z) 
- 3A(z)^2/z 
+ 11 A(z)/2 
- 13x/8 
- 2z 
\nonumber \\ &&
+ \deltaMSbar [(z-x) B(0, z)/2 + A(z)/2 -x/8 ] 
- m^2_{\epsilon} [2 A(z)/z + 2 B(0, z)]
,
\\
f_2 (x,0,z) &=&
(x - z)^2 M(0, 0, x, z, 0) 
+ (z^2 - x^2) M(0, z, x, 0, z) 
- x U(0, z, 0, x) 
+ (x - z) [U(x, 0, 0, 0) 
\nonumber \\ &&
- U(x, 0, z, z) 
-U(z, 0, 0, 0)/4]
+ (z - x/2) T(z, 0, 0) 
- S(0, 0, x)/2 
+ 5 S(0, 0, z)/4 
\nonumber \\ &&
+ S(x, z, z)/2 
+ [(1 - x/2z) A(z) + x/4 - 5 z/4] B(0, z) 
+ 3 A(z)^2/2z + A(x) - 13 A(z)/4 
\nonumber \\ &&
+ 3 x/16 
+ 2 z 
+ \deltaMSbar [3 (x - z) B(0, z)/4 - 3 A(z)/4 -x/16] 
+ m^2_{\epsilon} [A(z)/z + B(0, z)] 
,
\\
f_3 (x,0,z) &=&
(x - z)^2 M(0, 0, x, z, 0)
+ (x^2 - z^2) M(0, z, x, 0, z) 
- x U(0, z, 0, x) 
+ (x - z) [U(x, 0, 0, 0) 
\nonumber \\ &&
+ U(x, 0, z, z) - 9 U(z, 0, 0, 0)/4]
+ (x - z + z^2/x) T(z, 0, 0)/2 
- S(0, 0, x)/2 
- S(x, z, z)/2 
\nonumber \\ &&
+ (z/x - 5/4) S(0, 0, z) 
+ (9/2 - z/2x) I(0, x, z) 
+ [(x/2z - 1) A(z) + (3/2 + 5 z/2x) A(x) 
\nonumber \\ &&
- 7 x/4 + 3z/4] B(0,z)
+ A(z) [ 4 A(x)/x - 3 A(z)/2z -9/4 - z/2x]
- 6 A(x) 
+ z^2/2x - z/8 
\nonumber \\ &&
+ 87 x/16
+ \deltaMSbar [(5x +3 z) B(0, z)/4 + 3 A(z)/4 -3x/16] 
- m^2_{\epsilon} [A(z)/z + B(0, z)] 
.
\eeq

\subsection{Contributions from diagrams with two or more vector propagators}

We finally turn to the contributions from two-loop diagrams that contain
more than one vector (or ghost) propagator. Again it is useful to
organize the results in terms of common group theory factors. The
results for the self-energy functions are:
\beq
\Sigma^{(2,2)J}_I &=& \delta_I^J g_a^2 C_a(I) 
  \left [g_b^2 C_b(I) H_1(m_I^2) + g_a^2 C_a(G) H_2 (m_I^2) 
         + g_a^2 I_a(K) H_3 (m_I^2,m_K^2) 
         + g_a^2 I_a(i) H_4 (m_I^2,m_i^2) 
  \right ],
\\{}
\Omega^{(2,2)IJ} &=& m^{IJ} g_a^2 C_a(I) 
  \left [g_b^2 C_b(I) \overline{H}_1(m_I^2) 
       + g_a^2 C_a(G) \overline{H}_2 (m_I^2) 
         + g_a^2 I_a(K) \overline{H}_3(m_I^2,m_K^2) 
         + g_a^2 I_a(i) \overline{H}_4(m_I^2,m_i^2) 
\right ],
\phantom{xxx}
\eeq
where the required loop integral functions are:
\beq
%\mbox{CfCftotalA} 
H_1(x)
&=&
-4 x^2 M(0, x, x, 0, x) + 2 S(x, x, x) 
- (s + 4 x) T(x, 0, 0) 
+ 2 (s - x) U(x, 0, x, x)
+ (10x^2/s -2s 
\nonumber \\ &&
- 4x- 4x^3/s^2)  \ln^2 (1-s/x)
+[11s/2 + x - 29 x^2/2s + (4 - 3s/x + 11 x/s) A(x)] \ln (1-s/x)
\nonumber \\ &&
-73 s/8 
- 5x/2 + (3 + 9s/2x) A(x) + (6/x - s/x^2) A(x)^2
+ \deltaMSbar [(5 x^2/s -s)\ln (1-s/x) 
\nonumber \\ &&
+ 5x -s/2 -s A(x)/x]
+ (1-\xi) \bigl \lbrace 
4 s T(x,0,0) 
+ (3s - 11x^2/s + 8 x^3/s^2 )\ln^2 (1-s/x)
\nonumber \\ &&
+ [(2s/x - 14x/s) A(x) -5s 
- 6x + 19 x^2/s ]  \ln (1-s/x)
+3 s + 11 x - (s/x + 14) A(x) - s A(x)^2/x^2
\nonumber \\ &&
+ \deltaMSbar [ (s-5x^2/s) \ln (1-s/x) + s A(x)/x - 5x ]
\bigr \rbrace
+ (1-\xi)^2 \bigl \lbrace s T(x,0,0) 
+ (s-x) x^2/s^2 \ln^2 (1-s/x) 
\nonumber \\ &&
+ [3 x - s - 2 x^2/s + (x/s - s/x) A(x)] \ln (1-s/x)
+ 3s/2 - x + A(x) - s A(x)^2/x^2 \bigr \rbrace
,
\\
%
%\mbox{CfCftotalB} 
\overline{H}_1 (x)
&=& 
 - 2 (s + x) M(0, x, x, 0, x) 
 + 4 T(x, 0, 0) 
 - 8 U(x, 0, x, x)
 + 42 - 32 A(x)/x + 10 A(x)^2/x^2 
\nonumber \\ &&
 + (14 - 6x/s) (1 - x/s) \ln^2(1-s/x) 
 + [24 (1/x - 2/s) A(x) -36 + 52 x/s] \ln(1-s/x) 
\nonumber \\ &&
+ \deltaMSbar 8 [A(x)/x -1 + (1-2x/s)  \ln(1-s/x)]
+ (1-\xi) \lbrace 
- 4 T(x, 0, 0) + 8A(x)/x 
\nonumber \\ &&
- 2A(x)^2/x^2 -6 - 2x/s 
+ (8x/s - 6 - 2 x^3/s^3) \ln^2(1-s/x)
+ [(14/s - 8/x) A(x) + 12 
\nonumber \\ &&
- 12x/s - 4 x^2/s^2] \ln(1-s/x)
+ \deltaMSbar 2[1 - A(x)/x + (2x/s -1) \ln (1-s/x)]
\rbrace
\nonumber \\ &&
+ (1-\xi)^2 \lbrace 
-T(x,0,0) + A(x)^2/x^2 - A(x)/x + x/s -3/2 
+[(1/x - 1/s) A(x) -2x/s 
\nonumber \\ &&
+ 2 x^2/s^2]\ln(1-s/x) 
+ (x^3/s^3 - x/s) \ln^2(1-s/x)
\rbrace
,
\\
%
%\mbox{CGCftotalA} 
H_2(x)
&=& 
(x-s)(s+2x) M(0, 0, x, x, 0) + 2 x^2 M(0, x, x, 0, x) 
 + (x - s) U(x, 0, x, x) 
- (s + 2x) T(x, 0, 0) 
\nonumber \\ &&
- S(x, x, x) 
 - A(x)^2/x + (7 + 6s/x) A(x)
 -16 x - 17s/2
+ (1+2x/s) (x-s) \ln^2 (1-s/x)
\nonumber \\ &&
+ (x-s) [(3/s + 1/x) A(x) - 12x/s -7] \ln(1-s/x)
+ \deltaMSbar 13s/4
+ (1-\xi) \lbrace
s (s-x) M(0,0,x,x,0) 
\nonumber \\ &&
+ s T(x,0,0)
+ 2 s A(x)^2/x^2 - 5 (1+s/x) A(x) + 5 x + 11 s
+ (1-x/s) [(4x + 3s) \ln(1-s/x) 
\nonumber \\ &&
+ 5 (1+s/x) A(x) -5x -6s] \ln(1-s/x)
\rbrace/2
+ (1-\xi)^2 \lbrace
-s T(x,0,0) + (1 + s/2x) A(x) - 9s/4 -x/2
\nonumber \\ &&
+ (1-x/s) [(3s+x)/2 -(x+s) \ln(1-s/x) - (1+s/x) A(x)]\ln(1-s/x)
\rbrace/2
,
\\
%
%\mbox{CGCftotalB} 
\overline{H}_2(x)
&=& 
3 (s - x) M(0, 0, x, x, 0) 
+ (s + x) M(0, x, x, 0, x) 
+ 7 T(x, 0, 0) 
+ 4 U(x, 0, x, x)
+ A(x)^2/x^2 
\nonumber \\ &&
- 64 A(x)/3x + 112/3
+ (1-x/s) [8 \ln(1-s/x) + 9 A(x)/x - 85/3] \ln(1-s/x)
\nonumber \\ &&
+ \deltaMSbar [(x/s - 1) \ln(1-s/x) - A(x)/x - 7/2]
+ (1-\xi) 
\lbrace 
 (x-s) M(0,0,x,x,0) - T(x,0,0) 
\nonumber \\ &&
- 2 A(x)^2/x^2 + 10 A(x)/x - 16
+ (1-x/s) [(x/s -3) \ln (1-s/x)
+ 11 - 5 A(x)/x] \ln(1-s/x)
\rbrace/2
\nonumber \\ &&
+ (1-\xi)^2 \lbrace
T(x,0,0) - 3 A(x)/2x + 11/4 
+(1-x/s) [\ln(1-s/x) + A(x)/x - 5/2] \ln(1-s/x)
\rbrace/2
,
\phantom{xxxxx}
\\
%
%\mbox{VFVVFFtotA} 
H_3(x,y)
&=& 
\bigl \lbrace 
[-4s (s-x)^2 + 40 s^2 y + 4 s x y + 4 x^2 y +48 (x-2s) y^2] y T(y,y,x)
+ 2 x (s-x) [9 y (s+x) 
\nonumber \\ &&
-(s-x)^2 - 36 y^2] T(x,y,y)
+ [-2 (s-x)^3 + 22 s^2 y - 8 s x y - 14 x^2 y + 48 (x-2s) y^2] S(x,y,y)
\nonumber \\ &&
+ (s-x)^4 B(0,x)
+ [2 s^2 + 2 s x - 4x^2 + 24 (x-2s) y] A(y)^2 
+ [4 (s-x)^2 - 24 (s+x) y] A(x) A(y)
\nonumber \\ &&
+ [112 s x y -4 x (s-x)^2 - 92 s^2 y  
  + 28 x^2 y + 96 (2s - x) y^2] A(y)
+ [(s-x)^3 - 4 s^2 y - 10 s x y 
\nonumber \\ &&
+ 14 x^2 y + 24 (s-2x) y^2 ] A(x)
+ (5s/4 - 2x) (s-x)^3
-35s^3 y/2 + 77 s^2 x y/2 - 11 s x^2 y - 10 x^3 y
\nonumber \\ &&
 + 8s^2 y^2 +
 50 s x y^2 + 20 x^2 y^2 + 96 (x-2s) y^3
\bigr \rbrace/15 y (s-x)^2
+ \deltaMSbar 3s/4
,
\\
%
%\mbox{VFVVFFtotB} 
\overline{H}_3(x,y)
&=& 
\bigl \lbrace
[6(s-x)^2 + 12 x y + 16 y^2 - 28s y ]T(y,y,x)
+ 20 x (x-s) T(x,y,y)
+ (20x - 20s +16 y) S(x,y,y)
\nonumber \\ &&
+ (8 + 6 x/y - 6 s/y) A(y)^2 + 16 A(x) A(y)
+ (10s  - 10 x  + 8 y) A(x)
+ (48 s - 64 x - 32 y) A(y)
\nonumber \\ &&
+ 19 s^2/2 - 53 s x/2 + 17 x^2 
- 34 s y + 8 x y + 32 y^2
\bigr \rbrace/3(s-x)^2 
+ 2 (1 + A(y)/y) B(0,x)
- \deltaMSbar
,
\\
%
%\mbox{VFVVSStotA} 
H_4(x,y)
&=&
\bigl \lbrace
[48 (2s-x) y^2 -s(s-x)^2 -20 s^2 y + 16 s x y - 44 x^2 y] y T(y,y,x)
+ x (x-s) [(s-x)^2/2 
\nonumber \\ &&
-12 y (s + x + 6 y)] T(x,y,y)
+ [ y(13 s^2 - 2 s x - 11 x^2 + 96 s y - 48 x y)-(s-x)^3/2] S(x,y,y)
\nonumber \\ &&
+ [23 s^2 - 22 s x - x^2 + 48 s y - 24 x y] A(y)^2
+ [(s-x)^2 + 24 (s+x) y] A(x) A(y)
+ [(s-x)^3/4 
\nonumber \\ &&
+ (x-s) y (s + 11x) + 24 (2x -s) y^2] A(x)
+ [-x(s-x)^2 + 96 (x-2s) y^2 + 22 s^2 y - 92 s x y 
\nonumber \\ &&
+ 22 x^2 y] A(y)
+ (5s/16 -x/2) (s-x)^3 + 96 (2s-x) y^3 - 70 x y^2 (s+x)
-25 s^3 y/4 +  23 s^2 x y/2 
\nonumber \\ &&
+ 19 s x^2 y/4 - 10 x^3 y + 62 s^2 y^2
\bigr \rbrace/15 y (s-x)^2
+ (s-x)^2 B(0,x)/60 y
,
\\
%
%\mbox{VFVVSStotB} 
\overline{H}_4(x,y)
&=&
\bigl \lbrace
[3 (s-x)^2 - 8 s y + 24 x y - 16 y^2] T(y,y,x)
+ 16 x (x-s) T(x,y,y)
+ 16 (x-y-s) S(x,y,y)
\nonumber \\ &&
+ (12 x/y - 12 s/y - 8) A(y)^2
- 16 A(x) A(y)
+ 8 (s-x-y) A(x)
+ 8 (3s -x+4y) A(y)
+ 7 s^2 
\nonumber \\ &&
- 20 s x + 13 x^2 - 38 s y + 64 x y - 32 y^2
\bigr \rbrace/3 (s-x)^2
+ [1 + A(y)/y] B(0,x) .
\eeq
The corresponding contributions to the pole squared mass are therefore:
\beq 
\widetilde\Pi^{(2,2)}_{I} = g_a^2 C_a(I) \left[
g_b^2 C_b(I) F_1(m_I^2) +
g_a^2 C_a(G) F_2(m_I^2) +
g_a^2 I_a(K) F_3(m_I^2,m_K^2) +
g_a^2 I_a(i) F_4(m_I^2,m_i^2) \right ]
,
\eeq
where the required functions can be written compactly in terms of 
logarithms and dilogarithms:
\beq
F_1 (x) 
&=& \lim_{s \rightarrow x} \Bigl \lbrace 
  2 H_1(x) + 2 x \overline{H}_1(x) 
  + 4 [\propB_{FV}(x,0) + x \propB_{\Fbar V}(x,0) ]
      [ \propB_{FV}'(x,0) + x \propB_{\Fbar V}'(x,0) ]
\nonumber \\ &&
  - [\propB_{FV}(x,0)]^2/x
  + x [\propB_{\Fbar V}(x,0)]^2
\Bigr \rbrace
\\
&=& x \left [ 
41/4 + 10 \pi^2 - 27 \lnbar x + 18 \lnbar^2 x + 24 \zeta(3) - 16 \pi^2 \ln 2
+\deltaMSbar 12 (\lnbar x - 1) 
\right ]
,
\label{eq:Foneresult}
\\
F_2(x) 
&=& \lim_{s \rightarrow x} \left [2 H_2(x) + 2 x \overline{H}_2(x) 
\right ]
\\
&=&
x \left [
1093/12 - 8 \pi^2/3 - (179/3) \lnbar x   + 11 \lnbar^2 x 
- 12 \zeta(3) + 8 \pi^2 \ln 2 + \deltaMSbar (3/2 - 2 \lnbar x)
\right ]
,
\label{eq:Ftworesult}
\\
F_3(x,y) 
&=& \lim_{s \rightarrow x} \left [2 H_3(x) + 2 x \overline{H}_3(x) 
\right ]
\\
&=& -37x/3  - 12 y + (26x/3) \lnbar x  - 2 x \lnbar^2 x 
+ 4 y\ln (x/y) -2 (y^2/x)\ln^2(x/y) 
\nonumber \\ 
&&
+ 8 (x+y) f(\sqrt{y/x}) 
- 4 (x + y^2/x) [{\rm Li}_2(1-y/x) + \pi^2/6] + \deltaMSbar x/2     
,
\label{eq:Fthreeresult}
\\
F_4(x,y) 
&=& \lim_{s \rightarrow x} \left [2 H_4(x) + 2 x \overline{H}_4(x) 
\right ]
\\
&=&
-125 x/12 + 14 y + (19 x/3) \lnbar x - x \lnbar^2 x - 6 y \ln (x/y)  
+ (y^2/x) \ln^2(x/y) 
\nonumber \\ 
&&
+ 8 (x - y) f(\sqrt{y/x})  
+ 2 (y^2/x -x) [{\rm Li}_2(1-y/x) + \pi^2/6] ,
\eeq
\end{widetext}
with
\beq
f(r) &=&
r \lbrace {\rm Li}_2 ([1-r]/[1+r])
\nonumber \\ &&
- {\rm Li}_2 ([r-1]/[1+r])
+ \pi^2/4 \rbrace
.
\label{eq:definefunky}
\eeq
Here are some useful special cases:
\beq
F_3(x,0) &=& x \Bigl (
             \frac{26}{3} \lnbar x - 2 \lnbar^2 x 
-\frac{37}{3} - \frac{4 \pi^2}{3} 
\nonumber \\ &&
             + \frac{\deltaMSbar}{2} \Bigr )
,
\\
F_3(x,x) &=& x \Bigl (
             \frac{26}{3} \lnbar x - 2 \lnbar^2 x 
-\frac{73}{3} + \frac{8 \pi^2}{3} 
\nonumber \\ &&
             + \frac{\deltaMSbar}{2} \Bigr )
,
\\
F_4(x,x) &=& x \Bigl (\frac{43}{12} 
             + \frac{19}{3} \lnbar x - \lnbar^2 x  
             \Bigr )
,
\\ 
F_4(x,0) &=& x \Bigl (
             \frac{19}{3} \lnbar x - \lnbar^2 x  
-\frac{125}{12} - \frac{2 \pi^2}{3} 
             \Bigr )
,
\\
F_3(0,y) &=& F_4(0,y) = 0
.
\eeq
In the $\MSbar$ scheme with $\deltaMSbar = 1$, the expressions of 
eqs.~(\ref{eq:Foneresult}) above agree with those found originally in 
\cite{Gray:1990yh} (see also 
\cite{Fleischer:1998dw}). In the $\DRbar$ scheme with $\deltaMSbar = 0$,
the same equations are in agreement with the results of 
ref.~\cite{Avdeev:1997sz}.
In particular, the function $F_3(x,y)/2x$ found here was given in three
different mass expansions in eqs.~(18)-(20) of \cite{Avdeev:1997sz}.

\section{Examples and Applications}
\setcounter{equation}{0}
\setcounter{footnote}{1}

%\subsection{The Wess-Zumino Model}

In this section, I present some applications of the preceding general
results. These are all taken from the minimal supersymmetric standard
model (MSSM), and include all effects at one-loop order, but 
only terms that involve the strong gauge
coupling constant in the two-loop part. All of the couplings and masses
appearing below are tree-level running $\DRbarprime$ parameters in the
MSSM with no particles decoupled. 
(This means that contributions listed above containing $\deltaMSbar$
and $m^2_{\epsilon}$ are not present.)
The conventions used
here for these couplings and masses are identical to those found
in section II of ref.~\cite{effpotMSSM} and section II of
ref.~\cite{Martin:2004kr}, and will not be repeated here for the sake of
brevity. Note that the formalism allows for arbitrary CP violation within
the MSSM, and takes into account sfermion mixing within each generation,
but for simplicity the effects of possible sfermion mixing between 
generations
is neglected.
Throughout the following, each index that appears on the
right-hand side of an equation but not on the left-hand side is implicitly
summed over. The name of a particle is used in place of its renormalized,
tree-level squared mass when appearing as the argument of a loop 
integral function.

\begin{widetext}

\subsection{SUSYQCD corrections to the gluino mass}

In this section, I will study the specialization of the above
results to the case of the gluino mass.
(The results below partly overlap with recent independent results of
Y.~Yamada in \cite{Yamada:2005ua}.)
Applying the formulas of sections \ref{sec:oneloop} and
\ref{sec:twoloop},  I find that the two-loop gluino
pole mass, including all SUSYQCD effects, can be written as:
\beq
M^2_{\tilde g} - i \Gamma_{\tilde g} M_{\tilde g} 
&=& 
m^2_{\tilde g} 
+ \frac{g_3^2}{16\pi^2} \widetilde\Pi^{(1)}_{\tilde g}
+ \frac{g_3^4}{(16\pi^2)^2} \widetilde\Pi^{(2)}_{\tilde g}
,
\eeq
where $m_{\tilde g}$ is the tree-level running gluino mass (often seen
as $M_3$ in the literature), and
\beq
\widetilde\Pi^{(1)}_{\tilde g} &=&
C_G m_{\tilde g}^2 \bigl [ 10 - 6 \lnbar (m_{\tilde g}^2) \bigr ]
+
4 I_q \bigl  \lbrace \propB_{FS}(q, \tilde q_j) -
2{\rm Re}[L_{\tilde q_j} R_{\tilde q_j}^*] m_q m_{\tilde g}
\propB_{\Fbar S}(q, \tilde q_j) \bigr \rbrace
,
\\
\widetilde\Pi^{(2)}_{\tilde g} &=&
8 I_q (2 C_q - C_G) \Bigl \lbrace
L_{\tilde q_j} R_{\tilde q_j}^*L_{\tilde q_k}^* R_{\tilde q_k} 
   \propM_{SFFSF} (\tilde q_j, q, q, \tilde q_k, \tilde g)
- {\rm Re}[L_{\tilde q_k} R_{\tilde q_k}^*]  
   m_q m_{\tilde g}
   \propM_{SF\Fbar SF} (\tilde q_j, q, q, \tilde q_k, \tilde g)
\nonumber \\ &&
+\frac{1}{2} (|L_{\tilde q_j} L_{\tilde q_k}|^2 + 
   |R_{\tilde q_j} R_{\tilde q_k}|^2) 
     m_{\tilde g}^2 \propM_{SFFS\Fbar} (\tilde q_j, q, q, \tilde q_k, \tilde g)
+ |L_{\tilde q_j} R_{\tilde q_k}|^2  
   m_q^2
   \propM_{S\Fbar\Fbar SF} (\tilde q_j, q, q, \tilde q_k, \tilde g)
\nonumber \\ &&
- {\rm Re}[L_{\tilde q_j} R_{\tilde q_j}^*]
   m_q m_{\tilde g}
   \propM_{SF\Fbar S\Fbar} (\tilde q_j, q, q, \tilde q_k, \tilde g)
+ {\rm Re}[L_{\tilde q_j} R_{\tilde q_j}^*
            L_{\tilde q_k} R_{\tilde q_k}^*]  m_q^2 m_{\tilde g}^2
   \propM_{S\Fbar\Fbar S\Fbar} (\tilde q_j, q, q, \tilde q_k, \tilde g)
\Bigr \rbrace
\nonumber \\ &&
+ 8 I_q C_q \Bigl \lbrace
 (|L_{\tilde q_j} L_{\tilde q_k}|^2 +
  |R_{\tilde q_j} R_{\tilde q_k}|^2) 
  \propV_{SFFFS} (\tilde q_j,q,q,\tilde g, \tilde q_k)
- 2 {\rm Re}[L_{\tilde q_j} R_{\tilde q_j}^*] m_q m_{\tilde g} 
  \propV_{SF\Fbar FS} (\tilde q_j,q,q,\tilde g, \tilde q_k)
\nonumber \\ &&
+ 2 {\rm Re}[L_{\tilde q_j} R_{\tilde q_j}^*
           L_{\tilde q_k}^* R_{\tilde q_k}] m_{\tilde g}^2
  \propV_{SFF\Fbar S} (\tilde q_j,q,q,\tilde g, \tilde q_k)
+ (|L_{\tilde q_j} R_{\tilde q_k}|^2 +
   |R_{\tilde q_j} L_{\tilde q_k}|^2) m_q^2 
  \propV_{S\Fbar\Fbar FS} (\tilde q_j,q,q,\tilde g, \tilde q_k)
\nonumber \\ &&
- 2 {\rm Re}[L_{\tilde q_k} R_{\tilde q_k}^*] m_q m_{\tilde g} 
  \propV_{SF\Fbar \Fbar S} (\tilde q_j,q,q,\tilde g, \tilde q_k)
+ 2 {\rm Re}[L_{\tilde q_j} R_{\tilde q_j}^*
             L_{\tilde q_k} R_{\tilde q_k}^*] m_q^2 m_{\tilde g}^2
  \propV_{S\Fbar\Fbar\Fbar S} (\tilde q_j,q,q,\tilde g, \tilde q_k)
\nonumber \\ &&
+ \propV_{FSSFF} (q,\tilde q_j, \tilde q_j, q, \tilde g)
+ 2 {\rm Re}[L_{\tilde q_j} R_{\tilde q_k}^* 
             (R_{\tilde q_j}^* L_{\tilde q_k} + 
              L_{\tilde q_j}^* R_{\tilde q_k})] m_q^2 m_{\tilde g}^2 
  \propV_{\Fbar SS\Fbar\Fbar} (q,\tilde q_j, \tilde q_k, q, \tilde g)
\nonumber \\ &&
- 2 {\rm Re}[L_{\tilde q_j} R_{\tilde q_j}^*] m_q m_{\tilde g} 
  \propV_{\Fbar SSFF} (q,\tilde q_j, \tilde q_j, q, \tilde g)
- 2 {\rm Re}[L_{\tilde q_j} R_{\tilde q_j}^*] m_q m_{\tilde g} 
  \propV_{FSS\Fbar\Fbar} (q,\tilde q_j, \tilde q_k, q, \tilde g)
\nonumber \\ &&
+ \frac{1}{2} |L_{\tilde q_j} L_{\tilde q_k}^* - 
   R_{\tilde q_j} R_{\tilde q_k}^*|^2
  \propY_{FSSS}(q,\tilde q_j, \tilde q_j, \tilde q_k)
\nonumber \\ &&
- {\rm Re}[L_{\tilde q_j}^* R_{\tilde q_k}
  (L_{\tilde q_j} L_{\tilde q_n}^* -
   R_{\tilde q_j} R_{\tilde q_n}^*)
  (L_{\tilde q_k}^* L_{\tilde q_n} -
   R_{\tilde q_k}^* R_{\tilde q_n})] m_q m_{\tilde g}
  \propY_{\Fbar SSS}(q,\tilde q_j, \tilde q_k, \tilde q_n)
\Bigr \rbrace
\nonumber \\ &&
+ 4 I_q^2 \Bigl [
4 \lbrace
\propB_{FS}(q, \tilde q_j) - 2{\rm Re}[L_{\tilde q_j} R_{\tilde q_j}^*]
m_q m_{\tilde g} \propB_{\Fbar S}(q, \tilde q_j) \rbrace
\lbrace
\propB'_{FS}(Q, \tilde Q_k) - 2{\rm Re}[L_{\tilde Q_k} R_{\tilde Q_k}^*]
m_Q m_{\tilde g}
\propB'_{\Fbar S}(Q, \tilde Q_k) \rbrace
\nonumber \\ &&
-\propB_{FS}(q, \tilde q_j) \propB_{FS}(Q, \tilde Q_k)/m_{\tilde g}^2
+ 4 m_q m_Q 
  {\rm Re}[L_{\tilde q_j} R_{\tilde q_j}^*]
  {\rm Re}[L_{\tilde Q_k} R_{\tilde Q_k}^*]
\propB_{\Fbar S}(q, \tilde q_j) \propB_{\Fbar S}(Q, \tilde Q_k)
\Bigr ]
%
%\\
%
%\widetilde\Pi^{(2,1)}_{\tilde g} &=&
\nonumber \\ && +
(4 C_q - 2 C_G) I_q \bigl \lbrace
f_1(\tilde g, q, \tilde q_j)
- 2 {\rm Re}[L_{\tilde q_j} R_{\tilde q_j}^*] 
m_q m_{\tilde g}
f_4(\tilde g, q, \tilde q_j) \bigr \rbrace
\nonumber \\ &&
+ 4 C_G I_q \bigl \lbrace
f_3(\tilde g, q, \tilde q_j)
- 2 {\rm Re}[L_{\tilde q_j} R_{\tilde q_j}^*] 
m_q m_{\tilde g}
f_6(\tilde g, q, \tilde q_j) \bigr \rbrace
%
%\\
%
%\widetilde\Pi^{(2,2)}_{\tilde g} &=&
\nonumber \\ && +
C_G^2 \bigl [
F_1(\tilde g) + F_2(\tilde g) + F_3 (\tilde g, \tilde g)
\bigr ]
+ C_G I_q \bigl [2 F_3 (\tilde g, q)
+ F_4 (\tilde g, \tilde q_j) \bigr ]
.
\eeq
Here $C_q = 4/3$, $C_G = 3$,  and $I_q = 1/2$. The 
symbol
$q$ is summed over the 6 symbols $(u,d,c,s,t,b)$,
and the squark 
squared-mass eigenstate labels $j,k,n$ are summed over $1,2$.
In terms where $Q$ appears, it is also independently summed over
$(u,d,c,s,t,b)$.
The symbols $L_{\tilde q_j}$ and $R_{\tilde q_j}$ describe the squark 
mixing and CP violation; they
denote the left-handed and right-handed squark content amplitudes of each
squark mass eigenstate, as defined in ref.~\cite{effpotMSSM}.
The loop integral functions were listed in sections \ref{sec:oneloop}
and \ref{sec:twoloop} above, in terms of basis functions that
can be evaluated numerically using \cite{TSIL}. Each of them should be
evaluated with $s= m_{\tilde g}^2$ (the tree-level squared mass),
with an infinitesimal positive imaginary part. 
A computer program 
implementing this result is available from the author on request.

As a non-trivial check, I have verified
that this result for the gluino pole mass is 
invariant under changes in the renormalization scale governed by 
the two-loop SUSYQCD renormalization group equation for the running
gluino mass \cite{Martin:1993yx,Yamada:1994id}, \cite{Jack:1994kd}, 
up to consistently neglected terms
at three-loop order in SUSYQCD and two-loop order in the other couplings.

In the limit that squark mixing and quark masses are neglected, the 
expressions above simplify and can be given analytically
in terms of 
polylogarithms \cite{polylogs}.
The result for the pole squared mass is then:
\beq
M^2_{\tilde g} - i \Gamma_{\tilde g} M_{\tilde g} 
\label{eq:spolegluinonomix}
&=& m^2_{\tilde g} 
+ \frac{g_3^2}{16\pi^2} 
  [C_G \widetilde\Pi^{(1,a)}_{\tilde g}(\tilde g)
  + I_q \widetilde\Pi^{(1,b)}_{\tilde g}(\tilde g, \tilde q_j)]
\nonumber \\ &&
+ \frac{g_3^4}{(16\pi^2)^2}
\Bigl [C_G^2 \widetilde\Pi^{(2,a)}_{\tilde g}(\tilde g)
  +C_G I_q \widetilde\Pi^{(2,b)}_{\tilde g}(\tilde g,\tilde q_j)
  +C_q I_q \widetilde\Pi^{(2,c)}_{\tilde g}(\tilde g,\tilde q_j)
  +I_q^2 \widetilde\Pi^{(2,d)}_{\tilde g}(\tilde g,\tilde q_j,\tilde Q_k)
\Bigr ]
,
\eeq
where
\beq
\widetilde\Pi^{(1,a)}_{\tilde g}(x) &=& 
x[10 - 6 \lnbar x] 
,
\\
\widetilde\Pi^{(1,b)}_{\tilde g}(x,y) &=& 
2 x \bigl [(1-y/x)^2 \ln(1-x/y)+ \lnbar y - 2 + y/x \bigr ]
,
\\
\widetilde\Pi^{(2,a)}_{\tilde g}(x) &=& 
x[77 + 10 \pi^2 + 12 \zeta(3) - 8 \pi^2 \ln 2 - 78 \lnbar x + 27 \lnbar^2 x] 
,
\\
\widetilde\Pi^{(2,b)}_{\tilde g}(x,y)  &=&
(24 y - 18 x - 4 y^2/x) {\rm Li}_2(1-y/x)
+ 4 x (x-y) M(0,y,y,0,x)
+ 4 (x^2 - y^2) M(0,x,y,0,y)
\nonumber \\ &&
+ 2 (x-y)^2 [M(0,0,y,y,0)
+ 2 M(0,0,x,y,0) -(4/x) \ln^2(1-x/y)]
+ 8 (y-x) f(\sqrt{y/x})
\nonumber \\ &&
+ [(12 y^2/x - 12 y - 2 x) \lnbar x
+ (24 y - 16 x - 6 y^2/x) \lnbar y
+ 41 x - 40 y - y^2/x]\ln(1-x/y)
\nonumber \\ &&
+ (10 y -8x - 2 y^2/x) \ln^2(x/y)
- 18 x \lnbar x \lnbar y
+ (41 x - 14 y) \lnbar y
+ (9x + 20 y) \lnbar x
\nonumber \\ &&
+ 2y (1-y/x) \pi^2/3
-41 x - 9 y
,
\label{eq:pitwobgluino}
\\
\widetilde\Pi^{(2,c)}_{\tilde g}(x,y) 
&=&
8 x (y-x) M(0,y,y,0,x)
- 4 (x-y)^2 M(0,0,y,y,0)
+ (4x - 24 y + 24 y^2/x) {\rm Li}_2(1-y/x)
\nonumber \\ &&
- 2 (x-y)^2 (x/y^2 - 2/x) \ln^2(1-x/y)
+ [
(12x - 8 y + 4 x^3/y^2 - 8 x^2/y) \lnbar x
\nonumber \\ &&
+ (4x - 8 y - 4 x^3/y^2 + 8 x^2/y) \lnbar y
+4 (y-x) (x/y + 7 y/x)
]\ln(1-x/y)
\nonumber \\ &&
+ [12 y^2/x - 12 y + 4 x^2/y - 2 x^3/y^2 + 2 x]  \ln^2(x/y)
+ (4 x^2/y - 20x + 24 y) \ln (x/y)
\nonumber \\ &&
+ (4 x^2/y - 2 x^3/y^2 -4x/3 -8y/3 + 4 y^2/x) \pi^2
-22x + 4y
,
\\
\widetilde\Pi^{(2,d)}_{\tilde g}(x,y,z) 
&=&
\widetilde\Pi^{(1,b)}_{\tilde g}(x,y)
\bigl [(3+5z/x)(1-z/x) \ln(1-x/z) + 3 \lnbar z - 5z/x -2 \bigr ]/2 .
\label{eq:Pitwodgluino}
\eeq
The integral $M(0,x,y,0,y)|_{s=x}$ can be reduced using 
recurrence relations to results found in 
refs.~\cite{Fleischer:1998dw,Jegerlehner:2003py},
and was given in the present notation in \cite{TSIL}.
The analytic formulas for the master integral cases $M(0,0,x,y,0)$
and $M(0,y,y,0,x)|_{s=x}$ were also given 
in terms of polylogarithms in
\cite{TSIL}. 
The integral $M(0,0,y,y,0)$ was originally found in \cite{Broadhurst:1987ei},
and listed in the present notation in \cite{evaluation}.
The function $f$ appearing in eq.~(\ref{eq:pitwobgluino}) was 
defined in eq.~(\ref{eq:definefunky}) above.

To illustrate the numerical significance of the two-loop
correction, in fig.~\ref{fig:gluinopole} I show the fractional difference
between the real part of the gluino pole mass and the running renormalized 
mass (evaluated
at a renormalization scale equal to itself) as a function of the
ratio of the squark masses (assumed degenerate) to the tree-level
gluino mass. 
%%%%%%%%%%%%%%%%%%%%%%%%%%%%%%%%%%%%%%%%%%%%%%%%%%%%%%%%%%%%%%%%%%%%%%%%%%%%
\begin{figure}[t]
\centering
\mbox{\includegraphics[width=8.9cm]{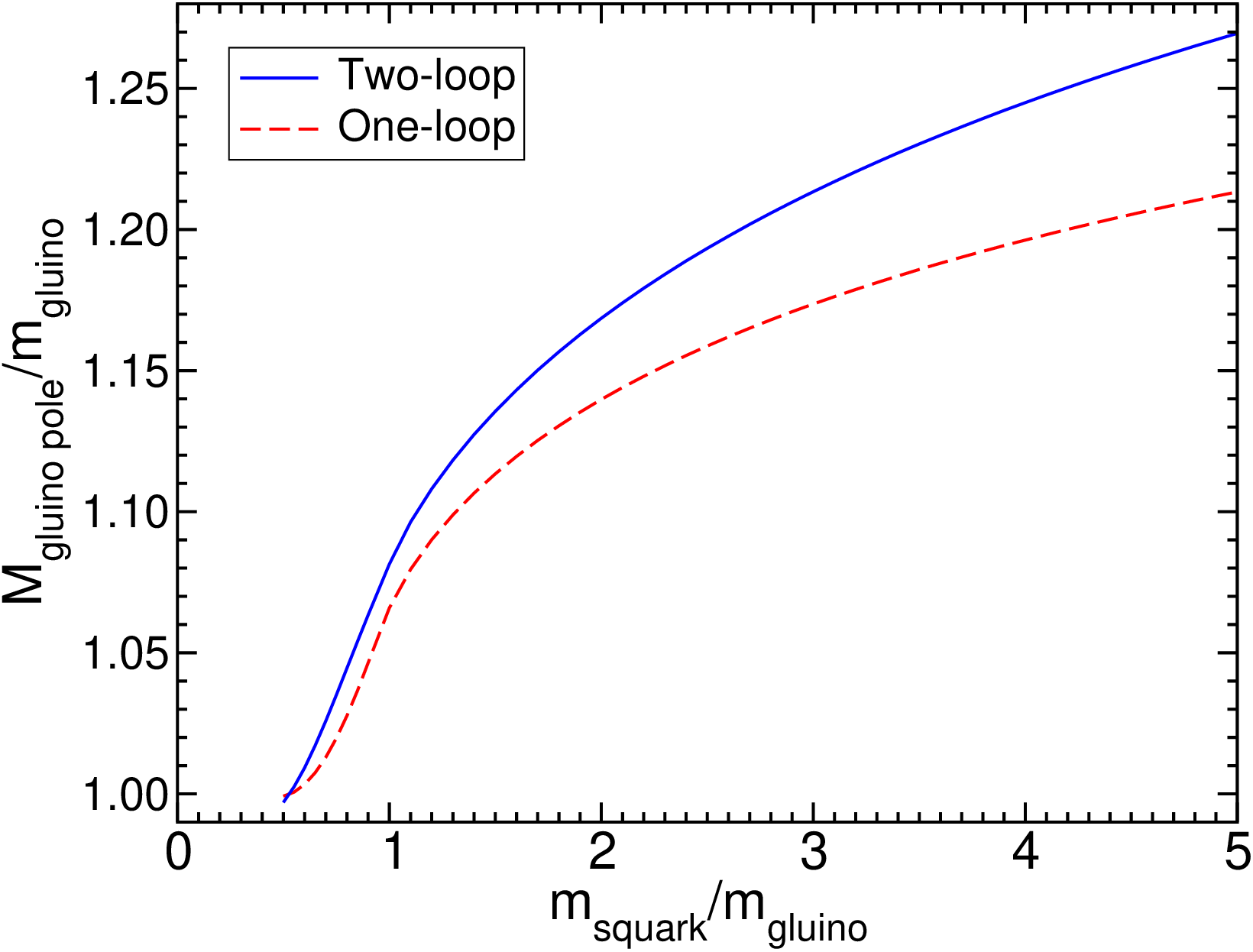}
~
\includegraphics[width=8.9cm]{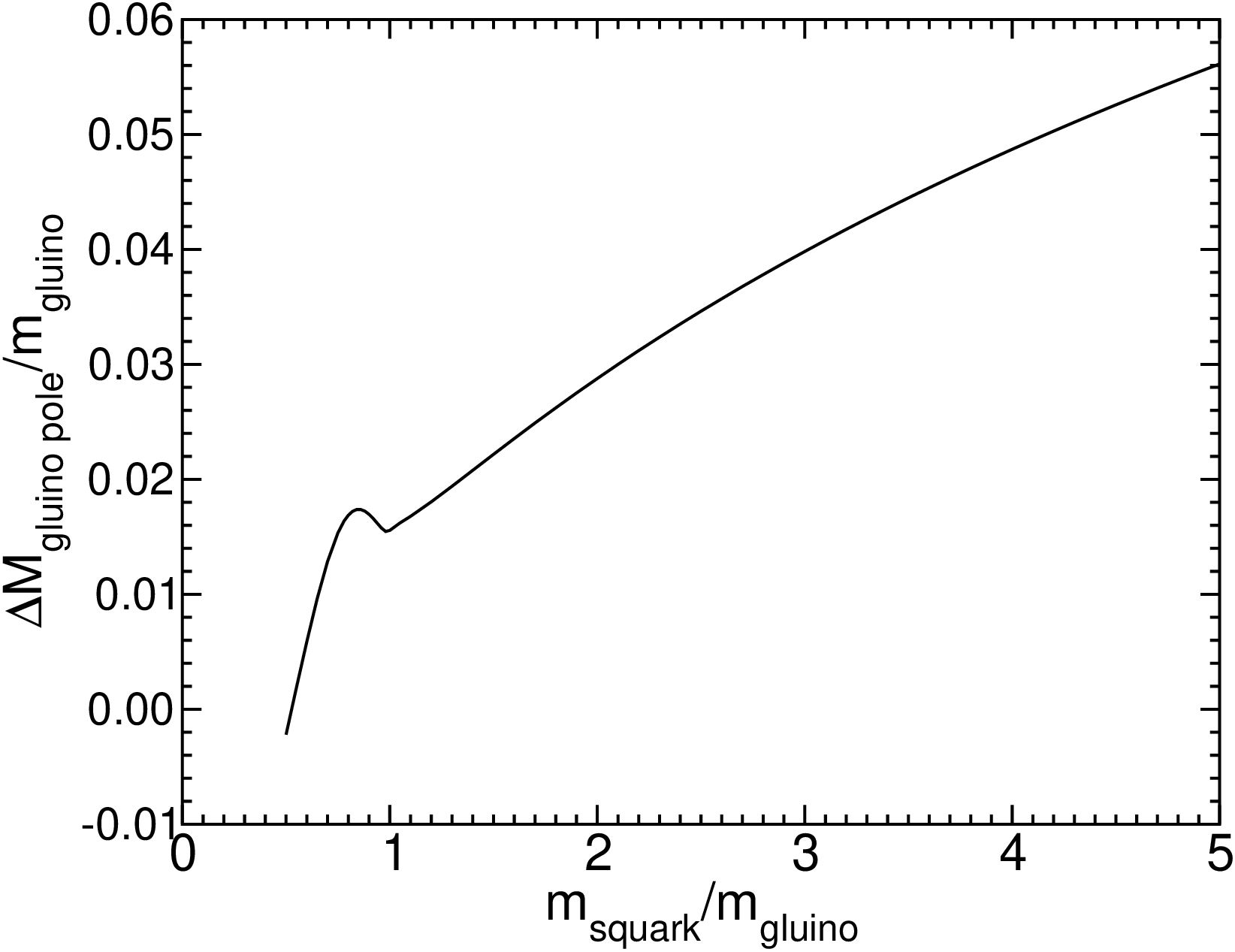}}
\caption{\label{fig:gluinopole}
The SUSYQCD corrections to the gluino pole mass $M_{\tilde g}$,
in the simplifying approximation of degenerate squarks, with no
squark mixing or quark masses and Yukawa and electroweak effects
neglected for simplicity. Here
 $m_{\rm squark} \equiv m_{\tilde q}$ and $m_{\rm gluino} \equiv
 m_{\tilde g}$ (often seen in the literature as $M_3$) are the
tree-level running $\DRbarprime$
squark and gluino mass parameters, all evaluated at a renormalization
scale $Q = m_{\tilde g}(Q)$,
and $M_{{\rm gluino}\,{\rm pole}} \equiv M_{\tilde g}$ is the 
square root of the real part of the pole mass.
In the first panel, 
the dashed line is the one-loop result, and the solid line is the two-loop
result, for the ratio of the gluino pole mass to the running gluino mass
evaluated at itself. 
The second panel shows the difference between the two-loop and one-loop
results.
The computations were done by specializing
eqs.~(\ref{eq:spolegluinonomix})-(\ref{eq:Pitwodgluino}) in the text,
using $\alpha_S(Q) = g_3^2/4\pi = 0.095$.}
\end{figure}
%%%%%%%%%%%%%%%%%%%%%%%%%%%%%%%%%%%%%%%%%%%%%%%%%%%%%%%%%%%%%%%%%%%%%%%%%%%%
Also for simplicity, the top quark mass is neglected. In most 
realistic
models of supersymmetry breaking, most of the squark masses are larger than
about $0.8 m_{\tilde g}$. Then the two-loop contribution to the gluino
pole mass is positive, and from 1\% to 2\% for comparable
gluino and squark masses, but it is larger when 
$m_{\tilde q} \gg m_{\tilde g}$. 

The scale-dependence of the calculated pole mass is shown in 
fig.~\ref{fig:gluinoscale}. 
%%%%%%%%%%%%%%%%%%%%%%%%%%%%%%%%%%%%%%%%%%%%%%%%%%%%%%%%%%%%%%%%%%%%%%%%%%%%
\begin{figure}[t]
\centering
\includegraphics[width=10.0cm]{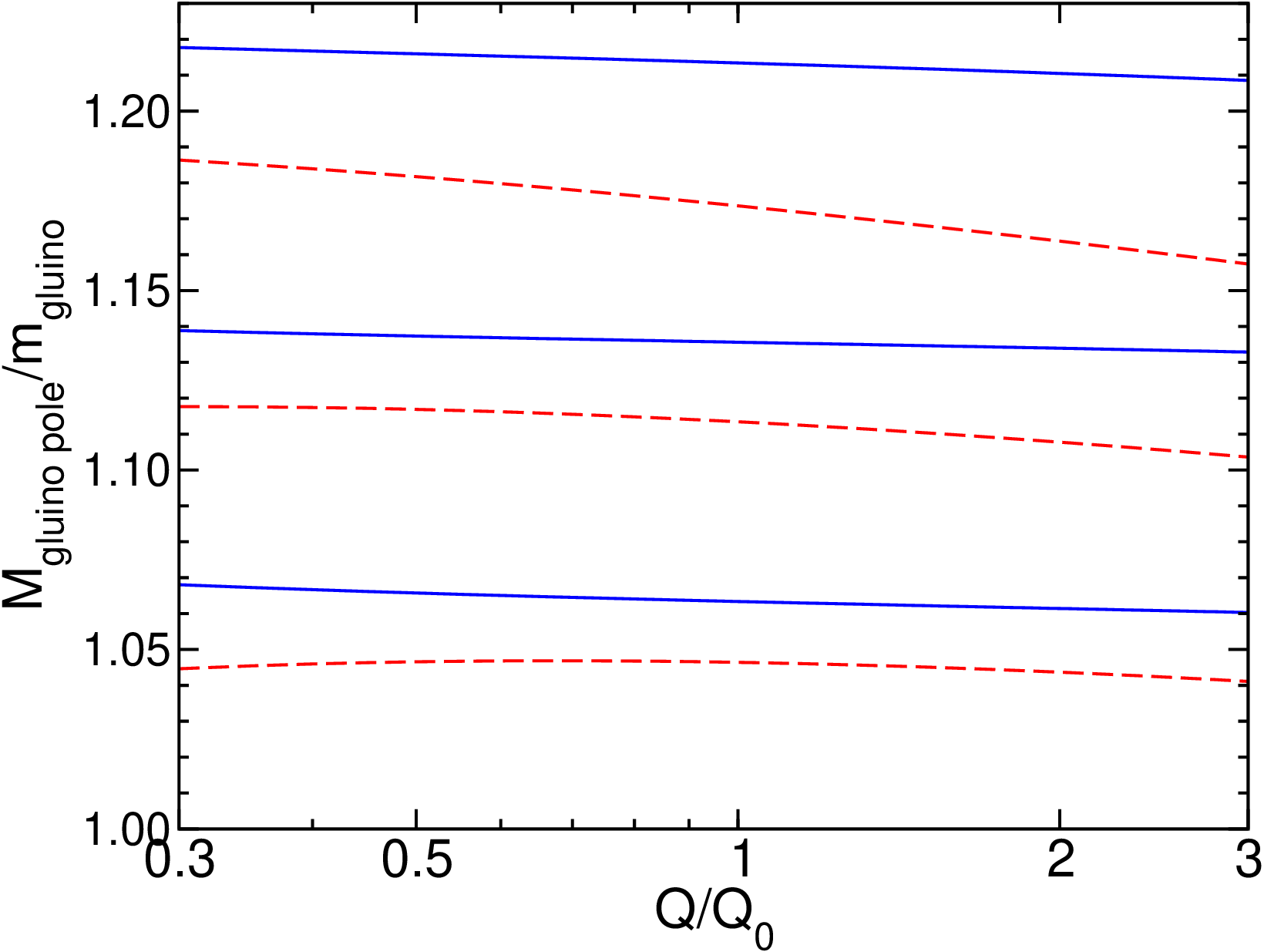}
\caption{\label{fig:gluinoscale}
The scale-dependence of the computed gluino pole 
mass $M_{{\rm gluino}\,{\rm pole}} \equiv
M_{\tilde g}$,
in the simplifying approximation of degenerate squarks with no
squark mixing or quark masses and Yukawa and electroweak effects. 
Here the top pair of lines are for
$m_{\tilde q}(Q_0)/m_{\tilde g}(Q_0) = 3$,
the middle pair for 
$m_{\tilde q}(Q_0)/m_{\tilde g}(Q_0) = 1.5$,
and the bottom pair for
$m_{\tilde q}(Q_0)/m_{\tilde g}(Q_0) = 0.9$,
where $m_{\tilde g}$ and $m_{\tilde q}$ are the running 
gluino and squark mass parameters evaluated at a reference
renormalization
scale $Q_0 = m_{\tilde g}(Q_0)$, with 
$\alpha_S(Q_0) = g_3^2/4\pi = 0.095$.
In each case, the solid line is the two-loop result, and the
dashed line is the one-loop result.
The parameters $m_{\tilde g}$, $m_{\tilde q}$, and $\alpha_S$ are each
run from the reference scale
$Q_0$ to the scale $Q$ using two-loop renormalization group equations, 
and the gluino pole mass is then recomputed.
The computations were done by specializing
eqs.~(\ref{eq:spolegluinonomix})-(\ref{eq:Pitwodgluino}) in the text.}
\end{figure}
%%%%%%%%%%%%%%%%%%%%%%%%%%%%%%%%%%%%%%%%%%%%%%%%%%%%%%%%%%%%%%%%%%%%%%%%%%%%
To make this graph, a reference renormalization scale $Q_0$ is chosen
such that the running gluino mass evaluated there is equal to it,
i.e. $Q_0 = m_{\tilde g}(Q_0)$. Then, for three different values of the
ratio $m_{\tilde q}/m_{\tilde g}$ at the scale $Q_0$, namely 0.9, 1.5 and 3.0,
the one-loop and two-loop gluino pole masses are computed as a function
of $Q$. To do this, the relevant running
parameters $m_{\tilde g}$, 
$m_{\tilde q}$,
and $\alpha_S = g_3^2/4\pi$ are evolved using their two-loop renormalization
group equations from $Q_0$ to $Q$, and then the pole mass is recomputed
using eqs.~(\ref{eq:spolegluinonomix})-(\ref{eq:Pitwodgluino}). 
In the ideal case, the lines shown would be exactly horizontal.
The scale dependence of the one-loop and two-loop results 
is about the same if the gluino mass is less than or about
equal to the squark masses (as for example in gaugino-mass dominated
or ``no-scale" models). For heavier squark masses, 
it is significantly improved by going to two-loop order.
Note that as usual, the scale-dependence of the one-loop approximation
is considerably less than the difference between the two-loop and one-loop
results. This strongly suggests 
that the scale-dependence should not be 
used
as an indicator of the accuracy of the two-loop approximation either.
A naive
estimate of the size of the three-loop SUSYQCD contribution to the gluino
pole mass can be obtained by considering the cube of the one-loop
fractional contribution, and so is perhaps of the order of a few tenths of 
a per cent.
 
As mentioned above, 
results equivalent to those above in the limit of no squark mixing have
previously been obtained independently 
by Y.~Yamada in \cite{Yamada:2005ua}, where
numerical results were given and the analytical form for the pole mass of
the gluino was shown explicitly in the limit 
$m_{\tilde q} \gg m_{\tilde g}$. 
I have checked that the results found 
here do agree with those in
ref.~\cite{Yamada:2005ua}. (When comparing the numerical results, it is 
useful
to note that small differences, formally of three-loop order, arise
due to the fact that the present paper works in terms of perturbative
corrections to the pole 
squared mass, while ref.~\cite{Yamada:2005ua} computed results for the
pole mass.)

Further accuracy can be obtained by including the contributions of Feynman
diagrams that are of order $\alpha_S$ times Yukawa or electroweak
couplings squared. Such corrections could be particularly important 
if the gluino is relatively light, since we are working in a 
non-decoupling scheme. These results are implicitly 
contained above in section
\ref{sec:twoloop}; obtaining their explicit form is only a matter of 
plugging
in the couplings and masses of the MSSM. I will not do that here,
because this paper already contains plenty of lengthy formulas,
but only note that it can be done straightforwardly by use of a
symbolic manipulation program, for example. 

\subsection{SUSYQCD corrections to quark masses in minimal supersymmetry}

As another application of the general results above, consider the relation
between the running and pole masses of the quarks in the MSSM,
found earlier in \cite{quarkpoleSUSYa,quarkpoleSUSYb,Bednyakov:2005kt}. 
Using the results of the present paper, I find for the top quark:
\beq
M^2_{t} - i \Gamma_t M_t &=&
m^2_{t}
+ \frac{1}{16\pi^2} \widetilde\Pi^{(1)}_{t}
+ \frac{g_3^4}{(16\pi^2)^2} \widetilde\Pi^{(2)}_{t}
\eeq
%See Mathematica file topgen and pages 128-134 of notes
where:
\beq
\widetilde\Pi^{(1)}_{t} &=&
g_3^2 C_q \bigl\lbrace
m_t^2( 10 - 6 \lnbar m_t^2) 
+ 2 \propB_{FS}(\tilde g, \tilde t_j) 
- 4 {\rm Re}[ L_{\tilde t_j} R_{\tilde t_j}^*] m_t m_{\tilde g}
 \propB_{\Fbar S}(\tilde g, \tilde t_j) 
\bigr \rbrace
+ 2 |Y_{t\overline t \phi_j^0}|^2 \propB_{FS}(t,\phi_j^0)
\nonumber \\ &&
+ 2 {\rm Re} [(Y_{t\overline t \phi_j^0})^2] \propB_{\Fbar S}(t,\phi_j^0)
+ (|Y_{\overline t b \phi_j^+}|^2 + |Y_{\overline b t \phi_j^-}|^2)
 \propB_{FS}(b,\phi_j^+)
+ 2 {\rm Re}[Y_{\overline t b \phi_j^+} Y_{\overline b t \phi_j^-}]
m_b m_t \propB_{\Fbar S} (b, \phi_j^+)
\nonumber \\ &&
+ (|Y_{t \tilde N_i \tilde t_j^*}|^2 + 
   |Y_{\overline t \tilde N_i \tilde t_j}|^2 ) 
   \propB_{FS}(\tilde N_i,\tilde t_j)
+ 2 {\rm Re}[ Y_{t \tilde N_i \tilde t_j^*} 
   Y_{\overline t \tilde N_i \tilde t_j}] m_t m_{\tilde N_i}
   \propB_{\Fbar S}(\tilde N_i,\tilde t_j)
\nonumber \\ &&
+ (|Y_{t \tilde C_i \tilde b_j^*}|^2 + 
   |Y_{\overline t \tilde C_i \tilde b_j}|^2 ) 
   \propB_{FS}(\tilde C_i,\tilde b_j)
+ 2 {\rm Re}[ Y_{t \tilde C_i \tilde b_j^*} 
   Y_{\overline t \tilde C_i \tilde b_j}] m_t m_{\tilde C_i}
   \propB_{\Fbar S}(\tilde C_i,\tilde b_j)
\nonumber \\ &&
+\frac{4}{9} e^2 m_t^2 [10 - 6 \lnbar m_t^2] 
+ \bigl [
(g^2 + g^{\prime 2})/4 - 2 Q_t T_3^t g^{\prime 2}
+ 2 Q_t^2 g^{\prime 4}/(g^2 + g^{\prime 2})
\bigr ] \propB_{FV}(t,Z)
\nonumber \\ &&
+ 2 Q_t g^{\prime 2} 
\bigl [Q_t g^{\prime 2}/(g^2 + g^{\prime 2}) - T_3^t \bigr ]
\propB_{\Fbar V}(t,Z)
+ g^2 \propB_{FV}(b,W)/2
,
\label{eq:toponeloop}
\\
\widetilde\Pi^{(2)}_{t} &=&
C_q (8 C_q - 4 C_G) \Bigl \lbrace
 L_{\tilde t_j} R_{\tilde t_j}^* L_{\tilde t_k}^* R_{\tilde t_k}
   \propM_{SFFSF} (\tilde t_j,\tilde g,\tilde g,\tilde t_k,t)
+ |L_{\tilde t_j} R_{\tilde t_k}|^2 m_t^2
   \propM_{SFFS\Fbar} (\tilde t_j,\tilde g,\tilde g,\tilde t_k,t)
\nonumber \\ &&
- {\rm Re}[L_{\tilde t_k} R_{\tilde t_k}^*] m_t m_{\tilde g}
   \propM_{SF\Fbar SF} (\tilde t_j,\tilde g,\tilde g,\tilde t_k,t)
+ \frac{1}{2} (|L_{\tilde t_j} L_{\tilde t_k}|^2  
             + |R_{\tilde t_j} R_{\tilde t_k}|^2) m_{\tilde g}^2
   \propM_{S\Fbar\Fbar SF} (\tilde t_j,\tilde g,\tilde g,\tilde t_k,t)
\nonumber \\ &&
- {\rm Re}[L_{\tilde t_j} R_{\tilde t_j}^*] m_t m_{\tilde g}
   \propM_{SF\Fbar S\Fbar} (\tilde t_j,\tilde g,\tilde g,\tilde t_k,t)
+ {\rm Re}[ L_{\tilde t_j} R_{\tilde t_j}^* 
            L_{\tilde t_k} R_{\tilde t_k}^*] m_t^2 m_{\tilde g}^2
   \propM_{S\Fbar\Fbar S\Fbar} (\tilde t_j,\tilde g,\tilde g,\tilde t_k,t)
\Bigr \rbrace
\nonumber \\ &&
+ 8 C_q I_q  \Bigl \lbrace
  \frac{1}{2} \propV_{SFFFS}(\tilde t_j, \tilde g, \tilde g, q, \tilde q_k)
+ 2 {\rm Re}[L_{\tilde t_j} R_{\tilde t_j}^* 
             L_{\tilde q_k}^* R_{\tilde q_k}] m_t m_q
  \propV_{SFF\Fbar S}(\tilde t_j, \tilde g, \tilde g, q, \tilde q_k)
\nonumber \\ &&
- 2 {\rm Re}[L_{\tilde t_j} R_{\tilde t_j}^* ] m_t m_{\tilde g}
  \propV_{SF\Fbar FS}(\tilde t_j, \tilde g, \tilde g, q, \tilde q_k)
+ \frac{1}{2} m_{\tilde g}^2 
  \propV_{S\Fbar\Fbar FS}(\tilde t_j, \tilde g, \tilde g, q, \tilde q_k)
\nonumber \\ &&
- 2 {\rm Re}[L_{\tilde q_k} R_{\tilde q_k}^* ] m_q m_{\tilde g}
  \propV_{SF\Fbar\Fbar S}(\tilde t_j, \tilde g, \tilde g, q, \tilde q_k)  
+ 2 {\rm Re}[L_{\tilde t_j} R_{\tilde t_j}^* 
             L_{\tilde q_k} R_{\tilde q_k}^* ] m_q m_t m^2_{\tilde g}
  \propV_{S\Fbar\Fbar\Fbar S}(\tilde t_j, \tilde g, \tilde g, q, \tilde q_k)  
\Bigr \rbrace
\nonumber \\ &&
+ 4 C_q^2 \Bigl \lbrace
  \propV_{FSSFF} (\tilde g, \tilde t_j, \tilde t_j, t, \tilde g)
- 2 {\rm Re}[L_{\tilde t_j} R_{\tilde t_j}^*] m_t m_{\tilde g}
  [\propV_{\Fbar SSFF} (\tilde g, \tilde t_j, \tilde t_j, t, \tilde g)
  +\propV_{FSS\Fbar\Fbar} (\tilde g, \tilde t_j, \tilde t_j, t, \tilde g)]
\nonumber \\ &&
+ 2 ({\rm Re}[L_{\tilde t_j} R_{\tilde t_j}^* 
            L_{\tilde t_k} R_{\tilde t_k}^*]
         + |L_{\tilde t_j} R_{\tilde t_k}|^2) m_t^2 m_{\tilde g}^2
  \propV_{\Fbar SS\Fbar\Fbar} (\tilde g, \tilde t_j, \tilde t_k, t, \tilde g)]
%\nonumber \\ &&
+ \frac{1}{2} |L_{\tilde t_j} L_{\tilde t_k}^*
              -R_{\tilde t_j} R_{\tilde t_k}^*|^2
  \propY_{FSSS}(\tilde g, \tilde t_j,  \tilde t_j, \tilde t_k)
\nonumber \\ &&
- {\rm Re}[L_{\tilde t_j}^* R_{\tilde t_k}
  (L_{\tilde t_j} L_{\tilde t_n}^*-R_{\tilde t_j} R_{\tilde t_n}^*)
  (L_{\tilde t_n} L_{\tilde t_k}^*-R_{\tilde t_n} R_{\tilde t_k}^*)]
  m_t m_{\tilde g}
  \propY_{\Fbar SSS}(\tilde g, \tilde t_j,  \tilde t_k, \tilde t_n)
\nonumber \\ &&
+ 
\bigl \lbrace \propB_{FS}(\tilde g, \tilde t_j) 
- 2 {\rm Re}[L_{\tilde t_j} R_{\tilde t_j}^*] 
m_t m_{\tilde g} \propB_{\Fbar S}(\tilde g, \tilde t_j)
\bigr \rbrace
\bigl \lbrace
\propB'_{FS}(\tilde g, \tilde t_k) 
- 2 {\rm Re}[L_{\tilde t_k} R_{\tilde t_k}^*]
m_t m_{\tilde g} \propB'_{\Fbar S}(\tilde g, \tilde t_k)
\bigr \rbrace
\nonumber \\ &&
+ {\rm Re}[L_{\tilde t_j} R_{\tilde t_j}^*]
  {\rm Re}[L_{\tilde t_k} R_{\tilde t_k}^*] m_{\tilde g}^2 
\propB_{\Fbar S}(\tilde g, \tilde t_j)
\propB_{\Fbar S}(\tilde g, \tilde t_k)
- |L_{\tilde t_j} R_{\tilde t_k}|^2 
  \propB_{FS}(\tilde g, \tilde t_j)
  \propB_{FS}(\tilde g, \tilde t_k)/m_t^2
\Bigr \rbrace
%
%\\
%
%\widetilde\Pi^{(2,1)}_{t} &=&
\nonumber \\ && +
C_G C_q f_1(t, \tilde g, \tilde t_j)
+ 2(C_G -C_q) C_q f_2(t, \tilde g, \tilde t_j)
+ 2 C_q^2 f_3(t, \tilde g, \tilde t_j) 
\nonumber \\ &&
- 2 C_q {\rm Re}[L_{\tilde t_j} R_{\tilde t_j}^*]
m_t m_{\tilde g} 
\bigl \lbrace 
C_G f_4(t, \tilde g, \tilde t_j) 
+ 2(C_G -C_q) f_5(t, \tilde g, \tilde t_j)
+ 2 C_q f_6(t, \tilde g, \tilde t_j)
\bigr \rbrace
%
%\\
%
%\widetilde\Pi^{(2,2)}_{t} &=&
\nonumber \\ && +
C_q^2 F_1(t) 
+ C_G C_q \bigl [F_2(t) + F_3 (t, \tilde g)\bigr ]
+ C_q I_q \bigl [2 F_3 (t, q) + F_4 (t, \tilde q_j) \bigr ]
.
\label{eq:toptwoloop}
\eeq
The gauge group constants are $C_q = 4/3$, $C_G = 3$, $I_q = 1/2$,
$Q_t = 2/3$ and $T_3^t = 1/2$. The results for the bottom quark can be
obtained by taking $t \leftrightarrow b$ everywhere, with $Q_t 
\rightarrow Q_b = -1/3$
and $T_3^t \rightarrow T_3^b = -1/2$.
The one-loop part given by eq.~(\ref{eq:toponeloop}) was given 
in a different notation in
\cite{PBMZ}. It includes all 
effects, including those due to the
exchange of virtual
neutral Higgs scalars [$\phi^0_j = (h^0, H^0, G^0, A^0)$ for $j=1,2,3,4$]
charged Higgs scalars [$\phi^\pm_j = (G^\pm, H^\pm)$ for $j=1,2$],
neutralinos [$\tilde N_j$ for  $j=1,2,3,4$],
and charginos [$\tilde C_j$ for  $j=1,2$] and top and bottom squarks. In 
each of the loop integral
functions in eqs.~(\ref{eq:toponeloop}) and (\ref{eq:toptwoloop}), one 
should take $s = m_t^2$ with 
an infinitesimal positive 
imaginary part. [Here I have only listed the pure SUSYQCD corrections
explicitly in the two-loop part, but 
the corrections involving Yukawa 
couplings and scalar trilinear couplings are also given 
implicitly by specializing the results of section
\ref{sec:twoloop}.]
 
I have checked that the result given above is consistent with the two-loop
renormalization group equations, in the sense that the pole mass is
invariant under changes in the renormalization scale given by the
two-loop SUSYQCD renormalization group equation for the top quark mass.
As another non-trivial
consistency check, I have verified that in the (clearly unrealistic)
supersymmetric limit, the top-quark pole mass given above is precisely
equal to the top-squark pole mass as found in eqs.~(5.28)-(5.30) of
ref.~\cite{Martin:2005eg}. [In the published and original preprint
versions of that paper, the term in eq.~(5.30)
proportional to $I_q$ was missing a factor of 2.]

The two-loop SUSYQCD contribution and Yukawa contributions had already 
been found in 
refs.~\cite{quarkpoleSUSYa,quarkpoleSUSYb,Bednyakov:2005kt},
using a method in which loop integrals are evaluated using an expansion in
small mass hierarchies.  In principle, the present paper generalizes this
somewhat, since here I use two-loop integral basis functions 
without mass expansions. However, as was
recently pointed out in \cite{Bednyakov:2005kt}, 
the actual top and bottom quark
masses are such that
the leading terms in the mass expansion already give an extremely 
good
approximation throughout most of the parameter space left available
to supersymmetry. 
I have also checked agreement with eqs.~(57)-(62) in 
ref.~\cite{quarkpoleSUSYa}.
References 
\cite{quarkpoleSUSYa,quarkpoleSUSYb,Bednyakov:2005kt} also 
include 
an extensive study of the impact of the
two-loop top and bottom quark mass corrections in the MSSM. 

\subsection{SUSYQCD corrections to neutralino and chargino masses}

In this section, 
I present the two-loop corrections to neutralino and chargino masses that
involve the strong coupling. These arise from gluon and gluino propagator
lines added to the one-loop Feynman diagrams that involve quarks and squarks,
and so are parametrically of order $\alpha_S y_t^2$, $\alpha_S y_b^2$,
$\alpha_S g^2$, $\alpha_S g g'$ and $\alpha_S g^{\prime 2}$. 
These two-loop contributions are evaluated by specializing the results above,
and do not require the neglect of $W$ and $Z$ boson masses.

Using the general results of sections
\ref{sec:oneloop} and \ref{sec:twoloop}, one can write the neutralino pole 
masses as:
\beq
M_{\tilde N_i}^2 - i \Gamma_{\tilde N_i} M_{\tilde N_i} 
= m_{\tilde N_i}^2
+ \frac{1}{16 \pi^2} \widetilde \Pi^{(1)}_{\tilde N_i}
+ \frac{1}{(16 \pi^2)^2} \widetilde \Pi^{(2)}_{\tilde N_i} ,
\eeq
for $i = 1,2,3,4$, where $m_{\tilde N_i}^2$ are the tree-level
squared mass eigenvalues, and the one-loop 
part is \cite{Pierce,PBMZ},
\beq
\widetilde \Pi^{(1)}_{\tilde N_i} &=& 
2 n_f \bigl (   |Y_{f \tilde N_i \tilde f_j^*}|^2 
  + |Y_{\overline f \tilde N_i \tilde f_j}|^2 \bigr )
\propB_{FS}(f, \tilde f_j)
+ 4 n_f {\rm Re}[ Y_{f \tilde N_i \tilde f_j^*}
             Y_{\overline f \tilde N_i \tilde f_j} ]
m_{\tilde N_i} m_f
\propB_{\Fbar S}(f, \tilde f_j)
\nonumber \\ &&
+ 2 |Y_{\tilde N_i \tilde N_j \phi^0_k}|^2 
\propB_{FS} (\tilde N_j,\phi^0_k)
+ 2 {\rm Re}[(Y_{\tilde N_i \tilde N_j \phi^0_k})^2] 
m_{\tilde N_i} m_{\tilde N_j} \propB_{\Fbar S} (\tilde N_j,\phi^0_k)
\nonumber \\ &&
+ 2 (|Y_{\tilde C_j^+ \tilde N_i \phi_k^-}|^2 
  +|Y_{\tilde C_j^- \tilde N_i \phi_k^+}|^2)
\propB_{FS} (\tilde C_j,\phi_k^+)
+ 4 {\rm Re} [
Y_{\tilde C_j^+ \tilde N_i \phi_k^-} Y_{\tilde C_j^- \tilde N_i \phi_k^+}]
m_{\tilde N_i} m_{\tilde C_j} \propB_{\Fbar S} (\tilde C_j,\phi_k^+)
\nonumber \\ &&
+ 2 (g^2 + g^{\prime 2}) \left \lbrace |O^{\prime\prime L}_{ij}|^2
  \propB_{FV} (\tilde N_j,Z)
- {\rm Re} [(O^{\prime\prime L}_{ij})^2]
  m_{\tilde N_i} m_{\tilde N_j}
  \propB_{\Fbar V} (\tilde N_j,Z) \right \rbrace
\nonumber \\ &&
+ 2 g^2 (|O^{L}_{ij}|^2 + |O^{R}_{ij}|^2) \propB_{FV} (\tilde C_j, W)
+ 4 g^2 {\rm Re}[O^{L}_{ij} O^{R*}_{ij} ] 
m_{\tilde N_i} m_{\tilde C_j} \propB_{\Fbar V} (\tilde C_j, W)
,
\label{eq:oneloopN}
\eeq
%
%\\
%
and the two-loop SUSYQCD part (i.e., the part involving 
the strong interactions) is
\beq
\widetilde \Pi^{(2)}_{\tilde N_i} &=& 
16 g_3^2 \Bigl \lbrace
\frac{1}{2} (|Y_{q \tilde N_i \tilde q_j^*}|^2 +
   |Y_{\overline q \tilde N_i \tilde q_j}|^2 )
   f_1(\tilde N_i, q, \tilde q_j)
   + {\rm Re} [Y_{q \tilde N_i \tilde q_j^*}
   Y_{\overline q \tilde N_i \tilde q_j}] m_{\tilde N_i} m_q 
   f_4(\tilde N_i, q, \tilde q_j) 
\nonumber \\ &&
-2 {\rm Re}[
Y_{q \tilde N_i \tilde q_j^*}
Y_{\overline q \tilde N_i \tilde q_k}^* 
R_{\tilde q_j} L_{\tilde q_k}
]
\propM_{SFFSF} (\tilde q_j, q,q, \tilde q_k, \tilde g)
%\nonumber \\ &&
-2 {\rm Re}[
Y_{q \tilde N_i \tilde q_j^*}
Y_{\overline q \tilde N_i \tilde q_k}^* 
L_{\tilde q_j} R_{\tilde q_k}
] m_q^2
\propM_{S\Fbar\Fbar SF} (\tilde q_j, q,q, \tilde q_k, \tilde g)
\nonumber \\ &&
+ {\rm Re}[
  Y_{q \tilde N_i \tilde q_j^*} Y_{q \tilde N_i \tilde q_k^*} 
  L_{\tilde q_j} L_{\tilde q_k}
+ Y_{\overline q \tilde N_i \tilde q_j} Y_{\overline q \tilde N_i \tilde q_k} 
  R_{\tilde q_j}^* R_{\tilde q_k}^*
] m_{\tilde N_i} m_{\tilde g}
\propM_{SFFS \Fbar} (\tilde q_j, q,q, \tilde q_k, \tilde g)
\nonumber \\ &&
- 2 {\rm Re}[
Y_{q \tilde N_i \tilde q_j^*} Y_{q \tilde N_i \tilde q_k^*} 
  L_{\tilde q_j} R_{\tilde q_k}
+ Y_{\overline q \tilde N_i \tilde q_j} Y_{\overline q \tilde N_i \tilde q_k} 
  R_{\tilde q_j}^* L_{\tilde q_k}^*
] m_{\tilde N_i} m_{q}
\propM_{SF\Fbar SF} (\tilde q_j, q,q, \tilde q_k, \tilde g)
\nonumber \\ &&
+2 {\rm Re}[
Y_{q \tilde N_i \tilde q_j^*} Y_{\overline q \tilde N_i \tilde q_k}^* 
  R_{\tilde q_j} R_{\tilde q_k}
+ Y_{\overline q \tilde N_i \tilde q_j}^* Y_{q \tilde N_i \tilde q_k^*} 
  L_{\tilde q_j} L_{\tilde q_k}
] m_q m_{\tilde g} 
\propM_{SF\Fbar S\Fbar} (\tilde q_j, q,q, \tilde q_k, \tilde g)
\nonumber \\ &&
+ {\rm Re}[
Y_{q \tilde N_i \tilde q_j^*} Y_{q \tilde N_i \tilde q_k^*} 
  R_{\tilde q_j} R_{\tilde q_k}
+ Y_{\overline q \tilde N_i \tilde q_j} Y_{\overline q \tilde N_i \tilde q_k} 
  L_{\tilde q_j}^* L_{\tilde q_k}^*
] m_{\tilde N_i} m_{\tilde g} m_q^2
\propM_{S\Fbar\Fbar S \Fbar} (\tilde q_j, q,q, \tilde q_k, \tilde g)
\nonumber \\ &&
+ 
(|Y_{q \tilde N_i \tilde q_j^*}|^2 
+ |Y_{\overline q \tilde N_i \tilde q_j}|^2)
\propV_{FSSFF}(q, \tilde q_j, \tilde q_j, \tilde g, q)
+ 2 {\rm Re}[Y_{q \tilde N_i \tilde q_j^*}
Y_{\overline q \tilde N_i \tilde q_j}]
m_{\tilde N_i} m_q 
\propV_{\Fbar SSFF}(q, \tilde q_j, \tilde q_j, \tilde g, q)
\nonumber \\ &&
- 2 {\rm Re}[ 
(Y_{q \tilde N_i \tilde q_j^*} Y_{q \tilde N_i \tilde q_k^*}^*
+Y_{\overline q \tilde N_i \tilde q_j}^* 
Y_{\overline q \tilde N_i \tilde q_k} )
R_{\tilde q_j} L_{\tilde q_k}^*]
m_q m_{\tilde g}
\propV_{FSS\Fbar\Fbar}(q, \tilde q_j, \tilde q_k, \tilde g, q)
\nonumber \\ &&
- 2 {\rm Re}[
Y_{q \tilde N_i \tilde q_j^*} Y_{\overline q \tilde N_i \tilde q_k} 
(R_{\tilde q_j} L_{\tilde q_k}^* + L_{\tilde q_j} R_{\tilde q_k}^*)]
m_{\tilde N_i} m_q^2 m_{\tilde g}
\propV_{\Fbar SS\Fbar\Fbar}(q, \tilde q_j, \tilde q_k, \tilde g, q)
\nonumber \\ &&
+ (|Y_{q \tilde N_i \tilde q_j^*} L_{\tilde q_k}|^2
+  |Y_{\overline q \tilde N_i \tilde q_j} R_{\tilde q_k}|^2)
\propV_{SFFFS} (\tilde q_j,q,q,\tilde g,\tilde q_k)
%\nonumber \\ &&
+ 2 {\rm Re}[
Y_{q \tilde N_i \tilde q_j^*} Y_{\overline q \tilde N_i \tilde q_j}
] m_{\tilde N_i} m_q
\propV_{SF\Fbar FS} (\tilde q_j,q,q,\tilde g,\tilde q_k)
\nonumber \\ &&
- 2 {\rm Re}[
Y_{q \tilde N_i \tilde q_j^*} Y_{\overline q \tilde N_i \tilde q_j}
L_{\tilde q_k}  R_{\tilde q_k}^*] m_{\tilde N_i} m_{\tilde g}
\propV_{SFF\Fbar S} (\tilde q_j,q,q,\tilde g,\tilde q_k)
\nonumber \\ &&
+ (|Y_{q \tilde N_i \tilde q_j^*} R_{\tilde q_k}|^2
+  |Y_{\overline q \tilde N_i \tilde q_j} L_{\tilde q_k}|^2)
m_q^2
\propV_{S\Fbar\Fbar FS} (\tilde q_j,q,q,\tilde g,\tilde q_k)
\nonumber \\ &&
-2  (
|Y_{q \tilde N_i \tilde q_j^*}|^2 +
|Y_{\overline q \tilde N_i \tilde q_j}|^2 ) 
{\rm Re}[L_{\tilde q_k}  R_{\tilde q_k}^*]
m_q m_{\tilde g} 
\propV_{SF\Fbar\Fbar S} (\tilde q_j,q,q,\tilde g,\tilde q_k)
\nonumber \\ &&
-2 {\rm Re}[Y_{q \tilde N_i \tilde q_j^*}Y_{\overline q \tilde N_i \tilde q_j}
 R_{\tilde q_k}L_{\tilde q_k}^*]
m_{\tilde N_i} m_q^2 m_{\tilde g}
\propV_{S\Fbar\Fbar\Fbar S} (\tilde q_j,q,q,\tilde g,\tilde q_k)
\nonumber \\ &&
+ \frac{1}{2} 
(L_{\tilde q_j} L_{\tilde q_n}^* -R_{\tilde q_j} R_{\tilde q_n}^*)
(L_{\tilde q_n} L_{\tilde q_k}^* -R_{\tilde q_n} R_{\tilde q_k}^*) 
\bigl [ 
(Y_{q\tilde N_i \tilde q_j^*} Y_{q\tilde N_i \tilde q_k^*}^*
+ Y_{\overline q \tilde N_i \tilde q_j}^* 
Y_{\overline q \tilde N_i \tilde q_k})
\propY_{FSSS}(q,\tilde q_j, \tilde q_k, \tilde q_n)
\nonumber \\ &&
+ (Y_{q\tilde N_i \tilde q_j^*} Y_{\overline q\tilde N_i \tilde q_k}
+  Y_{\overline q \tilde N_i \tilde q_j}^* Y_{q \tilde N_i \tilde q_k^*}^*)
m_{\tilde N_i} m_q
\propY_{\Fbar SSS}(q,\tilde q_j, \tilde q_k, \tilde q_n)
\bigr ]
\Bigr \rbrace .
\label{eq:twoloopN}
\eeq
In eq.~(\ref{eq:oneloopN}), $f$ is summed over the symbols
$e$, $\mu$, $\tau$, $\nu_e$, $\nu_\mu$, $\nu_\tau$,
$u$, $d$, $c$, $s$, $t$, $b$, with $n_f = 1$ for leptons and $3$ for
quarks, and in eq.~(\ref{eq:twoloopN}) the symbol $q$ is summed over
$u$, $d$, $c$, $s$, $t$, $b$. 
The 
indices $j,k,n$ are each summed over the appropriate ranges ($1,2$ 
for squarks, sleptons, charginos, and charged Higgs scalars, $1,2,3,4$
for neutralinos and neutral Higgs scalars, including the Goldstone
bosons) 
wherever they appear. The masses and couplings appearing on the
right-hand side are always running renormalized $\DRbarprime$ parameters.
In all 
of the self-energy functions appearing in eqs.~(\ref{eq:oneloopN}) and
(\ref{eq:twoloopN}), 
the external momentum invariant is $s = m^2_{\tilde N_i}$
with an infinitesimal positive imaginary part.

%\subsection{Two-loop SUSYQCD corrections to chargino masses}

Similarly, the pole masses for charginos in the MSSM can be written
as
\beq
M_{\tilde C_i}^2 -i \Gamma_{\tilde C_i} M_{\tilde C_i}
 = m_{\tilde C_i}^2
+ \frac{1}{16 \pi^2} \widetilde \Pi^{(1)}_{\tilde C_i}
+ \frac{1}{(16 \pi^2)^2} \widetilde \Pi^{(2)}_{\tilde C_i} ,
\eeq
for $i = 1,2$, where $ m_{\tilde C_i}^2$ are the tree-level
squared mass eigenvalues, and the one-loop
part is \cite{Pierce,PBMZ},
\beq
\widetilde \Pi^{(1)}_{\tilde C_i} &=& 
n_f
(|Y_{f\tilde C_i \tilde F_j^*}|^2 + |Y_{\overline f \tilde C_i \tilde F_j}|^2)
\propB_{FS} (f, \tilde F_j)
+ 2 n_f
{\rm Re}[Y_{f\tilde C_i \tilde F_j^*}Y_{\overline f \tilde C_i \tilde F_j}]
m_{\tilde C_i} m_f \propB_{\Fbar S} (f, \tilde F_j)
\nonumber \\ &&
+ (|Y_{\tilde C_i^+\tilde N_j \phi_k^-}|^2 
  +|Y_{\tilde C_i^-\tilde N_j \phi_k^+}|^2)
\propB_{FS} (\tilde N_j, \phi_k^+)
+ 2 {\rm Re}[Y_{\tilde C_i^+\tilde N_j \phi_k^-}
             Y_{\tilde C_i^-\tilde N_j \phi_k^+}]
m_{\tilde C_i} m_{\tilde N_j} \propB_{\Fbar S} (\tilde N_j, \phi_k^+)
\nonumber \\ &&
+ (|Y_{\tilde C_i^+\tilde C_j^- \phi_k^0}|^2 
  +|Y_{\tilde C_j^+\tilde C_i^- \phi_k^0}|^2)
\propB_{FS} (\tilde C_j, \phi_k^0)
+ 2 {\rm Re}[Y_{\tilde C_i^+\tilde C_j^- \phi_k^0}
             Y_{\tilde C_j^+\tilde C_i^- \phi_k^0}]
m_{\tilde C_i} m_{\tilde C_j} \propB_{\Fbar S} (\tilde C_j, \phi_k^0)
\nonumber \\ && 
+ e^2 m^2_{\tilde C_i} [10 - 6 \lnbar m^2_{\tilde C_i}]
+ g^2 (|O^L_{ji}|^2 + |O^R_{ji}|^2) \propB_{FV} (\tilde N_j, W)
+ 2 g^2 {\rm Re}[O^L_{ji} O^{R*}_{ji}] m_{\tilde C_i} m_{\tilde N_j}
\propB_{\Fbar V} (\tilde N_j, W)
\nonumber \\ && 
+ (g^2 + g^{\prime 2}) \bigl [
(|O^{\prime L}_{ij}|^2 + |O^{\prime R}_{ij}|^2) 
\propB_{FV}(\tilde C_j, Z)
+ 2 {\rm Re}[O^{\prime L}_{ji} O^{\prime R}_{ij}]
m_{\tilde C_i} m_{\tilde C_j}
\propB_{FV}(\tilde C_j, Z) \bigr ],
\label{eq:oneloopC}
\eeq
%
%\\
%
and the two-loop part involving the strong interaction is
\beq
\widetilde \Pi^{(2)}_{\tilde C_i} &=& 8 g_3^2 \Bigl \lbrace
\frac{1}{2} (
|Y_{q\tilde C_i \tilde Q_j^*}|^2 
+|Y_{\overline q\tilde C_i \tilde Q_j}|^2 ) f_1 (\tilde C_i,q,\tilde Q_j)
+ {\rm Re}[ Y_{q\tilde C_i \tilde Q_j^*}Y_{\overline q\tilde C_i \tilde Q_j}]
m_{\tilde C_i} m_q f_4(\tilde C_i,q,\tilde Q_j)
\nonumber \\ &&
- 2 {\rm Re} [
Y_{q\tilde C_i \tilde Q_j^*}Y_{\overline Q \tilde C_i \tilde q_k}^*
R_{\tilde Q_j} L_{\tilde q_k}] 
\propM_{SFFSF}(\tilde Q_j, Q, q, \tilde q_k, \tilde g)
\nonumber \\ &&
+ {\rm Re} [
Y_{q\tilde C_i \tilde Q_j^*} Y_{Q\tilde C_i \tilde q_k^*}
L_{\tilde Q_j} L_{\tilde q_k}
+ Y_{\overline q\tilde C_i \tilde Q_j} Y_{\overline Q\tilde C_i \tilde q_k}
R_{\tilde Q_j}^* R_{\tilde q_k}^*
] m_{\tilde C_i} m_{\tilde g}
\propM_{SFFS\Fbar}(\tilde Q_j, Q, q, \tilde q_k, \tilde g)
\nonumber \\ &&
- 2 {\rm Re}[
Y_{q\tilde C_i \tilde Q_j^*} Y_{Q\tilde C_i \tilde q_k^*}
L_{\tilde Q_j} R_{\tilde q_k}
+ Y_{\overline q\tilde C_i \tilde Q_j} Y_{\overline Q\tilde C_i \tilde q_k}
R_{\tilde Q_j}^* L_{\tilde q_k}^*]
m_{\tilde C_i} m_{q}
\propM_{SF\Fbar SF}(\tilde Q_j, Q, q, \tilde q_k, \tilde g)
\nonumber \\ &&
-2 {\rm Re} [
Y_{q\tilde C_i \tilde Q_j^*} Y_{\overline Q\tilde C_i \tilde q_k}^*
L_{\tilde Q_j} R_{\tilde q_k}]
m_q m_Q
\propM_{S\Fbar\Fbar SF}(\tilde Q_j, Q, q, \tilde q_k, \tilde g)
\nonumber \\ &&
+ 2 {\rm Re} [
Y_{q\tilde C_i \tilde Q_j^*} Y_{\overline Q\tilde C_i \tilde q_k}^*
R_{\tilde Q_j} R_{\tilde q_k}
+
Y_{\overline q\tilde C_i \tilde Q_j}^* Y_{Q\tilde C_i \tilde q_k^*}
L_{\tilde Q_j} L_{\tilde q_k}]
m_q m_{\tilde g}
\propM_{SF\Fbar S\Fbar}(\tilde Q_j, Q, q, \tilde q_k, \tilde g)
\nonumber \\ &&
+ {\rm Re}[
Y_{q\tilde C_i \tilde Q_j^*} Y_{Q\tilde C_i \tilde q_k^*}
R_{\tilde Q_j} R_{\tilde q_k}
+ Y_{\overline q\tilde C_i \tilde Q_j} Y_{\overline Q\tilde C_i \tilde q_k}
L_{\tilde Q_j}^* L_{\tilde q_k}^*
] m_{\tilde C_i} m_q m_Q m_{\tilde g}
\propM_{S\Fbar\Fbar S\Fbar}(\tilde Q_j, Q, q, \tilde q_k, \tilde g)
\nonumber \\ &&
+ (|Y_{q\tilde C_i \tilde Q_j^*}|^2 + |Y_{\overline q\tilde C_i \tilde Q_j}|^2)
\propV_{FSSFF} (q, \tilde Q_j, \tilde Q_j, \tilde g, Q)
+ 2 {\rm Re} [Y_{q\tilde C_i \tilde Q_j^*} Y_{\overline q\tilde C_i \tilde Q_j}]
m_{\tilde C_i} m_q
\propV_{\Fbar SSFF} (q, \tilde Q_j, \tilde Q_j, \tilde g, Q)
\nonumber \\ &&
- 2 {\rm Re} [
(Y_{q\tilde C_i \tilde Q_j^*} Y_{q\tilde C_i \tilde Q_k^*}^*
+Y_{\overline q\tilde C_i \tilde Q_j}^* Y_{\overline q\tilde C_i \tilde Q_k})
L_{\tilde Q_j} R^*_{\tilde Q_k}
]
m_Q m_{\tilde g} 
\propV_{FSS\Fbar\Fbar} (q, \tilde Q_j, \tilde Q_k, \tilde g, Q)
\nonumber \\ &&
- 2 {\rm Re} [
Y_{q\tilde C_i \tilde Q_j^*} Y_{\overline q\tilde C_i \tilde Q_k} 
(L_{\tilde Q_j} R^*_{\tilde Q_k} +
 R_{\tilde Q_j} L^*_{\tilde Q_k})]
m_{\tilde C_i} m_q m_Q m_{\tilde g} 
\propV_{\Fbar SS\Fbar\Fbar} (q, \tilde Q_j, \tilde Q_k, \tilde g, Q)
\nonumber \\ &&
+ (|Y_{q\tilde C_i \tilde Q_j^*} L_{\tilde q_k}|^2 
+ |Y_{\overline q\tilde C_i \tilde Q_j} R_{\tilde q_k}|^2 )
\propV_{SFFFS} (\tilde Q_j,q,q,\tilde g, \tilde q_k)
+ 2 {\rm Re} [
Y_{q\tilde C_i \tilde Q_j^*} Y_{\overline q\tilde C_i \tilde Q_j} ]
m_{\tilde C_i} m_q
\propV_{SF\Fbar FS} (\tilde Q_j,q,q,\tilde g, \tilde q_k)
\nonumber \\ &&
- 2 {\rm Re} [
Y_{q\tilde C_i \tilde Q_j^*} Y_{\overline q\tilde C_i \tilde Q_j} 
L_{\tilde q_k} R^*_{\tilde q_k}]
m_{\tilde C_i} m_{\tilde g}
\propV_{SFF\Fbar S} (\tilde Q_j,q,q,\tilde g, \tilde q_k)
\nonumber \\ &&
+ (|Y_{q\tilde C_i \tilde Q_j^*} R_{\tilde q_k}|^2
+ |Y_{\overline q\tilde C_i \tilde Q_j} L_{\tilde q_k}|^2) m_q^2
\propV_{S\Fbar\Fbar FS} (\tilde Q_j,q,q,\tilde g, \tilde q_k)
\nonumber \\ &&
- 2 (|Y_{q\tilde C_i \tilde Q_j^*} |^2
+ |Y_{\overline q\tilde C_i \tilde Q_j} |^2) {\rm Re}[
L_{\tilde q_k} R^*_{\tilde q_k}
]
m_q m_{\tilde g}
\propV_{SF\Fbar\Fbar S} (\tilde Q_j,q,q,\tilde g, \tilde q_k)
\nonumber \\ &&
- 2 {\rm Re} [
Y_{q\tilde C_i \tilde Q_j^*}
Y_{\overline q\tilde C_i \tilde Q_j}
L^*_{\tilde q_k} R_{\tilde q_k}]
m_{\tilde C_i} m_{\tilde g} m_q^2
\propV_{S\Fbar\Fbar\Fbar S} (\tilde Q_j,q,q,\tilde g, \tilde q_k)
\nonumber \\ &&
+ \frac{1}{2} 
(L_{\tilde Q_j} L_{\tilde Q_n}^* -R_{\tilde Q_j} R_{\tilde Q_n}^*)
(L_{\tilde Q_n} L_{\tilde Q_k}^* -R_{\tilde Q_n} R_{\tilde Q_k}^*) 
\bigl [ 
(Y_{q\tilde C_i \tilde Q_j^*} Y_{q\tilde C_i \tilde Q_k^*}^*
+ Y_{\overline q \tilde C_i \tilde Q_j}^* Y_{\overline q \tilde C_i \tilde Q_k})
\propY_{FSSS}(q,\tilde Q_j, \tilde Q_k, \tilde Q_n)
\nonumber \\ &&
+ (Y_{q\tilde C_i \tilde Q_j^*} Y_{\overline q\tilde C_i \tilde Q_k}
+  Y_{\overline q \tilde C_i \tilde Q_j}^* Y_{q \tilde C_i \tilde Q_k^*}^*)
m_{\tilde C_i} m_q
\propY_{\Fbar SSS}(q,\tilde Q_j, \tilde Q_k, \tilde Q_n)
\bigr ]
\Bigr \rbrace
.
\label{eq:twoloopC}
\eeq
\end{widetext}
In the one-loop part eq.~(\ref{eq:oneloopC}), 
$n_f = 1$ for leptons and $n_f = 3$ for quarks, and the symbols
$(f,F)$ are summed over the 12 ordered pairs: 
$(e,\nu_e)$, 
$(\nu_e, e)$, 
$(\mu,\nu_\mu)$, 
$(\nu_\mu, \mu)$, 
$(\tau,\nu_\tau)$, 
$(\nu_\tau, \tau)$, 
$(u,d)$, 
$(d,u)$, 
$(c,s)$, 
$(s,c)$, 
$(t,b)$, 
$(b,t)$, while in the two-loop part eq.~(\ref{eq:twoloopC})
the symbols $(q,Q)$ are summed over the last 6 of these.
The
indices $j,k,n$ are each summed over the appropriate ranges ($1,2$
for squarks, sleptons, charginos, and charged Higgs scalars, $1,2,3,4$
for neutralinos and neutral Higgs scalars, including the Goldstone
bosons)
wherever they appear.
In all of the self-energy functions appearing in eqs.~(\ref{eq:oneloopC})
and (\ref{eq:twoloopC}),
the external momentum invariant is $s = m^2_{\tilde C_i}$
with an infinitesimal positive imaginary part.

The numerical values of the two-loop neutralino and chargino pole
masses are rather sensitive to the values of the model parameters.
Most often, they can be expected to be at least several tenths of a per cent,
but larger in some regions of parameter space. A study of the
numerical significance of these results, and other contributions to
the neutralino and chargino masses that are implicitly
given in section \ref{sec:twoloop}, is deferred to future work.

\section{Outlook}
\setcounter{equation}{0}
\setcounter{footnote}{1}

In this paper, I have presented results for radiative corrections to the
self-energy functions and pole masses of fermions at two-loop order.
The main specific motivation for this work is to allow future experimental
data on superpartner masses to be connected to hypotheses for the mechanism
of supersymmetry breaking. However, the strategy used is designed to be
more flexible, with potential application to any semi-perturbative theory
that may appear, anticipated or not, at the TeV scale.

The application to the gluino mass may be particularly crucial,
because of the relative strength of the $SU(3)_C$ gauge coupling,
and the color octet representation of the gluino. 
When one extrapolates the soft-supersymmetry breaking parameters to
very high energies \cite{Blair:2000gy}-\cite{SPheno}
using two-loop or even three-loop \cite{threefreakingloops}
renormalization group equations, the gluino mass can also have a quite strong
effect on the determination and running of other parameters. 
It has been found that the combination of
the Large Hadron Collider and a future Linear Collider may be able to pin
down the gluino mass to 1\% or so \cite{LHCILC}. In that case, the
two-loop corrections to the gluino mass will definitely be required.
In general, the two-loop contributions for other fermions are parametrically
smaller, but still worth including on the grounds that theoretical
errors should be made negligible if at all practicable, in order to
cleanly isolate the experimental implications of new data. 

For reasons of relative simplicity (and not principle), the calculations
presented in this paper have neglected the masses
of vector bosons in the two-loop part. Although this is a quite
adequate approximation for many purposes, it 
can and should be improved on in future work. Also, the calculations
have been presented here in a general form, and require specialization;
this can be non-trivial for reasons related more to fatigue in
writing and typing than to
conceptual difficulty. However, this specialization of general
results seems well-suited to symbolic manipulation programs.
In any case, the more lengthy results for specific contributions to the 
neutralino and chargino masses, for example, 
might be better placed in the innards of
computer codes of the type described in
\cite{ISAJET}-\cite{SPheno} rather than explicitly on paper.

I am grateful to Dave Robertson for valuable conversations and
collaboration on the two-loop self-energy integral computer program TSIL
(ref.~\cite{TSIL}) and Howard Haber and Herbi Dreiner for useful 
comments on
section \ref{subsec:twocompself}. 
This work was supported by the National Science
Foundation under Grant No.~PHY-0140129.

\end{document}